\documentclass[a4paper,11pt]{article}
\pdfoutput=1 

\usepackage{jheppub} 

\usepackage{graphicx}
\usepackage{hyperref}
\usepackage{pdfpages}
\numberwithin{figure}{section}
\usepackage[section]{placeins}
\usepackage{epstopdf}

\usepackage[colorinlistoftodos]{todonotes}
\usepackage{epsfig}
\usepackage{graphicx,color}
\usepackage{cancel}
\usepackage[utf8]{inputenc}

\newcommand{\be}{\begin{equation}}
\newcommand{\ee}{\end{equation}}

\newcommand{\x}{{\bf x}}
\newcommand{\bk}{{\bf k}}
\newcommand{\br}{{\bf r}}
\newcommand{\bb}{{\bf b}}
\newcommand{\bp}{{\bf p}}

\newcommand{\y}{{\bf y}}
\newcommand{\z}{{\bf z}}

\newcommand{\rmd}{{\rm d}}

\newcommand{\sT}{{\scriptscriptstyle T}}
\newcommand{\xA}{{x_{\scriptscriptstyle A}}}

\newcommand{\beq}{\begin{eqnarray}}
\newcommand{\eeq}{\end{eqnarray}}

\definecolor{red}{rgb}{1,0,0}
\definecolor{gray}{rgb}{0.5,0.5,0.5}

\usepackage{bbold}
\usepackage{tensor}
\usepackage[normalem]{ulem} 
\renewcommand\sout{\bgroup  \ULdepth=-.5ex \ULset}

\def\bea{\begin{eqnarray}}
\def\eea{\end{eqnarray}}

\title{Photoproduction of three jets in the CGC:\break gluon TMDs and dilute limit}
\author[a]{Tolga Altinoluk,}
\author[b]{Renaud Boussarie,}
\author[c]{Cyrille Marquet,}
\author[c,d]{and Pieter Taels}

\affiliation[a]{National Centre for Nuclear Research, 02-093 Warsaw, Poland}
\affiliation[b]{Physics Department, Brookhaven National Laboratory, Upton, NY 11973, USA}
\affiliation[c]{Centre de Physique Th\'eorique, \'Ecole polytechnique, CNRS, I.P. Paris, F-91128 Palaiseau, France}
\affiliation[d]{INFN, Sezione di Cagliari, Cittadella Universitaria, I-09042 Monserrato, Cagliari, Italy}

\emailAdd{tolga.altinoluk@ncbj.gov.pl}
\emailAdd{renaud.boussarie@ifj.edu.pl}
\emailAdd{cyrille.marquet@polytechnique.edu}
\emailAdd{pieter.taels@polytechnique.edu}

\date{\today}

\preprint{???}

\abstract{
We study the process $\gamma A \to q \bar{q} g+X$ in the Color Glass Condensate (CGC) effective theory. After obtaining the cross section, we consider two kinematic limits which are encompassed in our result. In the so-called correlation limit, the vector sum of the transverse momenta of the three outgoing particles is small with respect to the individual transverse momenta; the cross section then simplifies considerably and can be written in a factorized form, sensitive to both the unpolarized and linearly-polarized Weizs\"acker-Williams transverse momentum dependent gluon distribution function (gluon TMD). The second limit of the CGC cross section that we consider is the dilute limit, which we obtain after performing a weak-field expansion; we recover a typical linear-regime expression, involving a single unintegrated gluon distribution function. Using numerical simulations of the small-$x$ QCD evolution of the TMDs, we investigate the rapidity dependence of the cross section in the correlation limit.}
\begin{document}

\maketitle

\section{Introduction}
Quantum Chromodynamics (QCD) at high energies has been the subject of intense study since many years. In particular, in certain processes such as deep-inelastic lepton-proton scattering, it is possible to probe the constituents of a proton or nucleus at small values of $x$, where $x$ is the fraction of the proton's longitudinal momentum carried by the struck parton. Since the basic laws of QCD favor the emission of gluons with arbitrary small energies, at low $x$ the proton structure is dominated by the gluon distribution. The power-like rise of this gluon distribution with decreasing $x$, as predicted by the Balitsky-Fadin-Kuraev-Lipatov (BFKL) evolution equations \cite{bfkl}, is expected to be damped by gluon recombination effects once a sufficiently dense regime is reached, characterized by the dynamically generated saturation scale $Q_s$. The ensuing non-linear low-$x$ evolution equations were established both from the point of view of the dilute projectile \cite{balitsky,kovchegov}, and from that of the highly dense proton or nucleus \cite{JIMWLK}. In the latter case, the evolution can be obtained from an effective theory known as the Color Glass Condensate (CGC) \cite{cgc}. Both approaches are equivalent and are often denoted BK-JIMWLK, after their main authors Balitsky, Kovchegov, Jalilian-Marian, Iancu, McLerran, Weigert, Leonidov and Kovner. Moreover, in the low-density regime, hence at momenta above the saturation scale, the results from linear BFKL evolution are recovered. In this work, we will use the most general semiclassical description of small-$x$ physics, and refer to it as the CGC framework.

More recently, in a series of papers \cite{firstlowxTMDs} the CGC was explored in reactions characterized by two strongly ordered scales, let us call them $P$ and $q$:
\beq
\sqrt{s} \gg P \gg q \sim Q_s \gtrsim \Lambda_\mathrm{QCD}\;.
\eeq 
The scales are required to be much smaller than the very high collision energy $\sqrt{s}$, but (a priori) larger than $\Lambda_\mathrm{QCD}$ and thus perturbative. A prime example for such a reaction is the forward production of a parton pair in lepton-proton or proton-proton collisions, in the so-called correlation limit where the total transverse momentum of the produced pair $q\sim|{\bf k}_1+{\bf k}_2|$ is much smaller than the typical momentum of the individual particles $P\sim|{\bf k}_1|\sim|{\bf k}_2|$. The crucial observation that was made in Refs.~\cite{firstlowxTMDs} is that in these kinematics, the small-$x$ limit of the proton's or nucleus' transverse momentum dependent gluon distributions (gluon TMDs) is probed \cite{gluonTMDs}. Remarkably, at least for $2\to2$ processes, taking the low-$x$ limit of the calculation in the TMD framework yields the same result as taking the correlation limit of the CGC calculation.

For the processes that have been studied, the CGC therefore generalizes two different QCD regimes at low $x$: the TMD region with two ordered hard scales $q_\sT \ll P$ on the one hand, and the linear BFKL regime at low enough gluon densities (hence $q_\sT \gg Q_s $) on the other. Inspired on this, a computation scheme dubbed `Improved TMD factorization' (ITMD) has been developed in Refs.~\cite{ITMD}, which is applicable to massless $2\to2$ processes and, just like the CGC, interpolates between the low-$x$ TMD and BFKL (sometimes called High-Energy Factorization, HEF) regimes. Its advantage, however, is that it is much more suitable for numerical implementation. The proof of this scheme as an all-order kinematic twist resummation was recently established in Refs.~\cite{proofITMD}.

In addition to the conceptual interest, understanding the TMD framework within the CGC makes it possible to apply the CGC machinery to TMDs, and vice versa. For instance, in \cite{Adrian2015,Marquet:2016cgx,Marquet:2017xwy} JIMWLK evolution was applied to unpolarized and linearly polarized gluon TMDs, using the nonperturbative McLerran-Venugopalan model \cite{MV} as an initial condition. Likewise, Sudakov logarithms, which govern the TMD evolution \cite{TMDevolution}, can be resummed in a consistent way at low $x$, as was shown in \cite{lowxSudakov} and carried out in e.g. \cite{ShuYi}.

At least to leading-order accuracy, the aforementioned correspondence between the CGC and TMD frameworks in their overlapping region of validity holds for $2\to2$ processes, also when masses are included \cite{massivedijet,Adrian2015,Marquet:2017xwy}, see \cite{Petreska:2018cbf} for an overview. The present work is part of our efforts to investigate whether this is also true for more complicated processes\footnote{We should remark that recently there was also made progress in this direction from the point of view of TMDs, see Ref.~\cite{Bury:2018kvg}.}, and can be regarded as a follow up of Ref.~\cite{monster}, in which we calculated the correlation limit of three final state particles (two jets and a photon) in proton-nucleus collisions\footnote{The correlation limit of two jets and a \emph{collinear} photon was obtained in \cite{Altinoluk:2018uax}.}. In this configuration, the total transverse momentum of the outgoing particles ${\bf q}_\sT={\bf k}_1+{\bf k}_2+{\bf k}_3$ is again required to be much smaller than their individual transverse momenta  (${\bf k}_1, {\bf k}_1, {\bf k}_3$). However, in contrast to the production of two final state particles, the $2\to3$ kinematics allow us to identify not one but two small transverse sizes in coordinate space. We showed that it is still possible to take the correlation limit, although the procedure becomes more involved, and that once again one arrives at an expression which is factorized in terms of TMDs. We will show in this paper that the same argument holds for three-jet photoproduction, with the additional feature that, due to the simple color structure of the process (since there is no initial-state radiation), only the unpolarized and linearly polarized gluon TMD of the Weizsäcker-Williams type play a role. Using the results from \cite{Marquet:2017xwy}, where the JIMWLK evolution of these TMDs was performed, allows us to do a numerical study. 

The remainder of the paper is organized as follows. In section \ref{sec:dilutedense}, we give a concise derivation of the CGC cross section for the process $\gamma A \to q \bar{q} g+X$, relegating most of the details to the appendices. We analytically calculate the correlation limit of our CGC result in section \ref{correlation}, and the dilute limit to obtain the high-energy factorization in section \ref{dilute}. Subsection \ref{TMDHEF} is devoted to an additional limit of the TMD result and the HEF result, in which both coincide. Furthermore, in section \ref{numerics}, we present our numerical results, and finally we summarize our conclusions and outlook in section \ref{outlook}.

\section{\label{sec:dilutedense}CGC cross section}
The usual approach to calculate a forward particle production cross section in the CGC is to go to a dipole frame, which is justified at large enough energies and in which the perturbative ‘dressing' of the photon state takes place long before the scattering with the highly boosted proton or nucleus target. In this frame, due to Lorentz contraction the projectile sees the target as a highly dense shockwave, and the scattering between both takes place almost instantaneous at a light-cone time which, without losing generality, we can set to be $x^+=0$. Due to the high energy, it is then justified to describe the scattering in the eikonal approximation, where the hard particles do not undergo a change in transverse position upon interacting with the target's gluon fields, but only exchange transverse momentum and color. Moreover, since due to the high density these gluon fields are semiclassical, i.e. $g_s A~\sim 1$, the interactions need to be resummed which is done by the use of Wilson lines. Finally, the way one averages over the semiclassical target gluon fields reflects the nonperturbative proton structure, which we leave unspecified.

We will work in the projectile light-cone gauge $A^+=0$, and use light-cone perturbation theory (LCPT), in which the partonic cross section is defined as the expectation value of the number operator calculated in the relevant component of the outgoing photon wave function:
\beq
\label{formal_partonic_X_section}
2k_1^+2k_2^+2k_3^+ (2\pi)^9  2\pi  \delta\big( p^+-\sum_{j=1}^3k_j^+\big) 2 p^+ \frac{\rmd\sigma^{\gamma A\to qg\bar q+X}}{\rmd^3\vec{k}_1\, \rmd^3\vec{k}_2 \, \rmd^3\vec{k}_3}\nonumber\\
=\frac{1}{2}\; \tensor[_{\rm out}]{\big{\langle}}{} (\boldsymbol{\gamma})
\big[\vec{p}\big]_\lambda\big| N_{q}(\vec{k}_1)N_g(\vec{k}_2)N_{\bar q}(\vec{k}_3) \big| (\boldsymbol{\gamma})
\big[\vec{p}\big]_\lambda\tensor[]{\big\rangle}{_{\rm out}}\;,
\eeq
where the factor $1/2$ on the right hand side stems from averaging over the photon polarization $\lambda$. The vectors $\vec{k}_i\equiv (k_i^+, {\bf k}_i)$ stand for the three-momenta of the produced particles, with $k_i^+$ being the longitudinal and ${\bf k}_i$ the transverse momenta ($\vec{k}_1$ for the quark, $\vec{k}_2$ for the gluon and $\vec{k}_3$ for the antiquark), while $p^+$ is the forward component of the incoming photon (note that we assume ${\bf p}=0$). Moreover, the number operators $N_i$ for the \textit{dressed} (this will be important later on) quark, gluon and antiquark, are defined in terms of the corresponding creation and annihilation operators. For example, the gluon number operator $N_g$ is defined as 
\beq
N_g(\vec{k}_2)= a^{\dagger a}_i(\vec{k}_2)\, a^a_i({\vec{k}_2})\;,
\eeq
with $a^{\dagger a}_i(\vec{k}_2)$ ($a^b_i({\vec{k}_2})$) the creation (annihilation) operator of a dressed gluon with color $a$, polarization $i$ and three-momentum $\vec{k}_2$. The action of these operators on dressed Fock states is defined in the standard way. For example, for the gluon we have 
\beq
a^a_i(\vec{k}_2)\big| ({\bf g})[\vec{q}_2]^b_j\big\rangle_{D}&=&2k_2^+ (2\pi)^3 \delta^{(3)}(\vec{k}_2-\vec{q}_2)\delta^{ab}\delta_{ij}\;,\nonumber\\
a^{\dagger a}_i(\vec{k}_2)\big| 0\big\rangle&=&\big| ({\bf g})\big[ \vec{k}_2\big]^{a}_i\big\rangle_D\;.
\eeq

In order to compute the partonic cross section, we need to calculate the explicit expression of the outgoing Fock state to the relevant order. Hence, for the process in this work, one needs the outgoing wave function for a quark-antiquark-gluon final state initiated by a real photon. Since the derivations of similar outgoing wave functions, i.e. $q\to qg\gamma$ and $g\to q\bar{q}\gamma$, have been presented in detail Ref.~\cite{monster}, we do not show its derivation here but rather give a summary in the appendices \ref{app:Fock} and \ref{app:outgoingFock}.
The order $g_s g_e$ contribution to the dressed outgoing wave function of a real photon with longitudinal momentum $p^+$ and vanishing transverse momentum is given by\footnote{We have introduced a shorthand notation for the two dimensional coordinate integrals: $\int_{\bf x}=\int \rmd^2{\bf x}$.}: 
\begin{align}
|(\boldsymbol{\gamma})[p^+,\mathbf{p}=0]_{\lambda}\rangle_{\mathrm{out}} & =g_{e}g_{s}\int\frac{\mathrm{d}k_{1}^{+}}{2\pi}\frac{\mathrm{d}k_{2}^{+}}{2\pi}\frac{1}{4k_{2}^{+}\sqrt{k_{1}^{+}k_{3}^{+}}}\int_{\mathbf{w}\mathbf{v}\mathbf{x}_{1}\mathbf{x}_{2}\mathbf{x}_{3}}\delta^{(2)}\big(\mathbf{w}- \xi_{i}\mathbf{x}_{i}\big)\nonumber\\
&\hspace{-5em}  \times\biggl\{\Psi_{\bar{s}s'}^{\lambda\bar{\lambda}}(\bar{\xi}_{3})\phi_{s\bar{s}}^{\eta\bar{\eta}}\bigl(\frac{\xi_{2}}{\bar{\xi}_{3}}\bigr)A^{\bar{\eta}}(\mathbf{x}_{1}-\mathbf{x}_{2})\delta^{(2)}\big(\mathbf{v}-\frac{\xi_{1}}{\bar{\xi}_{3}}\mathbf{x}_{1}-\frac{\xi_{2}}{\bar{\xi}_{3}}\mathbf{x}_{2}\big)\times\mathcal{G}_{q}\Bigl[\xi_{3},\frac{\xi_{1}}{\bar{\xi}_{3}};\mathbf{v},\mathbf{x}_{1},\mathbf{x}_{2},\mathbf{x}_{3}\Bigr]_{ij}^{c\bar{\lambda}}\nonumber\\
&\hspace{-5em} \quad- \Psi_{\bar{s}s}^{\lambda\bar{\lambda}}(\bar{\xi}_{1})\phi_{\bar{s}s'}^{\eta\bar{\eta}}\bigl(\frac{\xi_{2}}{\bar{\xi}_{1}}\bigr)A^{\bar{\eta}}(\mathbf{x}_{3}-\mathbf{x}_{2})\delta^{(2)}\big(\mathbf{v}-\frac{\xi_{2}}{\bar{\xi}_{1}}\mathbf{x}_{2}-\frac{\xi_{3}}{\bar{\xi}_{1}}\mathbf{x}_{3}\big)\times\mathcal{G}_{\bar{q}}\Bigl[\xi_{1},\frac{\xi_{3}}{\bar{\xi}_{1}};\mathbf{v},\mathbf{x}_{1},\mathbf{x}_{2},\mathbf{x}_{3}\Bigr]_{ij}^{c\bar{\lambda}}\nonumber\\
&\hspace{-5em} \quad +\varphi_{ss'}^{\lambda\eta}(\xi_{1},\xi_{2}) \delta^{(2)}(\mathbf{v})
\times\mathcal{G}_{C}\Bigl[\xi_{3},\xi_{1}\xi_{2};\mathbf{w},\mathbf{x}_{1},\mathbf{x}_{2},\mathbf{x}_{3}\Bigr]_{ij}^{c}\biggr\}\nonumber\\
&\hspace{-5em}\times|(\mathbf{q})[k_{1}^{+},\mathbf{x}_{1}]_{s}^{i};(\mathbf{g})[k_{2}^{+},\mathbf{x}_{2}]_{c}^{\eta};(\bar{\mathbf{q}})[k_{3}^{+},\mathbf{x}_{3}]_{s'}^{j}\rangle_{D}\;.
 \label{eq:fulloutgoingstate}
\end{align}
In the above, the quark with color $i$ and spin $s$ carries longitudinal momentum $k_1^+$ and is located at the transverse position ${\mathbf x}_1$. Likewise, the gluon with color 
$c$ and polarization $\eta$ has a longitudinal momentum $k_2^+$ and is located at the transverse position ${\mathbf x}_2$, and finally the antiquark with color $j$ and spin $s'$ carries longitudinal momentum $k_3^+=p^+-k_1^+-k_2^+$ and sits at the transverse position ${\mathbf x}_3$. The longitudinal momentum fractions that enter the splitting functions are defined as 
\begin{align}
\xi_1&=k_1^+/p^+ \; , \qquad \xi_2=k_2^+/p^+ \; , \qquad \xi_3= 1-\xi_1-\xi_2 \;, \nonumber\\
\bar\xi_1&=1-\xi_1 \; , \qquad \bar\xi_2= 1-\xi_2 \; , \qquad \bar\xi_3=\xi_1+\xi_2  \;.
\end{align}
\begin{figure}[t]
\begin{centering}
\includegraphics[clip,scale=0.18]{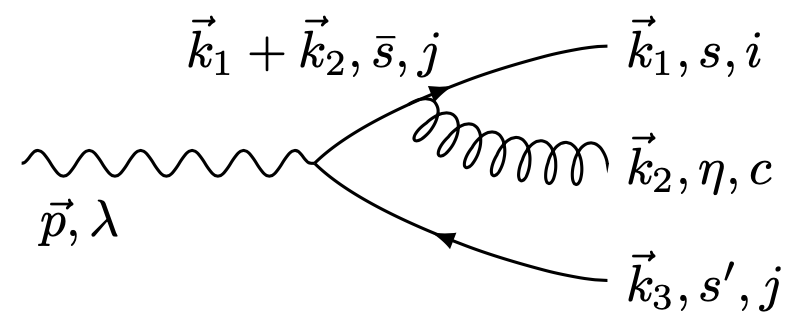}\;\includegraphics[clip,scale=0.18]{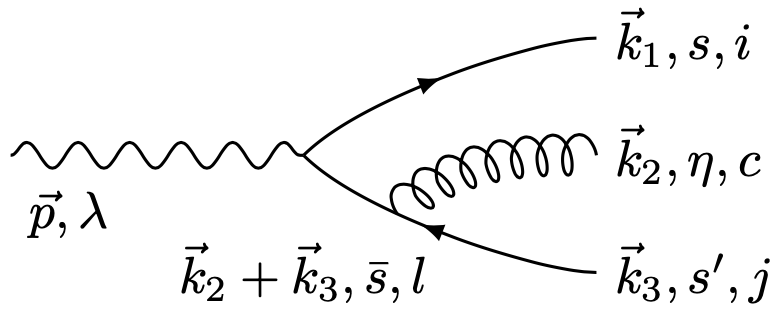}\;\includegraphics[clip,scale=0.28]{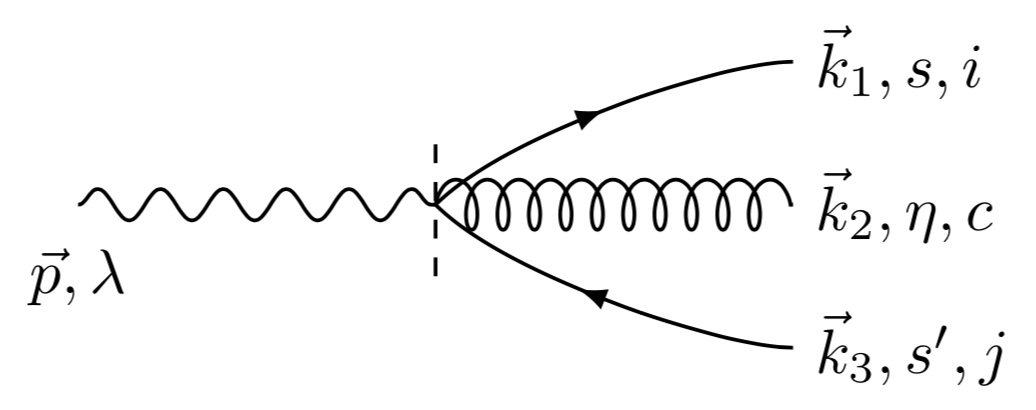}
\par\end{centering}
\caption{\label{fig:NLOsplitting}The three contributions to the $g_{s}g_{e}$
term in the photon's Fock state.}
\end{figure}
The outgoing state Eq.~\eqref{eq:fulloutgoingstate} contains combinations of three splitting functions, corresponding respectively to the splitting
of a photon into a quark-antiquark pair, the radiation of a gluon by
a quark, and the instantaneous splitting of a photon
into a gluon, quark and antiquark. Their explicit expressions are:
\begin{align}
\label{splitting_functions}
\Psi_{\alpha\beta}^{\lambda\bar{\lambda}}(z) & =\bigl(1-2z\bigr)\delta^{\lambda\bar{\lambda}}\delta_{\alpha,-\beta}-i\epsilon^{\lambda\bar{\lambda}}\sigma_{\alpha,-\beta}^{3}\;,\nonumber\\
\phi_{\alpha\beta}^{\lambda\bar{\lambda}}(z) & =(2-z)\delta^{\lambda\bar{\lambda}}\delta_{\alpha\beta}-iz\epsilon^{\lambda\bar{\lambda}}\sigma_{\alpha\beta}^{3}\;,\nonumber\\
\varphi_{\alpha\beta}^{\lambda\eta}(\xi_1,\xi_2) & =\frac{\xi_{1}\xi_{2}\xi_{3}}{\bar{\xi}_{1}\bar{\xi}_{3}}\big[(\bar{\xi}_{3}-\bar{\xi}_{1})\delta^{\lambda\eta}\delta_{\alpha,-\beta}+i(\bar{\xi}_{3}+\bar{\xi}_{1})\epsilon^{\lambda\eta}\sigma_{\alpha,-\beta}^{3}\big]\;,
\end{align}
where $\sigma^3$ is the third Pauli matrix, and with $\epsilon^{ij}$ the Levi-Civita tensor in two dimensions, with $\epsilon^{12}=+1$. Note that the splitting function for an antiquark emitting a gluon, is obtained from the corresponding splitting function for the quark by an overall multiplication with $-1$ and exchanging the spin indices.

The final ingredients of Eq.~\eqref{eq:fulloutgoingstate} are the functions ${\cal G}_q$, ${\cal G}_{\bar q}$ and ${\cal G}_C$, which incorporate the Wilson line structures of the three different amplitudes where the gluon is emitted from the quark, from the antiquark, or directly from the photon during its instantaneous splitting into a quark-antiquark-gluon state (see Fig. \ref{fig:NLOsplitting}). They are defined as:
\begin{align}
\label{G_q}
&
\mathcal{G}_{q}\Bigl[\xi_{3},\frac{\xi_{1}}{\bar{\xi}_{3}};\mathbf{v},\mathbf{x}_{1},\mathbf{x}_{2},\mathbf{x}_{3}\Bigr]_{ij}^{c\bar{\lambda}}
\nonumber\\ & 
=\big\{ \big[ 
S_{F}(\mathbf{x}_{1})S_{F}^{\dagger}(\mathbf{x}_{2})t^{c}S_{F}(\mathbf{x}_{2})S_{F}^{\dagger}(\mathbf{x}_{3})\big]_{ij}-t_{ij}^{c}\big\} \; 
\mathcal{A}^{\bar{\lambda}}\Big(\xi_{3},\mathbf{x}_{3}-\mathbf{v};\frac{\xi_{1}}{\bar{\xi}_{3}},\mathbf{x}_{1}-\mathbf{x}_{2}\Big)
\nonumber\\
& \qquad-\big\{ \big[t^{c}S_{F}(\mathbf{v})S_{F}^{\dagger}(\mathbf{x}_{3})\big]_{ij}-t_{ij}^{c} \big\} \;  
A^{\bar{\lambda}}(\mathbf{x}_{3}-\mathbf{v})\;,\\
\label{G_qbar} 
&\mathcal{G}_{\bar{q}}\Bigl[\xi_{1},\frac{\xi_{3}}{\bar{\xi}_{1}};\mathbf{v},\mathbf{x}_{1},\mathbf{x}_{2},\mathbf{x}_{3}\Bigr]_{ij}^{c\bar{\lambda}}
\nonumber\\
 &=
 \big\{ \big[ S_{F}(\mathbf{x}_{1})S_{F}^{\dagger}(\mathbf{x}_{2})t^{c}S_{F}(\mathbf{x}_{2})S_{F}^{\dagger}(\mathbf{x}_{3})\big]_{ij}-t_{ij}^{c} \big\} \;  
 \mathcal{A}^{\bar{\lambda}}\Big( \xi_{1},\mathbf{x}_{1}-\mathbf{v};\frac{\xi_{3}}{\bar{\xi}_{1}},\mathbf{x}_{3}-\mathbf{x}_{2}\Big) 
 \nonumber\\
 & \qquad- \big\{ \big[ S_{F}(\mathbf{x}_{1})S_{F}^{\dagger}(\mathbf{v})t^{c}\big]_{ij}-t_{ij}^{c}\big\} \; 
 A^{\bar{\lambda}}(\mathbf{x}_{1}-\mathbf{v})\;,\\
 \label{G_inst}
&\mathcal{G}_{C}\Bigl[\xi_{3},\xi_{1}\xi_{2};\mathbf{w},\mathbf{x}_{1},\mathbf{x}_{2},\mathbf{x}_{3}\Bigr]_{ij}^{c}\nonumber\\
 & =\big\{ \big[ S_{F}(\mathbf{x}_{1})S_{F}^{\dagger}(\mathbf{x}_{2})t^{c}S_{F}(\mathbf{x}_{2})S_{F}^{\dagger}(\mathbf{x}_{3})\big]_{ij}-t_{ij}^{c} \big\} \; \mathcal{C}\bigl(\xi_{3},\mathbf{x}_{3}-\mathbf{w};\xi_{1}\xi_{2},\mathbf{x}_{1}-\mathbf{x}_{2}\bigr)\;.
\end{align}
In these functions, the Wilson lines are defined in the standard way as the path ordered exponential of the semiclassical gluon field $\alpha_{a}^-(x^+,{\bf x})$ of the target: 
\beq
S_F({\bf x})\equiv {\cal P}\,  e^{ig\int \mathrm{d}x^+ t^a\alpha^-_a(x^+,{\bf x})}\;,
\label{eq:defWL}
\eeq
where $t^a$ is the $SU(N_c)$ generator in the fundamental representation, indicated with the subscript\footnote{In the intermediate steps of the calculation, relegated to the appendix, one will encounter Wilson lines where the generator is in the adjoint representation. These are indicated with the subscript $A$.}  $F$. 
Moreover, in the above expressions, $A^i({\bf x})$ stands for the standard non-Abelian Weizs\"acker-Williams field, which accounts for a single emission:
\begin{equation}
\begin{aligned}A^{i}(\mathbf{x}) & \equiv - \frac{1}{2\pi}\frac{x^{i}}{\mathbf{x}^{2}}\;,\end{aligned}
\label{eq:WWfield}
\end{equation}
On the other hand, ${\cal A}^i(\xi, {\bf x}; \chi, {\bf y})$ is the \emph{modified Weizs\"acker-Williams field}, responsible for two successive emissions, whose explicit expression is written as 
\begin{equation}
\begin{aligned}\mathcal{A}^{i}\bigl(\xi,\mathbf{x};\chi,\mathbf{y}\bigr) & \equiv - \frac{1}{2\pi}\frac{\xi x^{i}}{\xi\mathbf{x}^{2}+\chi\bar{\chi}\mathbf{y}^{2}}\;.\end{aligned}
\label{eq:modWWfield}
\end{equation}
Finally, we have introduced the \emph{Coulomb field} which accounts for the instantaneous emission, defined as:
\begin{equation}
\begin{aligned}\mathcal{C}\bigl(\xi,\mathbf{x};\chi,\mathbf{y}\bigr) & \equiv\frac{1}{(2\pi)^{2}}\frac{\bar{\xi}}{\xi \mathbf{x}^{2}+\chi\mathbf{y}^{2}}\;.\end{aligned}
\label{eq:Coulombfield}
\end{equation}

As mentioned earlier, the partonic cross section Eq.~\eqref{formal_partonic_X_section} is formally defined as the expectation value of the number operator in the outgoing wave function. This expectation value can be calculated in a straightforward way from the action of the creation/annihilation operators on the outgoing wave function given in Eq.~\eqref{eq:fulloutgoingstate}. The resulting cross section is then averaged over the configurations of the semiclassical target field $\alpha^-(x^+,{\bf x})$. We denote this averaging procedure by $\langle \cdots \rangle_{x_{\scriptscriptstyle{A}}}$, since it introduces an implicit dependence on $x_{\scriptscriptstyle{A}}$: the longitudinal momentum fraction of the gluons in the target wave function. 
The result is organized in the following way: 
\begin{align}
\label{X_section_CGC}
(2\pi)^{9}\frac{\mathrm{d}\sigma^{\gamma A\to q\bar{q}g+X}}{\mathrm{d}^{3}\vec{k}_{1}\mathrm{d}^{3}\vec{k}_{2}\mathrm{d}^{3}\vec{k}_{3}}&=g_{e}^{2}g_{s}^{2}\frac{1}{k^{+}_{2} p^{+}}2\pi\delta\big(p^{+}-\sum_{i=1}^3k_{i}^{+}\big) \nonumber\\
&\times \bigl\langle I_{qq}+I_{\bar{q}\bar{q}}+I_{CC}+2I_{q\bar{q}}+2I_{Cq}+2I_{C\bar{q}}\bigr\rangle_{x_{\scriptscriptstyle{A}}}\;.
\end{align}
Each contribution in Eq.~\eqref{X_section_CGC} can be studied separately. 

Let us start with what we call the ‘quark-quark' contribution $I_{qq}$, corresponding to a gluon emission from the quark (leftmost panel in Fig.~\ref{fig:NLOsplitting}) both in the amplitude and in the complex conjugate amplitude:
\begin{align}
\label{Iqq}
\langle \, I_{qq} \,  \rangle_{x_{\scriptscriptstyle{A}}}& =\mathcal{M}_{qq}^{\bar\lambda\bar{\lambda}';\bar\eta\bar{\eta}'}\Big(\bar{\xi}_{3},\frac{\xi_{2}}{\bar{\xi}_{3}}\Big)\int_{\mathbf{v}\mathbf{v}'}\prod_{i=1}^{3} \int_{\mathbf{x}_{i}\mathbf{x}'_{i}} e^{i\mathbf{k}_{i}\cdot(\mathbf{x}'_{i}-\mathbf{x}_{i})}\nonumber\\
 & \times 
 A^{\bar{\eta}}(\mathbf{x}_{1}-\mathbf{x}_{2})
 A^{\bar{\eta}'}(\mathbf{x}'_{1}-\mathbf{x}'_{2}) \, 
 \delta^{(2)}\Big(\mathbf{v}-\frac{\xi_{1}}{\bar{\xi}_{3}}\mathbf{x}_{1}-\frac{\xi_{2}}{\bar{\xi}_{3}}\mathbf{x}_{2}\Big)\delta^{(2)}\Big(\mathbf{v}'-\frac{\xi_{1}}{\bar{\xi}_{3}}\mathbf{x}'_{1}-\frac{\xi_{2}}{\bar{\xi}_{3}}\mathbf{x}'_{2}\Big)\nonumber\\
 & \times
 \Big\langle \mathrm{Tr} \Big\{ \mathcal{G}_{q}^{\dagger}\Bigl[\xi_{3},\frac{\xi_{1}}{\bar{\xi}_{3}};\mathbf{v}',\mathbf{x}'_{1},\mathbf{x}'_{2},\mathbf{x}'_{3}\Bigr]^{\bar{\lambda}'}\mathcal{G}_{q}\Bigl[\xi_{3},\frac{\xi_{1}}{\bar{\xi}_{3}};\mathbf{v},\mathbf{x}_{1},\mathbf{x}_{2},\mathbf{x}_{3}\Bigr]^{\bar{\lambda}}\Big\} \,  \Big\rangle_{x_{\scriptscriptstyle{A}}}\;,
\end{align}
where the function $\mathcal{M}_{qq}^{\bar\lambda\bar{\lambda}';\bar\eta\bar{\eta}'}$ is the product of the $\gamma \to q\bar{q}$ and $q\to qg$ splitting functions, defined in Eq.~\eqref{splitting_functions}, and reads 
\beq
\mathcal{M}_{qq}^{\bar\lambda\bar{\lambda}';\bar\eta\bar{\eta}'} \Big(\bar{\xi}_{3},\frac{\xi_{2}}{\bar{\xi}_{3}}\Big)
=
\frac{1}{8} \,
\Psi_{\tilde{s}s'}^{\lambda\bar{\lambda}'*}(\bar{\xi}_{3})  
\, 
\phi_{s\tilde{s}}^{\eta\bar{\eta}'*}\Big(\frac{\xi_{2}}{\bar{\xi}_{3}}\Big)
\,
\Psi_{\bar{s}s'}^{\lambda\bar{\lambda}}(\bar{\xi}_{3}) 
\, 
\phi_{s\bar{s}}^{\eta\bar{\eta}}\Big(\frac{\xi_{2}}{\bar{\xi}_{3}}\Big) \, . 
\eeq
It can be evaluated explicitly by using the expressions in Eq.~\eqref{splitting_functions}, yielding:
\begin{align}
\label{Mqq_def} 
\mathcal{M}_{qq}^{\bar\lambda\bar{\lambda}';\bar\eta\bar{\eta}'} \Big(\bar{\xi}_{3},\frac{\xi_{2}}{\bar{\xi}_{3}}\Big)
=
\big( \bar{\xi}_{3}^{2}+\xi_{3}^{2}\big) 
\Big[1+\Bigl(1-\frac{\xi_{2}}{\bar{\xi}_{3}}\Bigr)^{2}\Big]
\delta^{\bar{\eta}\bar{\eta}'}\delta^{\bar{\lambda}\bar{\lambda}'} \nonumber\\
-
\frac{\xi_{2}}{\bar{\xi}_{3}} \Big(2-\frac{\xi_{2}}{\bar{\xi}_{3}}\Big) 
\bigl(1-2\bar{\xi}_{3}\bigr)\epsilon^{\bar{\eta}\bar{\eta}'}\epsilon^{\bar{\lambda}\bar{\lambda}'}\,.
\end{align}
The color trace in Eq.~\eqref{Iqq}  can be performed by using the explicit expression of the function ${\cal G}_q$, given in Eq.~\eqref{G_q}, and using the Fierz identity 
\beq
t^a_{\alpha\beta}t^{a}_{\sigma\lambda}=\frac{1}{2}\Big( \delta_{\alpha\lambda}\delta_{\beta\sigma}-\frac{1}{N_c}\delta_{\alpha\beta}\delta_{\sigma\lambda}\Big)\;.
\eeq
The result of the color algebra can be written in a convenient way by introducing the following functions 
\begin{align}
\label{W_1}
W_{1}\big( \x_1,\x_2; \y_1,\y_2\big|\z_1,\z_2;{\bf v}_1,{\bf v}_2\big)
& = Q(\x_1,\x_2,\y_1,\y_2)Q(\z_1,\z_2,{\bf v}_1,{\bf v}_2)
\nonumber\\
&
- s(\x_1,\x_2)s(\z_1,\z_2)
- s(\y_1,\y_2)s({\bf v}_1,{\bf v}_2) 
+ 1\;, \\
\nonumber\\
\label{W_2}
W_{2}\big( \x_1,\x_2 \big| \y_1,\y_2; \z_1,\z_2\big)
&
= s(\x_1,\x_2)Q(\y_1,\y_2,\z_1,\z_2)
\nonumber\\
&
- s(\x_1,\x_2)s(\y_1,\y_2)
- s(\z_1,\z_2)
+ 1\;, \\
\nonumber\\
\label{W_3}
W_{3}\big(\x_1,\x_2;\y_1,\y_2\big)
&
= Q(\x_1,\x_2,\y_1,\y_2) - s(\x_1,\x_2) -s(\y_1,\y_2) +1\;,
\end{align}
where the dipole and the quadrupole operators are defined as 
\beq
s({\bf x}, {\bf y})&=&  \frac{1}{N_c} {\rm Tr}\big[ S_F({\bf x})S_F^{\dagger}({\bf y}) \big] \;,\nonumber \\
Q({\bf x}, {\bf y}, {\bf u}, {\bf v} )&=&   \frac{1}{N_c} {\rm Tr}\big[ S_F({\bf x})S_F^{\dagger}({\bf y}) S_F({\bf u}) S_F^{\dagger}({\bf v}) \big] \;.
\eeq
The final expression for the $I_{qq}$ contribution to the partonic cross section is:
\begin{align}
\label{qq_final}
\langle \, I_{qq}\, \rangle_{x_{\scriptscriptstyle{A}}}
&
= \frac{N_c^2}{2}\mathcal{M}_{qq}^{\bar\lambda\bar{\lambda}';\bar\eta\bar{\eta}'}\Big( \! \bar{\xi}_3, \frac{\xi_2}{\bar\xi_3} \Big) 
\int_{{\bf v}{\bf v}'} \prod_{i=1}^{3}\int_ {\x_i\x'_i} e^{i{\bf k}_i\cdot({\bf x}'_i-{\bf x}_i)}
\nonumber\\
&
\times
A^{\bar\eta}(\x_1-\x_2)A^{\bar\eta'}(\x'_1-\x'_2)
\, 
\delta^{(2)}\Big( \! {\bf v}-\frac{\xi_1}{\bar\xi_3}\x_1-\frac{\xi_2}{\bar\xi_3}\x_2 \! \Big)
\delta^{(2)}\Big( \! {\bf v}'- \frac{\xi_1}{\bar\xi_3}\x'_1-\frac{\xi_2}{\bar\xi_3}\x'_2 \! \Big)  
\, 
\nonumber\\
&
\times
\Big\langle
\Big[ 
W_{1}\big( \x_1.\x_2; \x'_2,\x'_1 \big| \x_2,\x_3;\x'_3,\x'_2\big)
- \frac{1}{N_c^2}
W_{3}\big(\x_1,\x_3;\x'_3,\x'_1\big)
\Big]
\nonumber\\
&
\hspace{4cm}
\times
{\cal A}^{\bar\lambda}\Big(\! \xi_3,\x_3-{\bf v}; \frac{\xi_1}{\bar\xi_3},\x_1-\x_2 \! \Big)
{\cal A}^{\bar\lambda'}\Big(\! \xi_3,\x'_3-{\bf v}'; \frac{\xi_1}{\bar\xi_3},\x'_1-\x'_2 \! \Big)
\nonumber\\
&
\hspace{0.3cm}
-
\Big[ 
W_{2}\big( \x_1,\x_2 \big| \x_2,\x_3; \x'_3,{\bf v}'\big)
- \frac{1}{N_c^2}
W_{3}\big( \x_1,\x_3;\x'_3,{\bf v}'\big)
\Big]
\nonumber\\
&
\hspace{4cm}
\times
{\cal A}^{\bar\lambda}\Big( \! \xi_3,\x_3-{\bf v}; \frac{\xi_1}{\bar\xi_3},\x_1-\x_2 \! \Big)
A^{\bar\lambda'}(\x'_3-{\bf v}')
\nonumber\\
&
\hspace{0.3cm}
-
\Big[ 
W_{2}\big( \x'_2,\x'_1 \big| {\bf v},\x_3; \x'_3,\x'_2 \big)
- \frac{1}{N_c^2}
W_{3}\big( {\bf v},\x_3;\x'_3,\x'_1\big)
\Big]
\nonumber\\
&
\hspace{4cm}
\times
A^{\bar\lambda}(\x_3-{\bf v})
{\cal A}^{\bar\lambda'}\Big( \! \xi_3,\x'_3-{\bf v}'; \frac{\xi_1}{\bar\xi_3}, \x'_1-\x'_2 \! \Big) 
\nonumber\\
&
\hspace{0.3cm}
-
\Big(1-\frac{1}{N_c^2}\Big) \,  W_{3}\big( {\bf v},\x_3; \x'_3,{\bf v}' \big)
\, A^{\bar\lambda}(\x_3-{\bf v}) A^{\bar\lambda'}(\x'_3-{\bf v}')
\Big\rangle_{\xA} \;.
\end{align}

The second contribution to the partonic cross section Eq.~\eqref{X_section_CGC} is the ‘antiquark-antiquark' term $I_{\bar q\bar q}$, corresponding to the emission of the gluon from the antiquark (see the middle panel in Fig.~\ref{fig:NLOsplitting}) both in the amplitude and in the complex conjugate amplitude. It reads:
\begin{align}
\label{barqbarq_first}
\langle \,  I_{\bar{q}\bar{q} }\,\rangle_{x_{\scriptscriptstyle{A}}} 
& =
\mathcal{M}_{\bar{q}\bar{q}}^{\bar{\lambda}\bar{\lambda}';\bar{\eta}\bar{\eta}'}\Big(\xi_{1},\frac{\xi_{2}}{\bar{\xi}_{1}}\Big)\int_{\mathbf{v}\mathbf{v}'}\prod_{i=1}^{3} \int_{\mathbf{x}_{i}\mathbf{x}'_{i}} 
e^{i\mathbf{k}_{i}\cdot(\mathbf{x}'_{i}-\mathbf{x}_{i})}
\nonumber\\
 &
 \times 
A^{\bar{\eta}}(\mathbf{x}_{3}-\mathbf{x}_{2}) 
A^{\bar{\eta}'}(\mathbf{x}'_{3}-\mathbf{x}'_{2}) \, 
\delta^{(2)}\Big(\mathbf{v}-\frac{\xi_{2}}{\bar{\xi}_{1}}\mathbf{x}_{2}-\frac{\xi_{3}}{\bar{\xi}_{1}}\mathbf{x}_{3}\Big) \delta^{(2)}\Big(\mathbf{v}'-\frac{\xi_{2}}{\bar{\xi}_{1}}\mathbf{x}'_{2}-\frac{\xi_{3}}{\bar{\xi}_{1}}\mathbf{x}'_{3}\Big)\nonumber\\
 & \times
\Big\langle 
\mathrm{Tr}\Big\{ 
\mathcal{G}_{\bar{q}}^{\dagger}\Bigl[\xi_{1},\frac{\xi_{3}}{\bar{\xi}_{1}};\mathbf{v}',\mathbf{x}'_{1},\mathbf{x}'_{2},\mathbf{x}'_{3}\Bigr]^{\bar{\lambda}'}
\mathcal{G}_{\bar{q}}\Bigl[\xi_{1},\frac{\xi_{3}}{\bar{\xi}_{1}};\mathbf{v},\mathbf{x}_{1},\mathbf{x}_{2},\mathbf{x}_{3}\Bigr]^{\bar{\lambda}}
\Big\} \, 
\Big\rangle_{x_{\scriptscriptstyle{A}}} \; . 
\end{align}
In order to arrive to the explicit expression of the antiquark-antiquark contribution, one computes the trace over color indices and performs the color algebra by using the explicit expression of the function ${\cal G}_{\bar q}$ given in Eq. \eqref{G_qbar}. The result can be written in terms of the functions defined in Eqs. \eqref{W_1}, \eqref{W_2} and \eqref{W_3}, and reads:
\begin{align}
\label{barqbarq_final}
\langle \, I_{\bar q\bar q} \, \rangle_{x_{\scriptscriptstyle{A}}} & =\frac{N_c^2}{2}
\mathcal{M}_{\bar{q}\bar{q}}^{\bar{\lambda}\bar{\lambda}';\bar{\eta}\bar{\eta}'}\Big(\! \xi_1, \frac{\xi_2}{\bar\xi_1}\!\Big)
\int_{{\bf v} {\bf v}'}  \prod_{i=1}^{3}   \int_{{\bf x}_i{\bf x}'_i} e^{i{\bf k}_i\cdot ({\bf x}'_i-{\bf x}_i)}
\nonumber\\
&
\times
A^{\bar\eta}(\x_3-\x_2) A^{\bar\eta'}(\x'_3-\x'_2)\, 
\delta^{(2)}\Big( \! {\bf v}-\frac{\xi_2}{\bar\xi_1}\x_2-\frac{\xi_3}{\bar\xi_1}\x_3 \! \Big)
\delta^{(2)}\Big( \! {\bf v}'-\frac{\xi_2}{\bar\xi_1}\x'_2-\frac{\xi_3}{\bar\xi_1}\x'_3 \! \Big)
\nonumber\\
&
\times\Big\langle
\Big[ W_1\big( \x_1,\x_2; \x'_2,\x'_1 \big| \x_2,\x_3; \x'_3,\x'_2 \big) - \frac{1}{N_c^2}W_3\big( \x_1,\x_3;\x'_3,\x'_1 \big) 
\Big]
\nonumber\\
&
\hspace{4cm}
\times
{\cal A}^{\bar\lambda}\Big(\! \xi_1,\x_1-{\bf v}; \frac{\xi_3}{\bar\xi_1}, \x_3-\x_2\!\Big)
{\cal A}^{\bar\lambda'}\Big(\! \xi_1,\x'_1-{\bf v}'; \frac{\xi_3}{\bar\xi_1}, \x'_3-\x'_2\!\Big)
\nonumber\\
&
\hspace{0.3cm}
- \Big[ 
W_2\big( \x_2,\x_3 \big| \x_1,\x_2;{\bf v}',\x'_1 \big) - \frac{1}{N_c^2}W_3\big( \x_1,\x_3; {\bf v}',\x'_1\big) 
\Big]
\nonumber\\
&
\hspace{4cm}
\times
{\cal A}^{\bar\lambda}\Big(\! \xi_1,\x_1-{\bf v}; \frac{\xi_3}{\bar\xi_1}, \x_3-\x_2\!\Big)
A^{\bar\lambda'}(\x'_1-{\bf v}')
\nonumber\\
&
\hspace{0.3cm}
- \Big[ 
W_2\big( \x'_3,\x'_2\big|\x'_2,\x'_1,\x_1,{\bf v}\big) - \frac{1}{N_c^2}W_3\big( x_1,{\bf v}; \x'_3,\x'_1\big)
\Big]
\nonumber\\
&
\hspace{4cm}
\times
A^{\bar\lambda}(\x_1-{\bf v})
{\cal A}^{\bar\lambda'}\Big(\! \xi_1,\x'_1-{\bf v}'; \frac{\xi_3}{\bar\xi_1}, \x'_3-\x'_2\!\Big)
\nonumber\\
&
\hspace{0.3cm}
-
\Big(1-\frac{1}{N_c^2}\Big) \,  W_{3}\big( \x_1,{\bf v};{\bf v}',\x'_1 \big)
\, A^{\bar\lambda}(\x_1-{\bf v}) A^{\bar\lambda'}(\x'_1-{\bf v}')
\Big\rangle_{\xA}\;.
\end{align}
Similarly to the quark-quark contribution, the function $\mathcal{M}_{\bar{q}\bar{q}}^{\bar{\lambda}\bar{\lambda}';\bar{\eta}\bar{\eta}'}$ encodes the products of the $\gamma\to qg$ and $\bar{q} \to \bar{q} g$ splitting functions:
\beq
\label{M_barqbarq_def}
 \mathcal{M}_{\bar{q}\bar{q}}^{\bar{\lambda}\bar{\lambda}';\bar{\eta}\bar{\eta}'}\Big(\xi_{1},\frac{\xi_{2}}{\bar{\xi}_{1}}\Big)
 &=&
\frac{1}{8}
\Psi_{s\tilde{s}}^{\lambda\bar{\lambda}'*}(\xi_{1})  \, 
\Big[-\phi_{\tilde{s}s'}^{\eta\bar{\eta}'*}\Big(\frac{\xi_{2}}{\bar{\xi}_{1}}\Big) \Big] \, 
\Psi_{s\bar{s}}^{\lambda\bar{\lambda}}(\xi_{1}) \, 
\Big[-\phi_{\bar{s}s'}^{\eta\bar{\eta}}\Big(\frac{\xi_{2}}{\bar{\xi}_{1}}\Big) \Big] \;,\nonumber\\
 &=& 
 \mathcal{M}_{qq}^{\bar{\eta}\bar{\eta}';\lambda\bar{\lambda}'}\Big(\bar{\xi}_{1},\frac{\xi_{2}}{\bar{\xi}_{1}}\Big)\;.
\eeq
Note that, due to the charge conjugation symmetry on the cross section level, $\mathcal{M}_{\bar{q}\bar{q}}^{\bar{\lambda}\bar{\lambda}';\bar{\eta}\bar{\eta}'}$ can be obtained from $\mathcal{M}_{qq}^{\lambda\bar{\lambda}';\bar{\eta}\bar{\eta}'}$ by swapping the quantum numbers of the quark and the antiquark. Alternatively, we can calculate it using the explicit expressions of the splitting functions given in Eq.~\eqref{splitting_functions}, yielding:
\begin{align}
\label{def_M_barqbarq}
\mathcal{M}_{\bar{q}\bar{q}}^{\bar{\lambda}\bar{\lambda}';\bar{\eta}\bar{\eta}'}\Big(\xi_{1},\frac{\xi_{2}}{\bar{\xi}_{1}}\Big)
 =
\big(\xi_{1}^{2}+\bar{\xi}_{1}^{2}\bigr) 
\Big[ 1+\Bigl(1-\frac{\xi_{2}}{\bar{\xi}_{1}}\Bigr)^{2}\Big] \delta^{\bar{\eta}'\bar{\eta}}\delta^{\bar{\lambda}\bar{\lambda}'}\nonumber\\-\frac{\xi_{2}}{\bar{\xi}_{1}}\Bigl(2-\frac{\xi_{2}}{\bar{\xi}_{1}}\Bigr)(1-2\bar{\xi}_{1})\epsilon^{\bar{\eta}\bar{\eta}'}\epsilon^{\bar{\lambda}\bar{\lambda}'}\;.
\end{align}

The next contribution that we consider is $I_{q\bar q}$, corresponding to the emission of the gluon from the antiquark in the amplitude and from the quark in the complex conjugate amplitude, or vice versa. This term can be written as:%
\begin{align}
\langle \, I_{q\bar{q}} \, \rangle_{x_{\scriptscriptstyle{A}}}
& =
\mathcal{M}_{q\bar{q}}^{\bar{\lambda}\bar{\lambda}';\bar{\eta}\bar{\eta}'}(\xi_{1},\xi_{2})
\int_{\mathbf{v}\mathbf{v}'}\prod_{i=1}^{3} \int_{\mathbf{x}_{i}\mathbf{x}'_{i}} e^{i\mathbf{k}_{i}\cdot(\mathbf{x}'_{i}-\mathbf{x}_{i})}
\nonumber\\
 & \times 
 A^{\bar{\eta}}(\mathbf{x}_{3}-\mathbf{x}_{2}) 
 A^{\bar{\eta}'}(\mathbf{x}'_{1}-\mathbf{x}'_{2}) \, 
 \delta^{(2)}\Big(\mathbf{v}-\frac{\xi_{2}}{\bar{\xi}_{1}}\mathbf{x}_{2}-\frac{\xi_{3}}{\bar{\xi}_{1}}\mathbf{x}_{3}\Big) \delta^{(2)}\Big(\mathbf{v}'-\frac{\xi_{1}}{\bar{\xi}_{3}}\mathbf{x}'_{1}-\frac{\xi_{2}}{\bar{\xi}_{3}}\mathbf{x}'_{2}\Big)\nonumber\\
 & \times
 \Big\langle 
 \mathrm{Tr}\Big\{ 
 \mathcal{G}_{q}^{\dagger}\Bigl[\xi_{3},\frac{\xi_{1}}{\bar{\xi}_{3}};\mathbf{v}',\mathbf{x}'_{1},\mathbf{x}'_{2},\mathbf{x}'_{3}\Bigr]^{\bar{\lambda}'}
 \mathcal{G}_{\bar{q}}\Bigl[\xi_{1},\frac{\xi_{3}}{\bar{\xi}_{1}};\mathbf{v},\mathbf{x}_{1},\mathbf{x}_{2},\mathbf{x}_{3}\Bigr]^{\bar{\lambda}}
 \Big\}
 \Big\rangle_{x_{\scriptscriptstyle{A}}}\;.
\end{align}
As in the previous two cases, to evaluate this contribution we use the explicit expressions for the functions ${\cal G}_q$ and ${\cal G}_{\bar q}$, given in Eqs.~\eqref{G_q} and \eqref{G_qbar}, respectively. This yields:
\begin{align}
\label{qbarq_final}
\langle \, I_{q\bar q} \, \rangle_{x_{\scriptscriptstyle{A}}} & =
\frac{N_c^2}{2} {\cal M}^{\bar\lambda\bar\lambda';\bar\eta\bar\eta'}_{q\bar q}(\xi_1,\xi_2)
\int_{{\bf v} {\bf v}'} \prod_{i=1}^{3} \int_{{\bf x}'_i{\bf x}_i} e^{i{\bf k}_i\cdot({\bf x}'_i-{\bf x}_i)} 
\nonumber\\
&
\times
A^{\bar\eta}(\x_3-\x_2)
A^{\bar\eta'}(\x'_1-\x'_2) \, 
\delta^{(2)}\Big( \! {\bf v}-\frac{\xi_2}{\bar\xi_1}\x_2-\frac{\xi_3}{\bar\xi_1}\x_3\! \Big)
\delta^{(2)}\Big( \! {\bf v}'-\frac{\xi_1}{\bar\xi_3}\x'_1-\frac{\xi_2}{\bar\xi_3}\x'_2\!\Big)
\nonumber\\
&
\times
\Big\langle
\Big[ 
W_1\big( \x_1,\x_2;\x'_2,\x'_1 \big| \x_2,\x_3;\x'_3,\x'_2 \big) -\frac{1}{N_c^2} W_3 \big( \x_1,\x_3; \x'_3,\x'_2 \big) 
\Big]
\nonumber\\
&
\hspace{4cm}
\times
{\cal A}^{\bar\lambda}\Big( \! \xi_1,\x_1-{\bf v}; \frac{\xi_3}{\bar\xi_1},\x_3-\x_2 \! \Big)
{\cal A}^{\bar\lambda'}\Big( \! \xi_3, \x'_3-{\bf v}'; \frac{\xi_1}{\bar\xi_3}, \x'_1-\x'_2 \! \Big) 
\nonumber\\
&
\hspace{0.3cm}
-
\Big[
W_2 \big( \x_1,\x_2 \big| \x_2,\x_3; \x'_3, {\bf v}' \big) - \frac{1}{N_c^2}W_3 \big( \x_1,\x_3; \x'_3, {\bf v}'\big) 
\Big]
\nonumber\\
&
\hspace{4cm}
\times
{\cal A}^{\bar\lambda}\Big( \! \xi_1,\x_1-{\bf v}; \frac{\xi_3}{\bar\xi_1},\x_3-\x_2 \! \Big)
A^{\bar\lambda'}(\x'_3-{\bf v}')
\nonumber\\
&
\hspace{0.3cm}
-
\Big[
W_2 \big( \x'_3,\x'_2 \big| \x'_2,\x'_1;\x_1,{\bf v} \big) - \frac{1}{N_c^2}W_3\big(\x_1,{\bf v}; \x'_3,\x'_1 \big) 
\Big]
\nonumber\\
&
\hspace{4cm}
\times 
A^{\bar\lambda}(\x_1 - {\bf v})
{\cal A}^{\bar\lambda'}\Big( \! \xi_3, \x'_3-{\bf v}'; \frac{\xi_1}{\bar\xi_3}, \x'_1-\x'_2 \! \Big)
\nonumber\\
&
\hspace{0.3cm}
+
\Big[ 
W_4\big( \x_1, {\bf v}; \x'_3, {\bf v}'\big) - \frac{1}{N_c^2}W_3\big( \x_1, {\bf v}; \x'_3, {\bf v}'\big)
\Big]
A^{\bar\lambda}(\x_1-{\bf v})A^{\bar\lambda'}(\x'_3-{\bf v}')
\Big\rangle_{\xA}\;,
\end{align}
where we have introduced a new function $W_4\big(\x_1,\x_2; \y_1,\y_2 \big)$ corresponding to the following combination of dipole operators:
\begin{align}
W_4\big(\x_1,\x_2; \y_1,\y_2 \big)= s(\x_1,\x_2)s(\y_1,\y_2)-s(\x_1,\x_2)-s(\y_1,\y_2)+1\;.
\end{align}
The product of splitting amplitudes ${\cal M}^{\bar\lambda\bar\lambda';\bar\eta\bar\eta'}_{q\bar q}$ for this contribution is:
\beq
{\cal M}^{\bar\lambda\bar\lambda';\bar\eta\bar\eta'}_{q\bar q}\left(\xi_{1},\xi_{2}\right)
&=& 
\frac{1}{8}
\Psi_{\tilde{s}s'}^{\lambda\bar{\lambda}'*}(\bar{\xi}_{3})\, 
\phi_{s\tilde{s}}^{\eta\bar{\eta}'*}\Big(\frac{\xi_{2}}{\bar{\xi}_{3}}\Big) \, 
\Psi_{s\bar{s}}^{\lambda\bar{\lambda}\dagger}(\bar{\xi}_{1}) \, 
\Big[-\phi_{\bar{s}s'}^{\eta\bar{\eta}}\Big(\frac{\xi_{2}}{\bar{\xi}_{1}}\Big) \Big]\;,
\eeq
and reads, when computed explicitly:
\begin{align}
\label{def_M_qbarq}
{\cal M}^{\bar\lambda\bar\lambda';\bar\eta\bar\eta'}_{q\bar q}\left(\xi_{1},\xi_{2}\right)
 =
 \big( \xi_{1}+\xi_{3}-2\xi_{1}\xi_{3}\big) 
 \Big( 2-\frac{\xi_{2}}{\bar{\xi}_{1}}-\frac{\xi_{2}}{\bar{\xi}_{3}}\Big) 
 \delta^{\bar{\lambda}\bar{\lambda}'}\delta^{\bar{\eta}\bar{\eta}'}\nonumber\\
 -
 \frac{\xi_{2}(\xi_{3}-\xi_{1})^{2}}{\bar{\xi}_{1}\bar{\xi}_{3}}\epsilon^{\bar{\lambda}\bar{\lambda}'}\epsilon^{\bar{\eta}\bar{\eta}'}\;.
\end{align}

The next step is to compute the contributions stemming from the instantaneous splitting of the photon into a quark, an antiquark and a gluon (rightmost panel in Fig.~\ref{fig:NLOsplitting}).
Let us start with the $I_{CC}$ contribution, which corresponds to the instantaneous splitting both in the amplitude and in the complex conjugate amplitude.
This contribution is written in terms of the function ${\cal G}_{C}$, defined in Eq.~\eqref{G_inst}, and reads 
\begin{align}
\langle \, I_{CC} \, \rangle_{x_{\scriptscriptstyle{A}}} 
& = 
\mathcal{M}_{CC}(\xi_1,\xi_2) 
\int_{\mathbf{w}\mathbf{w}'}\prod_{i=1}^{3} \int_{\mathbf{x}_{i}\mathbf{x}'_{i}} 
e^{i\mathbf{k}_{i}\cdot(\mathbf{x}'_{i}-\mathbf{x}_{i})}
\delta^{(2)}\big(\mathbf{w}- \xi_{i}\mathbf{x}_{i}\big)
\delta^{(2)}\big(\mathbf{w}'- \xi_{i}\mathbf{x}'_{i}\big)
\nonumber\\
& \times
\Big\langle
\mathrm{Tr}
\Big\{ 
\mathcal{G}_{C}^\dagger \Bigl[\xi_{3},\xi_{1}\xi_{2};\mathbf{w}',\mathbf{x}'_{1},\mathbf{x}'_{2},\mathbf{x}'_{3}\Bigr]\mathcal{G}_{C}\Bigl[\xi_{3},\xi_{1}\xi_{2};\mathbf{w},\mathbf{x}_{1},\mathbf{x}_{2},\mathbf{x}_{3}\Bigr]
\Big\}
\Big\rangle_{x_{\scriptscriptstyle{A}}}\;.
\label{eq:ICC}
\end{align}
After performing the color algebra, one obtains:
\begin{align}
\label{CC_final}
\langle \, I_{CC} \, \rangle_{x_{\scriptscriptstyle{A}}}&
 =
\frac{N_c^2}{2}\mathcal{M}_{CC}(\xi_1,\xi_2) 
\int_{\mathbf{w}\mathbf{w}'}\prod_{i=1}^{3} \int_{\mathbf{x}_{i}\mathbf{x}'_{i}} 
e^{i\mathbf{k}_{i}\cdot(\mathbf{x}'_{i}-\mathbf{x}_{i})}
\delta^{(2)}\big(\mathbf{w}- \xi_{i}\mathbf{x}_{i}\big)
\delta^{(2)}\big(\mathbf{w}'- \xi_{i}\mathbf{x}'_{i}\big) \, 
\nonumber\\
&
\times \, 
\Big\langle
W_1\big( \x_1,\x_2;\x'_2,\x'_1 \big| \x_2,\x_3;\x'_3,\x'_2 \big) -\frac{1}{N_c^2}W_3\big(\x_1,\x_3;\x'_3,\x'_1\big)
\Big\rangle_{\xA}
\nonumber\\
&
\times
{\cal C}\big( \xi_3,\x_3-{\bf w}; \xi_1\xi_2, \x_1-\x_2 \big) \, 
{\cal C}\big( \xi_3, \x'_3-{\bf w}'; \xi_1\xi_2, \x'_1-\x'_2 \big)\;.
\end{align}
The product of splitting functions for the $I_{CC}$ contribution reads
\beq
\label{def_M_CC}
\mathcal{M}_{CC}(\xi_1,\xi_2) =\frac{1}{8} \varphi_{ss'}^{\lambda\eta *}(\xi_{1},\xi_{2})\varphi_{ss'}^{\lambda\eta}(\xi_{1},\xi_{2})=\xi_{1}^{2}\xi_{2}^{2}\xi_{3}^{2}\Big( \frac{1}{\bar{\xi}_{1}^{2}}+\frac{1}{\bar{\xi}_{3}^{2}}\Big)\;.
\eeq

The remaining two contributions that need to be computed are $\langle \,  I_{Cq} \, \rangle_{x_{\scriptscriptstyle{A}}}$ and $\langle \,  I_{C\bar q} \, \rangle_{x_{\scriptscriptstyle{A}}}$, corresponding the interference between the gluon emission from the quark resp. antiquark in the amplitude, and the instantaneous splitting in the complex conjugate amplitude. The first one is given by:
\begin{align}
\langle \, I_{Cq} \, \rangle_{x_{\scriptscriptstyle{A}}} 
& =
\mathcal{M}_{Cq}^{\bar{\lambda}\bar{\eta}}\big(\xi_1,\xi_2 \big)
\int_{\mathbf{v}\mathbf{w}'} 
\prod_{i=1}^{3} \int_{\mathbf{x}_{i}\mathbf{x}'_{i}} 
e^{i\mathbf{k}_{i}\cdot(\mathbf{x}'_{i}-\mathbf{x}_{i})} 
\delta^{(2)}\Big(\mathbf{v}-\frac{\xi_{1}}{\bar{\xi}_{3}}\mathbf{x}_{1}-\frac{\xi_{2}}{\bar{\xi}_{3}}\mathbf{x}_{2}\Big)\delta^{(2)}\big(\mathbf{w}'- \xi_{i}\mathbf{x}'_{i}\big) 
\nonumber\\
& \times 
A^{\bar{\eta}}(\mathbf{x}_{1}-\mathbf{x}_{2})
\Big\langle 
\mathrm{Tr}\Big\{ 
\mathcal{G}_{C}^\dagger \Bigl[\xi_{3},\xi_{1}\xi_{2};\mathbf{w}',\mathbf{x}'_{1},\mathbf{x}'_{2},\mathbf{x}'_{3}\Bigr]\mathcal{G}_{q}\Bigl[\xi_{3},\frac{\xi_{1}}{\bar{\xi}_{3}};\mathbf{v},\mathbf{x}_{1},\mathbf{x}_{2},\mathbf{x}_{3}\Bigr]^{\bar{\lambda}} \Big\}
\Big\rangle_{x_{\scriptscriptstyle{A}}}\;.
\end{align}
As in the case of other contributions, using the explicit expressions of the functions ${\cal G}_q$ and ${\cal G}_C$ and performing the color algebra results in the following expression:%
\begin{align}
\label{Cq_final}
\langle \, I_{Cq} \, \rangle_{x_{\scriptscriptstyle{A}}} & = 
\frac{N_c^2}{2}\mathcal{M}_{Cq}^{\bar{\lambda}\bar{\eta}}\big(\xi_1,\xi_2 \big) \int_{\mathbf{v}\mathbf{w}'}\prod_{i=1}^{3} \int_{\mathbf{x}_{i}\mathbf{x}'_{i}} e^{i\mathbf{k}_{i}\cdot(\mathbf{x}'_{i}-\mathbf{x}_{i})}
\nonumber\\
&
\times
A^{\bar\eta}(\x_1-\x_2) \, 
\delta^{(2)}\Big( \! {\bf v}-\frac{\xi_1}{\bar\xi_3}\x_1-\frac{\xi_2}{\bar\xi_3}\x_2 \!\Big) \;  
\delta^{(2)}\big( {\bf w}'-\xi_i\x'_i \big) \; 
\nonumber\\
&
\times
\Big\langle
\Big[ W_1\big( \x_1,\x_2;\x'_2,\x'_1 \big| \x_2,\x_3;\x'_3,\x'_2 \big) -\frac{1}{N_c^2}W_3\big( \x_1,\x_3;\x'_3,\x'_1\big)
\Big]
\nonumber\\
&
\hspace{2cm}
\times
{\cal A}^{\bar\lambda}\Big(\xi_3, \x_3-{\bf v}; \frac{\xi_1}{\bar\xi_3}, \x_1-\x_2\Big) \; 
{\cal C}\big( \xi_3, \x'_3-{\bf w}'; \xi_1\xi_2, \x'_1-\x'_2 \big)
\nonumber\\
&
\hspace{0.3cm}
-
\Big[ W_2\big( \x'_2,\x'_1 \big| \x'_3,\x'_2;{\bf v},\x_3\big) -\frac{1}{N_c^2}W_3\big( {\bf v},\x_3;\x'_3,\x'_1 \big) 
\Big]
\nonumber\\
&
\hspace{2cm}
\times
A^{\bar\lambda}(\x_3-{\bf v}) \, 
{\cal C}\big(\xi_3, \x'_3-{\bf w}'; \xi_1\xi_2, \x'_1-\x'_2\big)
\Big\rangle_{\xA}\;,
\end{align}
with the product of splitting functions $\mathcal{M}_{Cq}^{\bar{\lambda}\bar{\eta}}$ equal to
\beq
\label{def_M_Cq}
\mathcal{M}_{Cq}^{\bar{\lambda}\bar{\eta}}\big(\xi_1,\xi_2 \big) 
=
\frac{1}{8} \, 
\varphi_{ss'}^{\lambda\eta *}(\xi_{1},\xi_{2}) \, 
\Psi_{\bar{s}s'}^{\lambda\bar{\lambda}}(\bar{\xi}_{3}) \, 
\phi_{s\bar{s}}^{\eta\bar{\eta}}\Big(\frac{\xi_{2}}{\bar{\xi}_{3}}\Big) 
=
-\xi_{1}\xi_{2}\xi_{3} \, \delta^{\bar{\lambda}\bar{\eta}} \, 
\Big(\frac{\bar{\xi}_{3}}{\bar{\xi}_{1}}+\frac{\xi_{1}\xi_{3}}{\bar{\xi}_{3}^{2}} \Big) \; . 
\eeq

Finally, the last contribution $\langle \, I_{C\bar q}\, \rangle_{x_{\scriptscriptstyle{A}}}$ is written in terms of the functions ${\cal G}_C$ and ${\cal G}_{\bar q}$ as follows: 
\begin{align}
\langle \, I_{C\bar{q}} \, \rangle_{x_{\scriptscriptstyle{A}}} & =\mathcal{M}_{C\bar{q}}^{\bar{\lambda}\bar{\eta}}\big(\xi_1,\xi_2 \big) \int_{\mathbf{v}\mathbf{w}'}\prod_{i=1}^{3} \int_{\mathbf{x}_{i}\mathbf{x}'_{i}} e^{i\mathbf{k}_{i}\cdot(\mathbf{x}'_{i}-\mathbf{x}_{i})} 
\delta^{(2)}\Big(\mathbf{v}-\frac{\xi_{2}}{\bar{\xi}_{1}}\mathbf{x}_{2}-\frac{\xi_{3}}{\bar{\xi}_{1}}\mathbf{x}_{3}\Big)\delta^{(2)}\big(\mathbf{w}'- \xi_{i}\mathbf{x}'_{i}\big)\nonumber\\
& \times 
A^{\bar{\eta}}(\mathbf{x}_{3}-\mathbf{x}_{2}) \, 
\Big\langle  \mathrm{Tr}\Big\{ 
\mathcal{G}_{C}^\dagger \Bigl[\xi_{3},\xi_{1}\xi_{2};\mathbf{w}',\mathbf{x}'_{1},\mathbf{x}'_{2},\mathbf{x}'_{3}\Bigr] \mathcal{G}_{\bar{q}}\Bigl[\xi_{1},\frac{\xi_{3}}{\bar{\xi}_{1}};\mathbf{v},\mathbf{x}_{1},\mathbf{x}_{2},\mathbf{x}_{3}\Bigr]^{\bar{\lambda}}
\Big\} \Big\rangle_{x_{\scriptscriptstyle{A}}}\;,
\end{align}
and yields, after performing the color algebra:
\begin{align}
\label{Cbarq_final}
\langle \, I_{C\bar{q}} \, \rangle_{x_{\scriptscriptstyle{A}}} & =\frac{N_c^2}{2}\mathcal{M}_{C\bar{q}}^{\bar{\lambda}\bar{\eta}}\big(\xi_1,\xi_2 \big) \int_{\mathbf{v}\mathbf{w}'}\prod_{i=1}^{3} \int_{\mathbf{x}_{i}\mathbf{x}'_{i}} e^{i\mathbf{k}_{i}\cdot(\mathbf{x}'_{i}-\mathbf{x}_{i})} 
\nonumber\\
&
\times
A^{\bar\eta}(\x_3-\x_2) \, 
\delta^{(2)} \Big( \! {\bf v}-\frac{\xi_2}{\bar\xi_1}\x_2-\frac{\xi_3}{\bar\xi_1}\x_3  \! \Big)
\delta^{(2)}\big( {\bf w}'-\xi_i\x'_i \big) \; 
\nonumber\\
&
\Big\langle
\Big[ 
W_1\big( \x_1,\x_2;\x'_2,\x'_1 \big| \x_2,\x_3; \x'_3,\x'_2 \big) - \frac{1}{N_c^2}W_3\big( \x_1,\x_3; \x'_3,\x'_1\big) 
\Big]
\nonumber\\
&
\hspace{2cm}
\times
{\cal A}^{\bar\lambda}\Big(\xi_1, \x_1-{\bf v}; \frac{\xi_3}{\bar\xi_1}, \x_3-\x_2 \Big)  \; 
{\cal C} \big( \xi_3, \x'_3-{\bf w}'; \xi_1\xi_2, \x'_1-\x'_2 \big) 
\nonumber\\
&
\hspace{0.3cm}
-
\Big[ 
W_2\big( \x'_3,\x'_2 \big| \x'_2,\x'_1;\x_1,{\bf v}\big) - \frac{1}{N_c^2}W_3\big(\x_1, {\bf v}; \x'_3,\x'_1 \big)
\Big]
\nonumber\\
&
\hspace{2cm}
\times
A^{\bar\lambda}(\x_1-{\bf v}) \; 
{\cal C} \big( \xi_3, \x'_3-{\bf w}'; \xi_1\xi_2, \x'_1-\x'_2 \big) 
\Big\rangle_\xA\;,
\end{align}
where the product of splitting functions $\mathcal{M}_{C\bar{q}}^{\bar{\lambda}\bar{\eta}}$ is:
\begin{align}
\label{def_M_Cbarq}
\mathcal{M}_{C\bar{q}}^{\bar{\lambda}\bar{\eta}}\big(\xi_1,\xi_2 \big)
=
\frac{1}{8} \, 
\varphi_{ss'}^{\lambda\eta *}(\xi_{1},\xi_{2}) \, 
\Psi_{\bar{s}s}^{\lambda\bar{\lambda}}(\bar{\xi}_{1}) \, 
\Big[-\phi_{s'\bar{s}}^{\eta\bar{\eta}}\Big(\frac{\xi_{2}}{\bar{\xi}_{1}}\Big) \Big]\nonumber\\
=
-\xi_{1}\xi_{2}\xi_{3} \, \delta^{\bar{\lambda}\bar{\eta}} \, 
\Big( \frac{\bar{\xi}_{1}}{\bar{\xi}_{3}}+\frac{\xi_{1}\xi_{3}}{\bar{\xi}_{1}^{2}} \Big) \;.
\end{align}

Let us summarize our findings for this section. The partonic cross section for the photoproduction of a quark, an antiquark, and a gluon, is given in Eq.~\eqref{X_section_CGC}. Each contribution to this cross section is calculated separately and the final results are given in Eqs. \eqref{qq_final}, \eqref{barqbarq_final}, \eqref{qbarq_final}, \eqref{CC_final}, \eqref{Cq_final} and \eqref{Cbarq_final}. As we would expect from the charge conjugation symmetry of QCD, the result is fully symmetric under the exchange of the quark with the antiquark.

Finally, the photoproduction cross section in $ep$ or in $eA$ collisions, or alternatively in ultra-peripheral $pp$ or $pA$ collisions (UPCs), can be obtained in a very straightforward way from the partonic cross section by using the equivalent photon approximation. This approximation consists in simply convolving the partonic cross section with the relevant photon flux $f_{e,P,A\to\gamma}(y,\mu^2)$:
\begin{align}
\label{convolution}
\sigma^{e,p,A+A\to q\bar{q}g+X}=\int \mathrm{d} y\, \hat{\sigma}^{\gamma A \to q\bar{q} g+X}(y) f_{e,p,A\to \gamma}(y,\mu^2)\;,
\end{align}
where $y=P\cdot p/ P\cdot \ell \simeq p^+/\ell^+$ is the longitudinal momentum fraction carried by the photon (with $\ell^\mu$ the momentum of the photon source) and where $\mu^2$ is the factorization scale. For example, if the photon source is an electron, the real photon flux $f_{e\to\gamma}(y,\mu^2)$ is given in the well-known Weizs\"acker-Williams approximation by the formula \cite{WW}:
\begin{align}
f_{e\to\gamma}(y,\mu^2)=\frac{\alpha}{2\pi}\Big(\frac{1+(1-y)^{2}}{y}\mathrm{ln}\frac{\mu^{2}(1-y)}{m_{e}^{2}y^{2}}+2m_{e}^{2}y\big(\frac{1}{\mu^{2}}-\frac{1-y}{m_{e}^{2}y^{2}}\big)\Big)\;,
\end{align}
where $m_e$ is the electron mass and $\alpha$ the fine-structure constant.

\section{\label{correlation}Correlation limit and gluon TMDs} 
In recent years, there has been a lot of activity in the study of gluon TMDs from CGC calculations. In Refs.~\cite{firstlowxTMDs}, it was shown that the dijet production cross section in forward $pA$ collisions can be written in terms of gluon TMDs, in a specific kinematic limit that is referred to as the correlation limit. In these specific kinematics, the total transverse momentum of the produced pair ($\sim|{\bf k}_1+{\bf k}_2|$) is assumed to be much smaller than the typical momentum of the jets ($\sim|{\bf k}_1|\sim|{\bf k}_2|$). In such a situation, the produced jets fly almost back-to-back in momentum space, which in coordinate space corresponds to a small transverse distance between the produced jet pair, usually referred to as the dipole size. 

Recently, the correlation limit of three final state particles (two jets and a photon) has been studied in Ref.~\cite{monster}. In this configuration the total transverse momentum ${\bf q}_\sT={\bf k}_1+{\bf k}_2+{\bf k}_3$ is again required to be much smaller than the individual transverse momenta of the produced jets (${\bf k}_1, {\bf k}_1, {\bf k}_3$). In contrast to the production of two final state particles, these kinematics allow us to identify not one but two small transverse sizes in coordinate space.

Let us turn to our process, and identify the small transverse sizes around which we can perform a Taylor expansion. These sizes are not necessarily the same for each subprocess, and can be identified by inspecting the denominators of the structures Eqs.~\eqref{eq:WWfield}, \eqref{eq:modWWfield} and \eqref{eq:Coulombfield}, which appear in the functions ${\cal G}_q$, ${\cal G}_{\bar q}$ and ${\cal G}_C$ defined in Eqs.~\eqref{G_q}, \eqref{G_qbar} and \eqref{G_inst}, respectively.

Firstly, let us consider the function ${\cal G}_q$, Eq. \eqref{G_q}, which is a part of the amplitude where the final state gluon is emitted from the quark. In the correlation limit, the small transverse distance for this contribution appears to be 
\begin{equation}
\label{dipole_q}
\mathbf{r}_{g}  =\mathbf{x}_{1}-\mathbf{x}_{2}\;\quad\mathrm{and}\quad\mathbf{r}_{\bar{q}}  =\mathbf{x}_{3}-\mathbf{v}\;,
\end{equation}
where ${\bf r}_g$ is identified as the size of the dipole formed by the final quark and gluon, and where ${\bf r}_{\bar q }$ is the dipole size of the final state antiquark and the intermediate quark. With the help of the delta function $\delta^{(2)}\big(\mathbf{v}-\frac{\xi_{1}}{\bar{\xi}_{3}}\mathbf{x}_{1}-\frac{\xi_{2}}{\bar{\xi}_{3}}\mathbf{x}_{2}\big)$, which accompanies the function ${\cal G}_q$ in the outgoing photon wave function Eq. \eqref{eq:fulloutgoingstate}, we can substitute the coordinates ${\bf x}_1$, ${\bf x}_3$ and $\bf v$ in favor of the above defined dipole sizes:
\begin{equation}
\label{covq} 
\mathbf{x}_{1}  =\mathbf{x}_{2}+\mathbf{r}_{g}\;,\quad\mathbf{x}_{3} =\mathbf{x}_{2}+\frac{\xi_{1}}{\bar{\xi}_{3}}\mathbf{r}_{g}+\mathbf{r}_{\bar{q}}\;,\quad\mathrm{and}\quad
\mathbf{v}=\mathbf{x}_{2}+\frac{\xi_{1}}{\bar{\xi}_{3}}\mathbf{r}_{g}\;.
\end{equation}
The small-dipole expansion of ${\cal G}_q$ is then straightforward, and reads:
\begin{align}
\label{expanded_Gq}
&\mathcal{G}_{q}\Bigl[\xi_{3},\frac{\xi_{1}}{\bar{\xi}_{3}};\mathbf{x}_{2},\mathbf{r}_{g},\mathbf{r}_{\bar{q}}\Bigr]_{ij}^{c\bar{\lambda}}  
\simeq 
\Big\{ \big[\partial_{k} S_{F}(\mathbf{x}_{2}) \big] S_{F}^{\dagger}(\mathbf{x}_{2}) t^{c}\Big\}_{ij} \, 
r_{g}^{k} \, \mathcal{A}^{\bar{\lambda}}\Big( \xi_{3},\mathbf{r}_{\bar{q}};\frac{\xi_{1}}{\bar{\xi}_{3}},\mathbf{r}_{g} \Big) \\
 & 
 +
 \Big\{ t^{c}S_{F}(\mathbf{x}_{2}) \big[ \partial_{k} S_{F}^{\dagger}(\mathbf{x}_{2})\big] \Big\}_{ij}
 \Big\{ 
 \frac{\xi_{1}}{\bar{\xi}_{3}} \, r_{g}^{k} \,
 \mathcal{A}^{\bar{\lambda}}\Big(\xi_{3},\mathbf{r}_{\bar{q}};\frac{\xi_{1}}{\bar{\xi}_{3}},\mathbf{r}_{g}\Big) 
 +
r_{\bar{q}}^{k} \Big[ 
\mathcal{A}^{\bar{\lambda}}\Big( \xi_{3},\mathbf{r}_{\bar{q}};\frac{\xi_{1}}{\bar{\xi}_{3}},\mathbf{r}_{g} \Big) - A^{\bar{\lambda}}(\mathbf{r}_{\bar{q}}) \Big] \Big\} \;,\nonumber
\end{align}
where we have used that $\big[ \partial_{k} S_{F}(\mathbf{x}_{2}) \big] S_{F}^{\dagger}(\mathbf{x}_{2}) = - S_{F}(\mathbf{x}_{2})\big[ \partial_{k} S_{F}^{\dagger}(\mathbf{x}_{2})\big] $.

Likewise, the dipole sizes appearing in the function ${\cal G}_{\bar q}$, Eq.~\eqref{G_qbar}, which corresponds to the emission of the final state gluon from the antiquark, are:
\begin{equation}
\mathbf{r}_{g}  =\mathbf{x}_{3}-\mathbf{x}_{2}\;\quad\mathrm{and}\quad\mathbf{r}_{q}  =\mathbf{x}_{1}-\mathbf{v}\;,
\label{eq:qbardipolesizes}
\end{equation}
where ${\bf r}_q$ is the dipole size of the final quark and the intermediate antiquark, and where in contrast to the previous case ${\bf r}_g$ is dipole size of the final gluon and \textit{anti}quark. Taking the delta function $\delta^{(2)}\big(\mathbf{v}-\frac{\xi_{2}}{\bar{\xi}_{1}}\mathbf{x}_{2}-\frac{\xi_{3}}{\bar{\xi}_{1}}\mathbf{x}_{3}\big)$ into account, which accompanies the function ${\cal G}_{\bar{q}}$ (see Eq.~\eqref{eq:fulloutgoingstate}), the transverse coordinates can be written as:
\begin{equation}
\mathbf{x}_{3}  =\mathbf{x}_{2}+\mathbf{r}_{g}\;,\quad\mathbf{x}_{1} =\mathbf{x}_{2}+\frac{\xi_{3}}{\bar{\xi}_{1}}\mathbf{r}_{g}+\mathbf{r}_{q}\;,\quad\mathrm{and}\quad
\mathbf{v}=\mathbf{x}_{2}+\frac{\xi_{3}}{\bar{\xi}_{1}}\mathbf{r}_{g}\;.
\end{equation}
The Taylor expansion of $\mathcal{G}_{\bar{q}}$ then yields:
\begin{align}
\label{expanded_Gbarq} 
&
\mathcal{G}_{\bar{q}}\Bigl[\xi_{1},\frac{\xi_{3}}{\bar{\xi}_{1}};\mathbf{x}_{2},\mathbf{r}_g,\mathbf{r}_{q}\Bigr]_{ij}^{c\bar{\lambda}} 
\simeq 
\Big\{ t^{c}S_{F}(\mathbf{x}_{2}) \big[ \partial_{k} S_{F}^{\dagger}(\mathbf{x}_{2}) \big] \Big\}_{ij} \, 
r_{g}^{k} \, 
\mathcal{A}^{\bar{\lambda}}\Big(\xi_{1},\mathbf{r}_{q};\frac{\xi_{3}}{\bar{\xi}_{1}},\mathbf{r}_{g}\Big) \\
 & +
 \Big\{ \big[ \partial_{k} S_{F}(\mathbf{x}_{2}) \big] S_{F}^\dagger(\mathbf{x}_{2})t^{c}\Big\}_{ij} 
 \Big\{ \frac{\xi_{3}}{\bar{\xi}_{1}} \, r_{g}^{k} \, 
 \mathcal{A}^{\bar{\lambda}}\Big(\xi_{1},\mathbf{r}_{q};\frac{\xi_{3}}{\bar{\xi}_{1}},\mathbf{r}_{g}\Big) 
 +
r_{q}^{k } 
\Big[ 
\mathcal{A}^{\bar{\lambda}}\Big(\xi_{1},\mathbf{r}_{q};\frac{\xi_{3}}{\bar{\xi}_{1}},\mathbf{r}_{g} \Big)
-A^{\bar{\lambda}}(\mathbf{r}_{q})\Big] 
\Big\}\;.\nonumber
\end{align}

Finally, the small transverse sizes that appear in the function ${\cal G}_C$, Eq. \eqref{G_inst}, corresponding to the instantaneous emission of the quark-gluon-antiquark from the incoming photon, are:
\begin{equation}
\mathbf{r}_{g}  =\mathbf{x}_{1}-\mathbf{x}_{2}\;\quad\mathrm{and}\quad\mathbf{r}_{\bar{q}}  =\frac{1}{\bar{\xi}_{3}}(\mathbf{x}_{3}-\mathbf{w})\;,
\label{eq:Cdipolesizes}
\end{equation}
with ${\bf r}_g$ the transverse size of the final quark-gluon dipole, and ${\bf r}_{\bar q}$ the size of the dipole formed by the final state antiquark with the incoming photon. Again, rewriting the transverse space coordinates in terms of these dipole sizes we get:
\begin{equation}
\mathbf{x}_{1}  =\mathbf{x}_{2}+\mathbf{r}_{g}\;,\quad\mathrm{and}\quad\mathbf{x}_{3}  =\mathbf{x}_{2}+\frac{\xi_{1}}{\bar{\xi}_{3}}\mathbf{r}_{g}+\mathbf{r}_{\bar{q}}\;.
\end{equation}
Then, the Taylor expansion of the function ${\cal G}_C$ reads
\begin{align} 
\label{expanded_GC} 
\mathcal{G}_{C}\Bigl[\xi_{1},\xi_{2};\mathbf{x}_{2},\mathbf{r}_g,\mathbf{r}_{\bar{q}}\Bigr]_{ij}^{c} 
 & \simeq 
 \bigg(
 r_{g}^{k}
 \big\{ \big[ \partial_{k} S_{F}(\mathbf{x}_{2})\big] S_{F}^{\dagger}(\mathbf{x}_{2})t^{c}\big\}_{ij} 
 \\
 &\hspace{.4cm}
 + 
 \Big(\frac{\xi_{1}}{\bar{\xi}_{3}}r_{g}^{k}+r_{\bar{q}}^{k} \Big) 
 \big\{ t^{c}S_{F}(\mathbf{x}_{2})\big[ \partial_{i} S_{F}^{\dagger}(\mathbf{x}_{2}) \big] \big\}_{ij}
 \bigg) \, 
\mathcal{C}\bigl(\xi_{3},\bar{\xi}_{3}\mathbf{r}_{\bar{q}};\xi_{1}\xi_{2},\mathbf{r}_{g}\bigr)\;. \nonumber
\end{align}
With the small-dipole expansions of the functions ${\cal G}_q$, ${\cal G}_{\bar q}$ and ${\cal G}_C$ at hand, we are now ready to compute the correlation limit of the partonic cross section Eq.~\eqref{X_section_CGC}.

Let us start with the $I_{qq}$ contribution, given in Eq.~\eqref{Iqq}. Using the change of variables that are introduced in Eq.~\eqref{covq} both in the amplitude and in the complex conjugate amplitude, as well as the Taylor-expanded expression of the function ${\cal G}_q$ given in Eq.~\eqref{expanded_Gq}, we obtain the following result:%
\beq
\label{Iqq_cont_first}
\langle \, I_{qq} \, \rangle_{x_{\scriptscriptstyle{A}}} 
&=&
-\frac{N_c}{2}{\cal M}^{\bar\lambda\bar\lambda';\bar\eta\bar\eta'}_{qq}\Big( {\bar\xi}_3, \frac{\xi_2}{{\bar \xi}_3}\Big)  \\
&\times&\int_{\x_2\x'_2} e^{i{\bf q}_\sT\cdot (\x'_2-\x_2)} 
\Big\langle {\rm Tr} \big\{ S_F(\x'_2)\big[ \partial_jS_F^\dagger(\x'_2)\big] S_F(\x_2)\big[ \partial_iS_F(\x_2)\big]\big\}\Big\rangle_{x_{\scriptscriptstyle{A}}} \nonumber\\
&\times&
\int_{{\bf r}_g {\bf r}'_g {\bf r}_{\bar q} {\bf r}'_{\bar q}}
e^{i{\bf Q}\cdot({\bf r}'_g-{\bf r}'_g)+i{\bf k}_3\cdot({\bf r}'_{\bar q}-{\bf r}_{\bar q})} 
A^{\bar\eta}({\bf r}_g) \, A^{\bar\eta'}({\bf r}'_g) 
\nonumber\\
&\times& 
\Big\{ 
\Big\lgroup 
r_{g}^{i} \, 
\mathcal{A}^{\bar{\lambda}}\Big(\xi_{3},\mathbf{r}_{\bar{q}};\frac{\xi_{1}}{\bar{\xi}_{3}},\mathbf{r}_{g}\Big)
\Big\rgroup 
\Big\lgroup 
r_{g}^{\prime j}  
\mathcal{A}^{\bar{\lambda}'}\Big( \xi_{3},\mathbf{r}'_{\bar{q}};\frac{\xi_{1}}{\bar{\xi}_{3}},\mathbf{r}'_{g} \Big)
\Big\rgroup
\nonumber\\
&&
\hspace{0cm}
+  
\Big\lgroup
r_{g}^{i} \, \frac{\xi_1}{\bar{\xi}_3}  \, 
\mathcal{A}^{\bar{\lambda}}\Big(\xi_{3},\mathbf{r}_{\bar{q}};\frac{\xi_{1}}{\bar{\xi}_{3}},\mathbf{r}_{g}\Big)
+
r_{\bar{q}}^{i} \, 
\Big[ 
\mathcal{A}^{\bar{\lambda}}\Big( \xi_{3},\mathbf{r}_{\bar{q}};\frac{\xi_{1}}{\bar{\xi}_{3}},\mathbf{r}_{g}\Big)
-A^{\bar{\lambda}}(\mathbf{r}_{\bar{q}})\Big]
\Big\rgroup
\nonumber\\
&&
\times
\Big\lgroup 
r_{g}^{\prime j} \, \frac{\xi_1}{\bar{\xi}_3}
 \mathcal{A}^{\bar{\lambda}'}\Big( \xi_{3},\mathbf{r}'_{\bar{q}};\frac{\xi_{1}}{\bar{\xi}_{3}},\mathbf{r}'_{g}\Big)
 +
 r_{\bar{q}}^{\prime j} 
 \Big[ 
 \mathcal{A}^{\bar{\lambda}'}\Big( \xi_{3},\mathbf{r}'_{\bar{q}};\frac{\xi_{1}}{\bar{\xi}_{3}},\mathbf{r}'_{g}\Big) 
 - A^{\bar{\lambda}'}(\mathbf{r}'_{\bar{q}})
 \Big] 
 \Big\rgroup
 \nonumber\\
&&
\hspace{-0.6cm}
- \frac{1}{N_c^2} 
\Big\lgroup 
 r_{g}^{i} \,  \frac{\xi_2}{\bar{\xi}_3}
 \mathcal{A}^{\bar{\lambda}}\Big(\xi_{3},\mathbf{r}_{\bar{q}};\frac{\xi_{1}}{\bar{\xi}_{3}},\mathbf{r}_{g}\Big)
 -
 r_{\bar{q}}^{i} 
 \Big[ 
 \mathcal{A}^{\bar{\lambda}}\Big( \xi_{3},\mathbf{r}_{\bar{q}};\frac{\xi_{1}}{\bar{\xi}_{3}},\mathbf{r}_{g} \Big) 
 -A^{\bar{\lambda}}(\mathbf{r}_{\bar{q}})
 \Big]
 \Big\rgroup\nonumber\\
&&
\times 
\Big\lgroup 
r_{g}^{\prime j} \, \frac{\xi_2}{\bar{\xi}_3}  
 \mathcal{A}^{\bar{\lambda}'}\Big( \xi_{3},\mathbf{r}'_{\bar{q}};\frac{\xi_{1}}{\bar{\xi}_{3}},\mathbf{r}'_{g} \Big)
 -
 r_{\bar{q}}^{\prime j}  
 \Big[ 
 \mathcal{A}^{\bar{\lambda}'}\Big( \xi_{3},\mathbf{r}'_{\bar{q}};\frac{\xi_{1}}{\bar{\xi}_{3}},\mathbf{r}'_{g} \Big) 
 -A^{\bar{\lambda}'}(\mathbf{r}'_{\bar{q}}) 
 \Big] 
 \Big\rgroup
 \Big\}\;,\nonumber
\eeq
where the combinations of transverse momenta $\mathbf{q}_{\sT}$ and $\mathbf{Q}$ are defined as
\begin{equation}
\label{eq:def_total_mom}
\mathbf{q}_{\sT}  =\mathbf{k}_{1}+\mathbf{k}_{2}+\mathbf{k}_{3}\;,
\end{equation}
\begin{equation}
\mathbf{Q}  =\mathbf{k}_{1}+\frac{\xi_{1}}{\bar{\xi}_{3}}\mathbf{k}_{3}\;,
\label{eq:def:Q_mom}
\end{equation}
and are ordered as $|{\bf q}_{\sT}| \ll |{\bf k}_{3}|, |{\bf Q}|$, corresponding to the dipole sizes to which they are conjugate being ordered as $|{\bf x}'_2 |, | {\bf x}_2| \gg |{\bf r}'_g|,|{\bf r}'_g|,|{\bf r}'_{\bar q}|,|{\bf r}_{\bar q}| $. The integrations over the dipole sizes ${\bf r}_g$, ${\bf r}'_g$, ${\bf r}_{\bar q}$ and ${\bf r}'_{\bar q}$ factorize from the Wilson line structure, and can be performed with the help of the following two integrals\footnote{The derivations of these two integrals can be found in \cite{monster}.}:
\begin{align}
\label{int_type_1}
\int_{\mathbf{r}_{g}\mathbf{r}_{\bar{q}}}
e^{-i\mathbf{Q} \cdot \mathbf{r}_{g} -i{\bf k}_3\cdot{\bf r}_{\bar q}  }
A^{\bar{\eta}}(\mathbf{r}_{g}) 
\, r_{g}^{i} \,  
\mathcal{A}^{\bar{\lambda}}\Big(\xi_{3},\mathbf{r}_{\bar{q}};\frac{\xi_{1}}{\bar{\xi}_{3}},\mathbf{r}_{g}\Big)  \nonumber\\
=
-i \frac{k_{3}^{\bar{\lambda}}}{{\bf k}_{3}^{2}} 
\frac{1}{{\bf Q}^{2}+c_{0}{\bf k}_{3}^{2}}
\Big( \delta^{\bar{\eta}i} - 2\frac{Q^{\bar{\eta}}Q^{i}}{{\bf Q}^{2}+c_{0}{\bf k}_{3}^{2}}\Big)\;,
\end{align}
and
\begin{align}
\label{int_type_2}
\int_{\mathbf{r}_{g}\mathbf{r}_{\bar{q}}}
e^{  -i{\bf Q}\cdot{\bf r}_g  -i{\bf k}_3\cdot{\bf r}_{\bar q}   }
A^{\bar{\eta}}(\mathbf{r}_{g}) \, 
r_{\bar{q}}^{i} \, 
\Big[ 
\mathcal{A}^{\bar{\lambda}}\Big( \xi_{3},\mathbf{r}_{\bar{q}};\frac{\xi_{1}}{\bar{\xi}_{3}},\mathbf{r}_{g}\Big) -A^{\bar{\lambda}}(\mathbf{r}_{\bar{q}}) 
\Big] \nonumber\\
= i  \frac{ Q^{\bar\eta} }{ {\bf Q}^2} 
\frac{1}{{\bf k}_3^2+c_0^{-1}{\bf Q}^2}
\Big( \delta^{\bar \lambda i} -2 \frac{ k_3^{\bar\lambda} k_3^{i} }{ {\bf k}_3^2+c_0^{-1}{\bf Q}^2 } \Big)\;,
\end{align}
where we introduced 
\beq
c_0=\frac{\xi_1\xi_2}{\xi_3\bar \xi_3^2}\;.
\eeq
Using the results Eqs.~\eqref{int_type_1} and \eqref{int_type_2}, Eq.~\eqref{Iqq_cont_first} can be simplified to:
\beq
\label{I_qq_intermediate}
\langle \, I_{qq}\, \rangle_{x_{\scriptscriptstyle{A}}}
&=& -\frac{N_c}{2} \, 
{\cal M}^{\bar\lambda\bar\lambda';\bar\eta\bar\eta'}_{qq}\Big( \bar\xi_3, \frac{\xi_2}{\bar\xi_3} \Big)
  \, \big[ {\rm H}_{qq} \big]^{\bar\lambda\bar\lambda'; \bar\eta\bar\eta'}_{ij}  \\
 &\times&\, 
  \int_{\x_2\x'_2} e^{i{\bf q}_T\cdot (\x'_2-\x_2)} 
\Big\langle {\rm Tr} \big\{ S_F(\x'_2)\big[ \partial_jS_F^\dagger(\x'_2)\big] S_F(\x_2)\big[ \partial_iS_F(\x_2)\big]\big\}\Big\rangle_{x_{\scriptscriptstyle{A}}}\;, \nonumber
\eeq
where the hard factor is defined as 
\beq
&&
\big[{\rm H}_{qq}\big]_{ij}^{\bar{\lambda}\bar{\lambda}^\prime ; \bar{\eta}\bar{\eta}^\prime}  
=  
\Pi^{\bar{\lambda};\bar{\eta}i}[\mathbf{k}_3,c_0,\mathbf{Q}]\; 
\Pi^{\bar{\lambda}^\prime;\bar{\eta}^\prime j}[\mathbf{k}_3,c_0,\mathbf{Q}] \label{eq:Hqq} \\
&&
+ \; 
\Big\lgroup 
\frac{\xi_1}{\bar{\xi}_3}
\Pi^{\bar{\lambda};\bar{\eta}i}[\mathbf{k}_3,c_0,\mathbf{Q}]
-
\Pi^{\bar{\eta};\bar{\lambda}i}[\mathbf{Q},c_0^{-1},\mathbf{k}_3]
\Big\rgroup
\Big\lgroup 
\frac{\xi_1}{\bar{\xi}_3}
\Pi^{\bar{\lambda}^\prime;\bar{\eta}^\prime j}[\mathbf{k}_3,c_0,\mathbf{Q}]
-
\Pi^{\bar{\eta}^\prime;\bar{\lambda}^\prime j}[\mathbf{Q},c_0^{-1},\mathbf{k}_3]
\Big\rgroup\nonumber\\
&&
-\frac{1}{N_c^2}
\Big\lgroup
\frac{\xi_2}{\bar{\xi}_3}
\Pi^{\bar{\lambda};\bar{\eta}i}[\mathbf{k}_3,c_0,\mathbf{Q}]
+
\Pi^{\bar{\eta};\bar{\lambda}i}[\mathbf{Q},c_0^{-1},\mathbf{k}_3]
\Big\rgroup 
\Big\lgroup
\frac{\xi_2}{\bar{\xi}_3}
\Pi^{\bar{\lambda}^\prime;\bar{\eta}^\prime j}[\mathbf{k}_3,c_0,\mathbf{Q}]
+
\Pi^{\bar{\eta}^\prime;\bar{\lambda}^\prime j}[\mathbf{Q},c_0^{-1},\mathbf{k}_3]\Big\rgroup \; ,
\nonumber
\eeq
and where we introduced the compact notation:
\begin{equation}
\label{def_pi}
\Pi^{i; jk}[\mathbf{p}, c_0, \mathbf{q}] 
\equiv 
\frac{p^i}{{\bf p}^2} \, \frac{1}{{\bf q}^2+c_0{\bf p}^2}\bigg(\delta^{jk}-2\frac{q^j q^k}{{\bf q}^2+c_0 {\bf p}^2}\bigg)\;.
\end{equation} 
The crucial observation is now that the remaining integrals of $\x_2$ and $\x'_2$ in Eq.~\eqref{I_qq_intermediate}, over the Wilson line structure, are nothing but the small-$x$ limit of the so-called Weizs\"acker-Williams (WW) gluon TMDs\footnote{Note that in our earlier work, these TMDs are usually written as $\mathcal{F}^{(3)}_{gg}$ and $\mathcal{H}^{(3)}_{gg}$.} (see e.g. Ref. \cite{Marquet:2017xwy}):
\begin{align}
\label{def_TMD}
& 
\int_{\mathbf{x}_{2}\mathbf{x}'_{2}} 
e^{i{\bf q}_T\cdot(\x'_2-\x_2)}
\Big\langle
\mathrm{Tr}\big\{ S_{F}(\mathbf{x}'_{2}) \big[ \partial_{j}S_{F}^{\dagger}(\mathbf{x}'_{2}) \big]
S_{F}(\mathbf{x}_{2}) \big[ \partial_{i} S_{F}^{\dagger}(\mathbf{x}_{2}) \big] \big\} 
\Bigr\rangle_{x_{\scriptscriptstyle{A}}}
\nonumber\\
 & 
 \hspace{2cm}
 = - g_s^2 \, (2\pi)^3 \, \frac{1}{4}  
 \bigg[\frac{1}{2}\delta^{ij}\mathcal{F}_{WW}(x_{\scriptscriptstyle{A}},{\bf q}_{\sT})+\frac{1}{2}\Bigl(2\frac{q_{\sT}^{i}q_{\sT}^{j}}{{\bf q}_{\sT}^{2}}-\delta^{ij}\Bigr)\mathcal{H}_{WW}(x_{\scriptscriptstyle{A}},{\bf q}_{\sT})\bigg]\;.
 \end{align}
In the above, ${\cal F}^{(3)}_{gg} (x_{\scriptscriptstyle{A}},{\bf q}_{\sT})$ is the unpolarized Weizs\"acker-Williams gluon TMD, and ${\cal H}^{(3)}_{gg}(x_{\scriptscriptstyle{A}},{\bf q}_{\sT})$ its linearly polarized partner. Substituting Eq.~\eqref{def_TMD} into Eq.~\eqref{I_qq_intermediate}, we can write the final result for $\langle \, I_{qq}\, \rangle_{x_{\scriptscriptstyle{A}}}$ in the following TMD-factorized form:
\begin{align}
\langle \, I_{qq} \, \rangle_{x_{\scriptscriptstyle{A}}} & = 
N_c \,  g_s^2 \, \pi^3 \, 
 \mathcal{M}_{qq}^{\bar\lambda\bar{\lambda}';\bar\eta\bar{\eta}'} \Big( \bar{\xi}_{3},\frac{\xi_{2}}{\bar{\xi}_{3}} \Big) \, 
 \big[ {\rm H}_{qq}\big]_{ij}^{\bar{\lambda}\bar{\lambda}^\prime ; \bar{\eta}\bar{\eta}^\prime}\nonumber\\
&\times
 \bigg[\frac{1}{2}\delta^{ij}\mathcal{F}_{WW}(x_{\scriptscriptstyle{A}},{\bf q}_{\sT})+\frac{1}{2}\Bigl(2\frac{q_{\sT}^{i}q_{\sT}^{j}}{{\bf q}_{\sT}^{2}}-\delta^{ij}\Bigr)\mathcal{H}_{WW}(x_{\scriptscriptstyle{A}},{\bf q}_{\sT})\bigg]\;.
 \label{eq:Iqq0}
\end{align}

Let us now calculate the correlation limit of the $\langle \, I_{\bar q\bar q}\, \rangle_{x_{\scriptscriptstyle{A}}}$ contribution to the partonic cross section, which corresponds to the gluon emission from the antiquark both in the amplitude and in the complex conjugate amplitude. The calculation is straightforward and can be performed following the same steps as in the previous case. Equivalently however, we can circumvent the calculation by observing that, since QCD preserves C-parity, $\langle \, I_{\bar q\bar q}\, \rangle_{x_{\scriptscriptstyle{A}}}$ has be equal to $\langle \, I_{qq} \, \rangle_{x_{\scriptscriptstyle{A}}}$ after exchanging the quark with the antiquark. In our notation, this corresponds to exchanging $1\leftrightarrow 3$, as well as swapping the relevant color indices (note that, with the introduction of the products of splitting functions $\mathcal{M}$, the spin indices are already contracted). Hence, we have:
\begin{align}
\mathcal{M}_{\bar{q}\bar{q}}^{\bar{\lambda}\bar{\lambda}';\bar{\eta}\bar{\eta}'}\Big(\xi_{1},\frac{\xi_{2}}{\bar{\xi}_{1}}\Big) &\overset{1\leftrightarrow3}{=} \mathcal{M}_{qq}^{\bar\lambda\bar{\lambda}';\bar\eta\bar{\eta}'} \Big( \bar{\xi}_{3},\frac{\xi_{2}}{\bar{\xi}_{3}} \Big)\;,\nonumber\\
\mathcal{G}_{\bar{q}}\Bigl[\xi_{1},\frac{\xi_{3}}{\bar{\xi}_{1}};\mathbf{x}_{2},\mathbf{r}_g,\mathbf{r}_{q}\Bigr]_{ij}^{c\bar{\lambda}} &\overset{1\leftrightarrow3}{=} \mathcal{G}_{q}\Bigl[\xi_{3},\frac{\xi_{1}}{\bar{\xi}_{3}};\mathbf{x}_{2},\mathbf{r}_{g},\mathbf{r}_{\bar{q}}\Bigr]_{ji}^{c\bar{\lambda}}\;,\nonumber\\
e^{i{\bf q}_\sT\cdot (\x'_2-\x_2)+i{\bf Q}\cdot({\bf r}'_g-{\bf r}'_g)+i{\bf k}_3\cdot({\bf r}'_{\bar q}-{\bf r}_{\bar q})} &\overset{1\leftrightarrow3}{=} e^{i\mathbf{q}_{\sT} \cdot (\mathbf{x}'_{2}-\mathbf{x}_{2}) + i{\bf K}\cdot({\bf r}'_g-{\bf r}_g) +i{\bf k}_1\cdot ({\bf r}'_q-{\bf r}_q)}\;,
\end{align} 
where the new combination of transverse momentum $\mathbf{K}$ is defined as
\begin{align}
\label{def:eq:K_mom}
\mathbf{K} & =\mathbf{k}_{3}+\frac{\xi_{3}}{\bar{\xi}_{1}}\mathbf{k}_{1}\;,
\end{align}
and $|{\bf q}_{\sT}| \ll |{\bf k}_{1}| \sim |{\bf K}|$. Thus, one can immediately write down the following result:
\begin{align} 
\label{def_hard_barqbarq}
\langle \, I_{\bar{q}\bar{q}} \, \rangle_{x_{\scriptscriptstyle{A}}} 
&  = \langle \, I_{qq} \, \rangle_{x_{\scriptscriptstyle{A}}} (1\leftrightarrow 3)
=
N_c\, g_s^2 \, \pi^3 \, 
\mathcal{M}_{\bar{q}\bar{q}}^{\bar{\lambda}\bar{\lambda}';\bar{\eta}\bar{\eta}'}\Big(\xi_{1},\frac{\xi_{2}}{\bar{\xi}_{1}}\Big)\big[{\rm H}_{\bar{q}\bar{q}}\big]_{ij}^{\bar{\lambda}\bar{\lambda}^\prime ; \bar{\eta}\bar{\eta}^\prime} \nonumber\\
&\times
\bigg[ 
\frac{1}{2}\delta^{ij}\mathcal{F}_{WW}(x_{\scriptscriptstyle{A}},{\bf q}_{\sT})+\frac{1}{2}\Big(2 \frac{q_{\sT}^{i}q_{\sT}^{j}}{{\bf q}_{\sT}^{2}}-\delta^{ij}\Bigr)\mathcal{H}_{WW}(x_{\scriptscriptstyle{A}},{\bf q}_{\sT})\bigg]\;,
\end{align}
where the hard part is defined as 
\begin{align}
&
\big[{\rm H}_{\bar{q}\bar{q}}\big]_{ij}^{\bar{\lambda}\bar{\lambda}^\prime ; \bar{\eta}\bar{\eta}^\prime}  =  
\big[{\rm H}_{qq}\big]_{ij}^{\bar{\lambda}\bar{\lambda}^\prime ; \bar{\eta}\bar{\eta}^\prime}(1\leftrightarrow 3)
\label{eq:Hqbarqbar}
= 
\Pi^{\bar{\lambda};\bar{\eta}i}[\mathbf{k}_1,\tilde{c}_0,\mathbf{K}]\; \Pi^{\bar{\lambda}^\prime;\bar{\eta}^\prime j}[\mathbf{k}_1,\tilde{c}_0,\mathbf{K}] \\
&+\Big\lgroup 
 \frac{\xi_3}{\bar{\xi}_1}\Pi^{\bar{\lambda};\bar{\eta}i}[\mathbf{k}_1,\tilde{c}_0,\mathbf{K}]-\Pi^{\bar{\eta};\bar{\lambda}i}[\mathbf{K},\tilde{c}_0^{-1},\mathbf{k}_1]
 \Big\rgroup
 \Big\lgroup \frac{\xi_3}{\bar{\xi}_1}\Pi^{\bar{\lambda}^\prime;\bar{\eta}^\prime j}[\mathbf{k}_1,\tilde{c}_0,\mathbf{K}]-\Pi^{\bar{\eta}^\prime;\bar{\lambda}^\prime j}[\mathbf{K},\tilde{c}_0^{-1},\mathbf{k}_1]
 \Big\rgroup 
 \nonumber\\
&
-\frac{1}{N_c^2}
\Big\lgroup  
\frac{\xi_2}{\bar{\xi}_1}\Pi^{\bar{\lambda};\bar{\eta}i}[\mathbf{k}_1,\tilde{c}_0,\mathbf{K}]+\Pi^{\bar{\eta};\bar{\lambda}i}[\mathbf{K},\tilde{c}_0^{-1},\mathbf{k}_1]
\Big\rgroup
\Big\lgroup
\frac{\xi_2}{\bar{\xi}_1}\Pi^{\bar{\lambda}^\prime;\bar{\eta}^\prime j}[\mathbf{k}_1,\tilde{c}_0,\mathbf{K}]+\Pi^{\bar{\eta}^\prime;\bar{\lambda}^\prime j}[\mathbf{K},\tilde{c}_0^{-1},\mathbf{k}_1]
\Big\rgroup
\;,\nonumber
\end{align}
with ${\tilde c}_0=c_0(1\leftrightarrow 3)=\frac{\xi_2 \xi_3}{\xi_1 \bar{\xi}_1^2}$. 
 
The interference contribution $\langle \, I_{q\bar q}\, \rangle_{x_{\scriptscriptstyle{A}}}$ originates from the gluon emission from the antiquark in the amplitude and from the quark in the complex conjugate amplitude, or vice versa. The small transverse dipole sizes that appear in this contribution are given by Eqs.~\eqref{covq} and \eqref{eq:qbardipolesizes}. When written in terms of these new variables, this contribution takes the following form:
\begin{align}
\langle \, I_{q\bar{q}} \, \rangle_{x_{\scriptscriptstyle{A}}} 
& =
\mathcal{M}_{q\bar{q}}^{\bar{\lambda}\bar{\lambda}';\bar{\eta}\bar{\eta}'}(\xi_{1},\xi_{2})
\int_{\mathbf{x}_{2}\mathbf{x}'_{2}\mathbf{r}_{g}\mathbf{r}'_{g}\mathbf{r}_{q}\mathbf{r}'_{\bar{q}}} 
e^{i{\bf q}_{\sT}\cdot(\x'_2-\x_2) + i{\bf Q}\cdot {\bf r}'_g +i{\bf k}_3\cdot{\bf r}'_{\bar q} -i{\bf K}\cdot{\bf r}_g -i {\bf k}_1\cdot{\bf r}_q} 
 A^{\bar{\eta}}(\mathbf{r}_{g}) A^{\bar{\eta}'}(\mathbf{r}'_{g}) 
\nonumber\\
 & \times 
 \Big\langle  
 \mathrm{Tr}\Big\{ \mathcal{G}_{q}^{\dagger}\Bigl[\xi_{3},\frac{\xi_{1}}{\bar{\xi}_{3}};\mathbf{x}'_{2},\mathbf{r}'_{g},\mathbf{r}'_{\bar{q}}\Bigr]^{\bar{\lambda}'}\mathcal{G}_{\bar{q}}\Bigl[\xi_{1},\frac{\xi_{3}}{\bar{\xi}_{1}};\mathbf{x}_{2},\mathbf{r}_{g},\mathbf{r}_{q}\Bigr]^{\bar{\lambda}}\Big\}
 \Big\rangle_{x_{\scriptscriptstyle{A}}}\;.
\end{align}
After introducing the expanded expressions for the functions ${\cal G}_q$ and ${\cal G}_{\bar q}$ given in Eqs. \eqref{expanded_Gq} and \eqref{expanded_Gbarq}, respectively, one can easily perform the transverse integrals. The final expression for the $\langle \, I_{q\bar q}\, \rangle_{x_{\scriptscriptstyle{A}}}$ contribution in the correlation limit reads:
\beq
\label{def_hard_qbarq}
 \langle \, I_{q\bar q} \, \rangle_{x_{\scriptscriptstyle{A}}} &=&  
 N_c \, g_s^2 \, \pi^{3} \, 
\mathcal{M}_{q\bar{q}}^{\bar{\lambda}\bar{\lambda}';\bar{\eta}\bar{\eta}'}(\xi_{1},\xi_{2})
\big[{\rm H}_{q\bar{q}}\big]_{ij}^{\bar{\lambda}\bar{\lambda}^\prime ; \bar{\eta}\bar{\eta}^\prime} \nonumber\\
&\times& \bigg[\frac{1}{2}\delta^{ij}\mathcal{F}_{WW}(x_{\scriptscriptstyle{A}},{\bf q}_{\sT})+\frac{1}{2}\Big(2 \frac{q_{\sT}^{i}q_{\sT}^{j}}{{\bf q}_{\sT}^{2}}-\delta^{ij}\Bigr)\mathcal{H}_{WW}(x_{\scriptscriptstyle{A}},{\bf q}_{\sT})\bigg]\;,
\eeq
where the hard part is equal to:
\begin{align}
\label{eq:Hqqbar}
&\big[ {\rm H}_{q\bar q} \big]^{\bar\lambda\bar\lambda';\bar\eta\bar\eta'}_{ij}  = 
\Pi^{\bar\lambda';\bar\eta' i}\big[ {\bf k}_3; c_0, {\bf Q} \big] 
\Big\lgroup \frac{\xi_3}{\bar\xi_1} \Pi^{\bar\lambda; \bar\eta j}\big[ {\bf k}_1; \tilde{c}_0, {\bf K}\big] - \Pi^{\bar\eta; \bar\lambda j}\big[ {\bf K}; \tilde{c}_0^{-1}, {\bf k}_1\big] \Big\rgroup
\\
& 
\hspace{1.9cm}
+ \Big\lgroup \frac{\xi_1}{\bar\xi_3} \Pi^{\bar\lambda'; \bar\eta' i}\big[ {\bf k}_3, c_0, {\bf Q}\big] - \Pi^{\bar\eta'; \bar\lambda' i}\big[ {\bf Q}; c_0^{-1}, {\bf k}_3\big] \Big\rgroup 
 \Pi^{\bar\lambda; \bar\eta j}\big[ {\bf k}_1; \tilde{c}_0, {\bf K}\big]\nonumber\\
&+\frac{1}{N_c^2} 
\Big\lgroup \frac{\xi_2}{\bar\xi_3} \Pi^{\bar\lambda'; \bar\eta' i}\big[ {\bf k}_3, c_0, {\bf Q}\big] + \Pi^{\bar\eta'; \bar\lambda' i}\big[ {\bf Q}; c_0^{-1}, {\bf k}_3\big] \Big\rgroup
\Big\lgroup \frac{\xi_2}{\bar\xi_1} \Pi^{\bar\lambda; \bar\eta j}\big[ {\bf k}_1; \tilde{c}_0, {\bf K} \big] + \Pi^{\bar\eta; \bar\lambda j}\big[ {\bf K}; \tilde{c}_0^{-1}, {\bf k}_1\big] \Big\rgroup\;.\nonumber
\end{align}

We can now continue with those contributions to the cross section that include the instantaneous emission of the quark-antiquark-gluon final state from the incoming photon. The first of these contributions is $\langle \, I_{CC} \, \rangle_{x_{\scriptscriptstyle{A}}}$, corresponding to the instantaneous emission both in the amplitude and in the complex conjugate amplitude. After introducing the change of variables given in Eq. \eqref{eq:Cdipolesizes}, and using the expanded expression of the function ${\cal G}_C$, Eq.~\eqref{expanded_GC}, $\langle \, I_{CC} \, \rangle_{x_{\scriptscriptstyle{A}}}$ can be written as:
\beq
\langle \, I_{CC} \, \rangle_{x_{\scriptscriptstyle{A}}} &=&-\frac{N_c}{2} \mathcal{M}_{CC}(\xi_1,\xi_2)
\int_{\x_2\x'_2} e^{i{\bf q}_{\sT}\cdot(\x'_2-\x_2)}
\Big\langle {\rm Tr} \big\{ S_F(\x'_2)\big[ \partial_jS_F^\dagger(\x'_2)\big] S_F(\x_2)\big[ \partial_iS_F(\x_2)\big]\big\}\Big\rangle_{x_{\scriptscriptstyle{A}}} \nonumber\\
&&
\hspace{-1cm}
\times \;  
\int_{{\bf r}_g {\bf r}'_g {\bf r}_{\bar q} {\bf r}'_{\bar q}} 
e^{i{\bf Q}\cdot({\bf r}'_g-{\bf r}_g) \, +\, i{\bf k}_3\cdot({\bf r}'_{\bar q}-{\bf r}_{\bar q})} \, 
{\cal C}\big( \xi_3, \bar\xi_3\, {\bf r}'_{\bar q}; \xi_1\xi_2, {\bf r}'_g \big)  \, 
{\cal C}\big( \xi_3, \bar\xi_3\, {\bf r}_{\bar q}; \xi_1\xi_2, {\bf r}_g \big)
\nonumber\\
&&
\hspace{-1cm}
\times\;  
\bigg\{ 
\Bigg\lgroup \frac{\xi_1}{\bar\xi_3}r'^{j}_g +r'^j_{\bar q} \Bigg\rgroup 
\Bigg\lgroup \frac{\xi_1}{\bar\xi_3}r^{i}_g +r^i_{\bar q} \Bigg\rgroup 
+ r'^j_g r^i_g
-\frac{1}{N_c^2}
\Big\lgroup \frac{\xi_2}{\bar\xi_3}r'^j_g- r'^j_{\bar q} \Big\rgroup
\Big\lgroup \frac{\xi_2}{\bar\xi_3}r^i_g- r^i_{\bar q}\Big\rgroup
\bigg\}\;.
\eeq
The integrals over the dipole sizes ${\bf r}_g$, ${\bf r}'_g$, ${\bf r}_{\bar q}$ and ${\bf r}'_{\bar q}$ can be performed thanks to the following expressions
\begin{align}
&\int_{\mathbf{r}_{g}\mathbf{r}_{\bar{q}}}
e^{- i{\bf Q}\cdot{\bf r}_g \, - \,i {\bf k}_3\cdot{\bf r}_{\bar q}} \, 
r_{g}^{i} \, 
{\cal C}\bigl(\xi_{3},\bar{\xi}_{3} \, \mathbf{r}_{\bar{q}};\xi_{1}\xi_{2},\mathbf{r}_{g}\bigr)  
= 
- \frac{i}{\xi_3 \bar{\xi}_3}\,  \frac{1}{c_0} \, \Pi^{i;jj}[{\bf Q}; c_0^{-1},{\bf k}_3]\;,\nonumber\\
&\int_{\mathbf{r}_{g}\mathbf{r}_{\bar{q}}}
e^{- i{\bf Q}\cdot{\bf r}_g \, - \,i {\bf k}_3\cdot{\bf r}_{\bar q}} \, 
r_{\bar{q}}^{i} \, 
{\cal C}\bigl(\xi_{3},\bar{\xi}_{3}\mathbf{r}_{\bar{q}};\xi_{1}\xi_{2},\mathbf{r}_{g}\bigr) 
= - 
\frac{i}{\xi_3 \bar{\xi}_3} \, \Pi^{i;jj}[{\bf k}_3; c_0,{\bf Q}]\;.
\end{align}
Finally, one obtains:
\beq
\label{def_hard_CC}
\langle \, I_{CC} \, \rangle_{x_{\scriptscriptstyle{A}}} &=& N_c \, g_s^2 \, \pi^3\, 
{\cal M}_{CC}(\xi_1,\xi_2)  \, \big[ {\rm H}_{CC} \big]_{ij} 
\nonumber\\
&\times& \bigg[\frac{1}{2}\delta^{ij}\mathcal{F}_{WW}(x_{\scriptscriptstyle{A}},{\bf q}_{\sT})+\frac{1}{2}\Big(2 \frac{q_{\sT}^{i}q_{\sT}^{j}}{{\bf q}_{\sT}^{2}}-\delta^{ij}\Bigr)\mathcal{H}_{WW}(x_{\scriptscriptstyle{A}},{\bf q}_{\sT})\bigg]\;,
\eeq
with the hard part equal to
\begin{align}  &\big[{\rm H}_{CC}\big]_{ij} =\frac{1}{\xi_3^2\bar\xi_3^2} \Bigg\{ \frac{1}{c_{0}^{2}} \, 
\Pi^{i;kk}\big[{\bf Q};c_{0}^{-1},{\bf k}_{3}\big] \; 
\Pi^{j;kk}\big[{\bf Q};c_{0}^{-1},{\bf k}_{3}\big] \label{eq:HCC}\\
 &+
 \Big\lgroup \!
\frac{\xi_{3}\bar{\xi}_{3}}{\xi_{2}}\Pi^{i;kk}\big[{\bf Q};c_{0}^{-1},{\bf k}_{3}\big]+\Pi^{i;kk}\big[{\bf k}_{3};c_{0},{\bf Q}\big]\Big\rgroup \!\!\!
\Big\lgroup \!
\frac{\xi_{3}\bar{\xi}_{3}}{\xi_{2}}\Pi^{j;kk}\big[{\bf Q};c_{0}^{-1},{\bf k}_{3}\big]+\Pi^{j;kk}\big[{\bf k}_{3};c_{0},{\bf Q}\big]\Big\rgroup \nonumber\\
 & -\frac{1}{N_{c}^{2}} \!
 \Big\lgroup \!
 \frac{\xi_{3}\bar{\xi}_{3}}{\xi_{1}}\Pi^{i;kk}\big[{\bf Q};c_{0}^{-1},{\bf k}_{3}\big]-\Pi^{i;kk}\big[{\bf k}_{3};c_{0},{\bf Q}\big] \!\! \Big\rgroup \!\!\!
 \Big\lgroup \!
 \frac{\xi_{3}\bar{\xi}_{3}}{\xi_{1}}\Pi^{j;kk}\big[{\bf Q};c_{0}^{-1},{\bf k}_{3}\big]-\Pi^{j;kk}\big[{\bf k}_{3};c_{0},{\bf Q}\big] \!\!\Big\rgroup \Bigg\}
 \;.\nonumber
\end{align}

The last two contributions to the cross section are the terms with the instantaneous emission in the conjugate amplitude, and the gluon emission from the quark resp. antiquark in the amplitude. Their calculation is completely analogous to the previous cases, and for $\langle \, I_{Cq} \, \rangle_{x_{\scriptscriptstyle{A}}}$ one obtains:
\begin{align}
\langle \, I_{Cq} \, \rangle_{x_{\scriptscriptstyle{A}}}
& =
 N_c \, g_s^2 \, \pi^3 \, 
{\cal M}_{Cq}^{\bar{\lambda}\bar{\eta}}\big(\xi_{1},\xi_2 \big) \, 
 \big[ {\rm H}_{Cq} \big]_{ij}^{\bar\lambda\bar\eta}\nonumber\\
&\times
\bigg[\frac{1}{2}\delta^{ij}\mathcal{F}_{WW}(x_{\scriptscriptstyle{A}},{\bf q}_{\sT})+\frac{1}{2}\Big(2 \frac{q_{\sT}^{i}q_{\sT}^{j}}{{\bf q}_{\sT}^{2}}-\delta^{ij}\Bigr)\mathcal{H}_{WW}(x_{\scriptscriptstyle{A}},{\bf q}_{\sT})\bigg]\;,
\end{align}
with 
\begin{align} 
\label{eq:HCq}
 &\big[{\rm H}_{Cq}\big]_{ij}^{\bar{\lambda}\bar{\eta}}
 =
\frac{1}{\xi_3 \, \bar \xi_3} \Bigg\{
 \frac{1}{c_{0}} \, 
 \Pi^{\bar{\lambda};\bar{\eta} i}\big[{\bf k}_{3};c_{0},{\bf Q}\big] \, 
 \Pi^{j;kk}\big[{\bf Q};c_{0}^{-1},{\bf k}_{3}\big]\\
 &+
 \Big\lgroup \!
 \frac{\xi_{1}}{\bar{\xi}_{3} }\Pi^{\bar{\lambda};\bar{\eta} j}\big[{\bf k}_{3};c_{0},{\bf Q}\big]
 -
 \Pi^{\bar{\eta};\bar{\lambda} j}\big[{\bf Q};c_{0}^{-1},{\bf k}_{3}\big]\!
 \Big\rgroup \!\!
 \Big\lgroup \!
 \frac{\bar{\xi}_{3}\xi_{3}}{\xi_{2}}\Pi^{i;kk}\big[{\bf Q};c_{0}^{-1},{\bf k}_{3}\big]+\Pi^{i;kk}\big[{\bf k}_{3};c_{0},{\bf Q}\big]\!\Big\rgroup
\nonumber\\
 & -
\frac{1}{N_{c}^{2}}
\Big\lgroup \!
\frac{\xi_{2}}{\bar{\xi}_{3}}\Pi^{\bar{\lambda};\bar{\eta} i}\big[{\bf k}_{3};c_{0},{\bf Q}\big]
+
\Pi^{\bar{\eta};\bar{\lambda} i}\big[{\bf Q};c_{0}^{-1},{\bf k}_{3}\big] \!
\Big\rgroup\!\!
\Big\lgroup \!
\frac{\bar{\xi}_{3}\xi_{3}}{\xi_{2}}\Pi^{j;kk}\big[{\bf Q};c_{0}^{-1},{\bf k}_{3}\big]-\Pi^{j;kk}\big[{\bf k}_{3};c_{0},{\bf Q}\big]\!\Big\rgroup \Bigg\}
\;.\nonumber
\end{align}
Likewise, the result for $\langle \, I_{C\bar q} \, \rangle_{x_{\scriptscriptstyle{A}}}$ is:
\begin{align}
\langle \, I_{C\bar q} \, \rangle_{x_{\scriptscriptstyle{A}}}
& =
 N_c \, g_s^2 \, \pi^3 \, 
{\cal M}_{C\bar q}^{\bar{\lambda}\bar{\eta}}\big(\xi_{1},\xi_2 \big) \, 
 \big[ {\rm H}_{C\bar q} \big]_{ij}^{\bar\lambda\bar\eta}\nonumber\\
&\times
\bigg[\frac{1}{2}\delta^{ij}\mathcal{F}_{WW}(x_{\scriptscriptstyle{A}},{\bf q}_{\sT})+\frac{1}{2}\Big(2 \frac{q_{\sT}^{i}q_{\sT}^{j}}{{\bf q}_{\sT}^{2}}-\delta^{ij}\Bigr)\mathcal{H}_{WW}(x_{\scriptscriptstyle{A}},{\bf q}_{\sT})\bigg]\;,
\end{align}
where the hard part reads:
\beq
\label{eq:HCbarq}
\big[ {\rm H}_{C\bar q} \big]_{ij}^{\bar\lambda\bar\eta}&&= \frac{1}{\xi_3 \, \bar \xi_3} \, \Bigg\{
\Big\lgroup 
\frac{\xi_1}{\bar\xi_3} \frac{1}{c_0} 
\Pi^{j; kk}\big[ {\bf Q}; c_0^{-1}, {\bf k}_3 \big] + \Pi^{j, kk}\big[ {\bf k}_3; c_0, {\bf Q}\big] 
\Big\rgroup
\Pi^{\bar\lambda; \bar\eta i}\big[ {\bf k}_1; \tilde{c}_0, {\bf K}\big] 
\\
&&+ \; 
\frac{1}{c_0}\, \Pi^{j; kk}\big[ {\bf Q}; c_0^{-1}, {\bf k}_3\big] 
\Big\lgroup 
\frac{\xi_3}{\bar\xi_1} \Pi^{\bar\lambda; \bar\eta i}\big[ {\bf k}_1; \tilde{c}_0, {\bf K}\big] 
-
\Pi^{\bar\eta; \bar\lambda i}\big[ {\bf K}; \tilde{c}_0^{-1}, {\bf k}_1\big] 
\Big\rgroup 
\nonumber\\
&&
\hspace{-1.7cm}
-\frac{1}{N_c^2}
\Big\lgroup 
\Pi^{j, kk}\big[ {\bf k}_3; c_0; {\bf Q}\big] 
- \frac{\xi_2}{\bar\xi_3}\, \frac{1}{c_0} \, \Pi^{j, kk}\big[ {\bf Q}; c_0^{-1}, {\bf k}_3\big] \!
\Big\rgroup\!\!
\Big\lgroup 
\frac{\xi_2}{\bar\xi_1}\Pi^{\bar\lambda; \bar\eta i}\big[ {\bf k}_1; \tilde{c}_0; {\bf K}\big] 
+
\Pi^{\bar\eta; \bar\lambda i}\big[{\bf K}; \tilde{c}_0^{-1}, {\bf k}_1\big] \! 
\Big\rgroup \Bigg\}\;. \nonumber
\eeq
Combining all separate contributions, we reach the main result of this section, i.e. the TMD factorized expression for the photoproduction of three jets in the correlation limit at low-$x$:
\beq
\label{total_Fact}
(2\pi)^{9}\frac{\mathrm{d}\sigma^{\gamma A\to q\bar{q}g+X}}{\mathrm{d}^{3}\vec{k}_{1}\mathrm{d}^{3}\vec{k}_{2}\mathrm{d}^{3}\vec{k}_{3}}\bigg|_{\rm corr. \,  limit}
&&= 
2\pi\delta\big(p^{+}-\sum_{i=1}^3k_{i}^{+}\big) 
\, \big[ {\rm H} \big]_{ij}^{\rm total} \\
&&
\times
\bigg[\frac{1}{2}\delta^{ij}\mathcal{F}_{WW}(x_{\scriptscriptstyle{A}},{\bf q}_{\sT})+\frac{1}{2}\Big(2 \frac{q_{\sT}^{i}q_{\sT}^{j}}{{\bf q}_{\sT}^{2}}-\delta^{ij}\Bigr)\mathcal{H}_{WW}(x_{\scriptscriptstyle{A}},{\bf q}_{\sT})\bigg]\;,\nonumber
\eeq
where $\mathcal{F}_{WW}(x_{\scriptscriptstyle{A}},{\bf q}_{\sT})$ and $\mathcal{H}_{WW}(x_{\scriptscriptstyle{A}},{\bf q}_{\sT})$ are the unpolarized and linearly polarized Weizs\"acker-Williams gluon TMDs defined in Eq.~\eqref{def_TMD}, and where the hard factor is found to be:
\beq
\big[ {\rm H} \big]_{ij}^{\rm total} &&= N_c \, g_e^2 \, g_s^4\, \pi^3 \, \frac{1}{k_2^+ p^+ } 
\Big\{ 
\mathcal{M}_{qq}^{\bar\lambda\bar{\lambda}';\bar\eta\bar{\eta}'} \Big( \bar\xi_3, \frac{\xi_2}{\bar\xi_3} \Big) 
\big[ {\rm H}_{qq} \big]_{ij}^{\bar\lambda\bar\lambda'; \bar\eta\bar\eta'}
+
{\cal M}^{\bar\eta\bar\eta'; \bar\lambda\bar\lambda'}_{\bar q\bar q} \Big( \xi_1, \frac{\xi_2}{\bar\xi_1} \Big) 
\big[ {\rm H}_{\bar q\bar q} \big]_{ij}^{\bar\lambda\bar\lambda'; \bar\eta\bar\eta'}
\nonumber\\
&&
+ \,2 \, {\cal M}^{\bar\eta\bar\eta'; \bar\lambda\bar\lambda'}_{q\bar q} \big( \xi_1, \xi_2\big) 
\big[ {\rm H}_{q\bar q} \big]_{ij}^{\bar\lambda\bar\lambda'; \bar\eta\bar\eta'}
+
{\cal M}_{CC}(\xi_1,\xi_2)
\big[ {\rm H}_{CC} \big]_{ij}
\nonumber\\
&&
+\,
2\, {\cal M}_{Cq}^{\bar\lambda\bar\eta}(\xi_1,\xi_2)\, \big[ {\rm H}_{Cq} \big]^{\bar\lambda\bar\eta}_{ij}
+  
2 \, {\cal M}^{\bar\lambda\bar\eta}_{C\bar q}(\xi_1,\xi_2) \, \big[ {\rm H}_{C\bar q} \big]^{\bar\lambda\bar\eta}_{ij}
\Big\}\;.
\eeq
The explicit expressions of the products of splitting functions and of the hard factors are given in Eqs.~\eqref{Mqq_def}-\eqref{eq:Hqq}, \eqref{def_M_barqbarq}-\eqref{eq:Hqbarqbar}, \eqref{def_M_qbarq}-\eqref{eq:Hqqbar}, \eqref{def_M_CC}-\eqref{eq:HCC}, \eqref{def_M_Cq}-\eqref{eq:HCq} and \eqref{def_M_Cbarq}-\eqref{eq:HCbarq}. We note that, as expected, the whole cross section is symmetric under the exchange of the outgoing quark with the antiquark. In particular, this symmetry even holds on the level of the hard parts: 
\beq
\mathcal{M}_{qq}^{\bar\lambda\bar{\lambda}';\bar\eta\bar{\eta}'} \Big( \bar\xi_3, \frac{\xi_2}{\bar\xi_3} \Big) 
\big[ {\rm H}_{qq} \big]_{ij}^{\bar\lambda\bar\lambda'; \bar\eta\bar\eta'}
&\overset{1\leftrightarrow3}{=}&
{\cal M}^{\bar\eta\bar\eta'; \bar\lambda\bar\lambda'}_{\bar q\bar q} \Big( \xi_1, \frac{\xi_2}{\bar\xi_1} \Big) 
\big[ {\rm H}_{\bar q\bar q} \big]_{ij}^{\bar\lambda\bar\lambda'; \bar\eta\bar\eta'}\;,\nonumber\\
{\cal M}_{Cq}^{\bar\lambda\bar\eta}(\xi_1,\xi_2)\, \big[ {\rm H}_{Cq} \big]^{\bar\lambda\bar\eta}_{ij}
&\overset{1\leftrightarrow3}{=}&
{\cal M}^{\bar\lambda\bar\eta}_{C\bar q}(\xi_1,\xi_2) \,   \big[ {\rm H}_{C\bar q} \big]^{\bar\lambda\bar\eta}_{ij}\;,\nonumber\\
{\cal M}^{\bar\eta\bar\eta'; \bar\lambda\bar\lambda'}_{q\bar q} \big( \xi_1, \xi_2\big) 
\big[ {\rm H}_{q\bar q} \big]_{ij}^{\bar\lambda\bar\lambda'; \bar\eta\bar\eta'}
&\overset{1\leftrightarrow3}{=}&
{\cal M}^{\bar\eta\bar\eta'; \bar\lambda\bar\lambda'}_{q\bar q} \big( \xi_1, \xi_2\big) 
\big[ {\rm H}_{q\bar q} \big]_{ij}^{\bar\lambda\bar\lambda'; \bar\eta\bar\eta'}\;,\nonumber\\
{\cal M}_{CC}(\xi_1,\xi_2)
\big[ {\rm H}_{CC} \big]_{ij}
&\overset{1\leftrightarrow3}{=}&
{\cal M}_{CC}(\xi_1,\xi_2)
\big[ {\rm H}_{CC} \big]_{ij}\;. \label{eq:hardpartssymmetry}
\eeq
Moreover, in accordance with TMD factorization, the hard parts are collinear in the sense that they \emph{do not} have any dependence on the transverse momentum $q_{\scriptscriptstyle{T}}$ of the incoming gluon, since such a dependence would only come at higher twists in the expansion $|q_{\scriptscriptstyle{T}}|/|k_i|\ll1$. Inside the hard parts, the total transverse momentum is therefore strictly zero: $\mathbf{k}_1+\mathbf{k}_2+\mathbf{k}_3=0$, and the $q_{\scriptscriptstyle{T}}$-dependence is only contained in the gluon correlator (\ref{def_TMD}) which parametrizes the target hadron in terms of gluon TMDs.

\section{\label{dilute}Weak-field approximation in the CGC and the HEF limit}

As is well known, the high-energy factorization (HEF) framework is compassed within the CGC formalism, and can be extracted by performing the weak-field approximation for the gauge fields of the target. In this limit, the Wilson line structure in the amplitude and in the complex conjugate amplitude should be expanded to second order in powers of the background field $\alpha^-_a(x^+,{\bf x})$ of the target. However, the structure of the functions ${\cal G}_q$, ${\cal G}_{\bar q}$ and ${\cal G}_{ C}$, which incorporate the Wilson lines at the amplitude level, ensure that it is enough to perform the expansion to the first order in powers of the background field since the zeroth order terms vanish (see Eqs. \eqref{G_q}, \eqref{G_qbar} and \eqref{G_inst}).
The standard weak-field expansion of a fundamental Wilson line reads
\beq
\label{expansion}
S_{F}({\bf x})_{ij}=\delta_{ij}+ig_s\, \int dx^+ t^a_{ij}\, \alpha^-_a(x^+,{\bf x})\, +\mathcal{O}\big[ (\alpha^-_a)^2\big]\;.
\eeq
In order to compute the dilute limit of the cross section, we will follow the same strategy as in the correlation limit, i.e.: we first perform the expansion on the level of the amplitude for the quark, antiquark and instantaneous contributions, and then combine the results on the cross section level. To do so, we introduce the \textit{reduced quark amplitude}
\beq
\big[ {\cal T}_q\big]^{c,\bar\eta\bar\lambda}_{ij}=\prod_{i=1}^3 \int_{\x_i} 
e^{-i\bk_i\cdot\x_i}
A^{\bar\eta}(\x_1-\x_2)  \, 
{\cal G}_q\bigg[ \xi_3,\frac{\xi_1}{\bar\xi_3}; \frac{\xi_1\x_1+\xi_2\x_2}{\bar\xi_3},\x_1,\x_2,\x_3\bigg]^{c\bar\lambda}_{ij}\;.
\eeq
As is clear from the above expression, we define the reduced amplitude as the regular amplitude, stripped from the tensor structure of the product of the splitting functions ${\cal M}^{\bar\lambda\bar\lambda';\bar\eta\bar\eta'}$, such that, taking the example of the $I_{qq}$ contribution to the cross section \eqref{Iqq}:
\beq
\langle I_{qq} \rangle_{x_{\scriptscriptstyle{A}}}=\mathcal{M}_{qq}^{\bar{\eta}\bar{\eta}';\bar{\lambda}\bar{\lambda}'}\big(\bar{\xi}_3,\frac{\xi_2}{\bar{\xi}_3}\big)\big\langle \mathrm{Tr} \big[ {\cal T}_q^\dagger\big]^{\bar\eta'\bar\lambda'} \big[ {\cal T}_q\big]^{\bar\eta\bar\lambda} \big\rangle_{x_{\scriptscriptstyle{A}}}\;.
\eeq
After introducing the following change of variables
\beq
{\bf r}=\x_1-\x_2\;, 
\qquad
{\bf b}=\frac{1}{2}(\x_1+\x_2)\;,
\qquad
\bar{\x}_3= \x_3-\frac{\xi_1}{\bar\xi_3}\x_1-\frac{\xi_2}{\bar\xi_3}\x_2\;,
\eeq
and using Eq.~\eqref{expansion} for the expansion of the Wilson lines in the function ${\cal G}_q$, the reduced quark amplitude can be cast into the following compact form:
\begin{align}
\label{T_q_int}
\big[ {\cal T}_q\big]^{c,\bar\eta\bar\lambda}_{ij}&=ig_s\, \int_{\bb}e^{-i{\bf q}_{\sT}\cdot\bb}\int \mathrm{d}x^+\, \alpha^-_a(x^+,\bb)
\\
&\times
\int_{\br \bar\x_3}
\bigg\{
(t^at^c)_{ij} \, 
A^{\bar\eta}(\br) \, {\cal A}^{\bar\lambda}\Big( \xi_3,\bar\x_3;\frac{\xi_1}{\bar\xi_3},\br\Big) \,
\Big[ e^{-i\br\cdot({\bf Q}-{\bf q}_{\sT})-i{\bf k}_3\cdot{\bf \bar x}_3} - e^{-i\br\cdot{\bf Q}-i{\bf k}_3\cdot{\bf \bar x}_3} \Big]
\nonumber\\
&
\hspace{0.2cm}
+
(t^ct^a)_{ij}  \, 
A^{\bar\eta}(\br) \, {\cal A}^{\bar\lambda}\Big( \xi_3,\bar\x_3;\frac{\xi_1}{\bar\xi_3},\br\Big) \,
\Big[ e^{-i\br\cdot{\bf Q}-i{\bf k}_3\cdot{\bf \bar x}_3} - e^{-i\br\cdot\left({\bf Q}-\frac{\xi_1}{\bar\xi_3}{\bf q}_{\sT}\right) -i{\bf \bar x}_3\cdot({\bf k}_3-{\bf q}_{\sT})} \Big]
\nonumber\\
&
\hspace{0.2cm}
+
(t^ct^a)_{ij}  \, 
A^{\bar\eta}(\br) \, A^{\bar\lambda}(\bar\x_3) \,
\Big[ e^{-i\br\cdot\left({\bf Q}-\frac{\xi_1}{\bar\xi_3}{\bf q}_{\sT}\right) -i{\bf \bar x}_3\cdot({\bf k}_3-{\bf q}_{\sT})} 
- e^{-i\br\cdot\left({\bf Q}-\frac{\xi_1}{\bar\xi_3}{\bf q}_{\sT}\right) -i{\bf \bar x}_3\cdot({\bf k}_3-{\bf q}_{\sT})} \Big] \bigg\}.
\nonumber
\end{align} 
In the above expression, ${\bf q}_{\sT}$ is the total transverse momentum defined in Eq.~\eqref{eq:def_total_mom} and momentum ${\bf Q}$ is defined in Eq.~\eqref{eq:def:Q_mom}.

In Eq.~\eqref{T_q_int} the integrals over $\br$ and $\bar\x_3$ are factorized from the rest of the expression and can be performed in straightforward manner. The reduced quark amplitude then becomes:
\begin{align}
\label{T_q_final}
\big[ {\cal T}_q\big]^{c,\bar\eta\bar\lambda}_{ij}&=ig_s\, \int_{\bb}e^{-i{\bf q}_\sT \cdot\bb}
\int \mathrm{d}x^+\, \alpha^-_a(x^+,\bb)\\
&\times
\Big\{ 
(t^at^c)_{ij}\, {\tilde A}^{\bar\lambda}(\bk_3)
\Big[ {\cal {\tilde A}}^{\bar \eta}({\bf Q}-{\bf q}_\sT;c_0,\bk_3) - {\cal \tilde A}^{\bar \eta}({\bf Q};c_0,\bk_3)\Big]
\nonumber\\
&
\hspace{0.3cm}
+
(t^ct^a)_{ij}
\Big\lgroup 
{\tilde A}^{\bar\lambda}(\bk_3) 
\Big[ {\cal A}^{\bar \eta}({\bf Q};c_0,\bk_3) - {\tilde A}^{\bar\eta}\Big({\bf Q}-\frac{\xi_1}{\bar\xi_3}{\bf q}_\sT\Big) \Big]
\nonumber\\
&
\hspace{1.7cm}
+
{\tilde A}^{\bar\lambda}(\bk_3-{\bf q}_{\sT} ) 
\Big[ {\tilde A}^{\bar\eta}\Big({\bf Q}-\frac{\xi_1}{\bar\xi_3}{\bf q}_\sT\Big) - {\cal \tilde A}^{\bar\eta}\Big({\bf Q}-\frac{\xi_1}{\bar\xi_3}{\bf q}_\sT:c_0, \bk_3-{\bf q}_\sT\Big)\Big] \Big\rgroup\Big\}\, , 
\nonumber
\end{align}   
where ${\tilde A}^{\lambda}(\bk)$ and $\tilde{\cal A}^{\lambda}(\bk;c,\bp)$ are the standard and modified  Weizs\"acker-Williams fields in momentum space, given by: 
\beq
{\tilde A}^{\lambda}(\bk)=\frac{k^\lambda}{\bk^2} \hspace{0.5cm} \text {and} \hspace{0.5cm}
 \tilde{\cal A}^{\lambda}(\bk;c,\bp)=\frac{k^{\lambda}}{\bk^2+c\bp^2}\;,
\eeq
and the combination of forward momentum fractions $c_0=(\xi_1\xi_2) / (\bar\xi_3^2\xi_3)$ which was defined earlier. 

In a similar way, the reduced antiquark amplitude is defined as 
\beq
\big[ {\cal T}_{\bar q}\big]^{c,\bar\eta\bar\lambda}_{ij}=\prod_{i=1}^3 \int_{\x_i} 
e^{-i\bk_i\cdot\x_i}
A^{\bar\eta}(\x_3-\x_2)  \, 
{\cal G}_{\bar q}\bigg[ \xi_1,\frac{\xi_3}{\bar\xi_1}; \frac{\xi_2}{\bar \xi_1}\x_2+\frac{\xi_3}{\bar\xi_1}\x_3\x_1,\x_2,\x_3\bigg]^{c\bar\lambda}_{ij}\;.
\eeq
Following the same steps as for the reduced quark amplitude, the weak-field limit of the reduced antiquark amplitude can be computed and reads:
\begin{align}
\label{T_qbar_final}
&
\big[ {\cal T}_{\bar q}\big]^{c,\bar\eta\bar\lambda}_{ij} = ig_s\, \int_{\bb}e^{-i{\bf q}_\sT\cdot\bb}
\int \mathrm{d}x^+\, \alpha^-_a(x^+,\bb)
\\
&
\times
\Big\{ 
(t^ct^a)_{ij}\, {\tilde A}^{\bar\lambda}(\bk_1)
\Big[ {\cal \tilde A}^{\bar\eta}({\bf K};\tilde c_0, \bk_1) - {\cal \tilde A}^{\bar\eta}({\bf K}-{\bf q}_\sT;\tilde c_0, \bk_1)\Big]
\nonumber\\
&
\hspace{0.3cm}
+
(t^at^c)_{ij}\Big\lgroup {\tilde A}^{\bar\lambda}(\bk_1) 
\Big[ {\tilde A}^{\bar\eta}\Big({\bf K}-\frac{\xi_3}{\bar\xi_1}{\bf q}_\sT\Big) -{\cal \tilde A}^{\bar\eta}({\bf K};\tilde c_0, \bk_1)\Big] 
\nonumber\\
&
\hspace{1.7cm}
+
{\tilde A}^{\bar\lambda}(\bk_1-{\bf q}_\sT)
\Big[ {\cal \tilde A}^{\bar\eta}\Big({\bf K}-\frac{\xi_3}{\bar\xi_1}{\bf q}_\sT;\tilde c_0, \bk_1-{\bf q}_\sT\Big)-{\tilde A}^{\bar\eta}\Big({\bf K}-\frac{\xi_3}{\bar\xi_1}{\bf q}_\sT\Big)\Big]\Big\rgroup\Big\}\, , 
\nonumber
\end{align}
 where momentum ${\bf K}$ is defined in Eq.~\eqref{def:eq:K_mom} and the combination $\tilde{c}_0=(\xi_2\xi_3)/(\bar\xi_1^2\xi_1)$ which was also defined earlier. 

Finally, we define the reduced amplitude for the instantaneous contribution:
\begin{align}
\big[ {\cal T}_{C}\big]^{c,\bar\lambda}_{ij}=\prod_{i=1}^3 \int_{\x_i} 
e^{-i\bk_i\cdot\x_i}
\, 
{\cal G}_{C}\big[ \xi_3,\xi_1\xi_2; \bar\xi_3\x_3-\xi_1\x_1-\xi_2\x_2,\x_1,\x_2,\x_3\big]^{c\bar\lambda}_{ij}\;,
\end{align}
which, after the weak-field expansion, takes the following form
\begin{align}
\label{T_inst_final}
\big[ {\cal T}_{C}\big]^{c,\bar\lambda}_{ij}&=ig_s\, \int_{\bb}e^{-i{\bf q}_\sT\cdot\bb}\int dx^+\alpha^-_a(x^+,\bb)
\\
&
\times\bigg\{ 
(t^at^c)_{ij}
\Big\lgroup {\tilde C}\big[ \xi_3, \bar\xi_3({\bf Q}-{\bf q}_\sT);\xi_1\xi_2, \bk_3\big] 
- {\tilde C}\big[ \xi_3, \bar\xi_3{\bf Q};\xi_1\xi_2, \bk_3\big] \Big\rgroup
\nonumber
 \\
 &
 \hspace{0.3cm}
 +
 (t^ct^a)_{ij}
 \Big\lgroup {\tilde C}\big[ \xi_3, \bar\xi_3{\bf Q};\xi_1\xi_2, \bk_3\big]
 -
 {\tilde C}\Big[ \xi_3,\bar\xi_3\Big({\bf Q}-\frac{\xi_1}{\bar\xi_3}{\bf q}_\sT\Big);\xi_1\xi_2, \bk_3-{\bf q}_\sT\Big]\Big\rgroup\bigg\}\, , 
 \nonumber
\end{align}
where the function ${\tilde {\cal C}}(\xi_3,\bar\xi_3\bp;\xi_1\xi_2,\bk)$ is the Fourier transform of the Coulomb field  defined in Eq.~\eqref{eq:Coulombfield} and reads:
\beq
\label{tildeC_def}
{\tilde {\cal C}}(\xi_3,\bar\xi_3\bp;\xi_1\xi_2,\bk)=\frac{\bar{\xi}_3}{\xi_3\bar\xi_3^2\bp^2+\xi_1\xi_2\bk^2}\;.
\eeq
 
We will now use the explicit expressions of the reduced amplitudes in the weak-field limit: Eq.~\eqref{T_q_final}, \eqref{T_qbar_final}, and \eqref{T_inst_final}, to calculate the HEF expression for the photoproduction cross section.

Let us start with the $\langle \,  I_{qq} \, \rangle_{x_{\scriptscriptstyle{A}}}$ contribution to the partonic cross section. Written in terms of the weak-field expansion of the reduced quark amplitude Eq.~\eqref{T_q_final}, this contribution can be cast into the following form: 
\begin{align}
\langle\, I_{qq}\, \rangle_{x_{\scriptscriptstyle{A}}}&=g_s^2 
\frac{N_c}{4} \delta^{aa'}  {\cal W}^{\bar\lambda\bar\lambda';  \bar\eta\bar\eta'}_{qq}
{\cal M}^{\bar\lambda\bar\lambda';\bar\eta\bar\eta'}_{qq}\Big(\bar\xi_3,\frac{\xi_2}{\bar\xi_3}\Big)\, 
\nonumber\\
&
\times \, \int_{\bb\bb'}e^{-i{\bf q}_\sT\cdot (\bb-\bb')}
\int \mathrm{d}x^+ \mathrm{d}x'^+ \big\langle \alpha_{a'}^-(x'^+,\bb')\alpha^-_a(x^+,\bb)\big\rangle_{x_{\scriptscriptstyle{A}}}\,  \end{align}
where the Weizs\"acker-Williams field structure ${\cal W}^{\bar\lambda\bar\lambda';  \bar\eta\bar\eta'}_{qq}$ for this contribution is computed as 
\begin{align}
\label{def_WW_qq}
{\cal W}^{\bar\lambda\bar\lambda';  \bar\eta\bar\eta'}_{qq}&=  
\Big\lgroup
{\tilde A}^{\bar\lambda}(\bk_3)
\Big[ {\cal \tilde A}^{\bar\eta}({\bf Q}-{\bf q}_\sT;c_0, \bk_3)-{\cal \tilde A}^{\bar\eta}({\bf Q};c_0, \bk_3)\Big]\Big\rgroup
\\
&
\hspace{2.4cm}
\times
\Big\lgroup 
{\tilde A}^{\bar\lambda'}(\bk_3)
\Big[ {\cal \tilde A}^{\bar\eta'}({\bf Q}-{\bf q}_\sT;c_0,\bk_3)-{\cal \tilde A}^{\bar\eta'}({\bf Q};c_0,\bk_3)\Big]\Big\rgroup  
\nonumber\\
&
+
\Big\lgroup
{\tilde A}^{\bar\lambda}(\bk_3)\Big[ {\cal \tilde A}^{\bar\eta}({\bf Q};c_0,\bk_3)-{\tilde A}^{\bar\eta}\Big({\bf Q}-\frac{\xi_1}{\bar\xi_3}{\bf q}_\sT\Big)\Big]
\nonumber\\
&
\hspace{2.4cm}
+
{\tilde A}^{\bar\lambda}(\bk_3-{\bf q}_\sT)
\Big[ {\tilde A}^{\bar\eta}\Big({\bf Q}-\frac{\xi_1}{\bar\xi_3}{\bf q}_\sT\Big)-{\cal \tilde A}^{\bar\eta}\Big( {\bf Q}-\frac{\xi_1}{\bar\xi_3}{\bf q}_\sT;c_0, \bk_3-{\bf q}_\sT\Big)\Big]\Big\rgroup
\nonumber\\
&
\times
\Big\lgroup
{\tilde A}^{\bar\lambda'}(\bk_3)\Big[ {\cal \tilde A}^{\bar\eta'}({\bf Q};c_0,\bk_3)-{\tilde A}^{\bar\eta'}\Big({\bf Q}-\frac{\xi_1}{\bar\xi_3}{\bf q}_\sT\Big)\Big]
\nonumber\\
&
\hspace{2.4cm}
+
{\tilde A}^{\bar\lambda'}(\bk_3-{\bf q}_\sT)
\Big[ {\tilde A}^{\bar\eta'}\Big({\bf Q}-\frac{\xi_1}{\bar\xi_3}{\bf q}_\sT\Big)-{\cal \tilde A}^{\bar\eta'}\Big( {\bf Q}-\frac{\xi_1}{\bar\xi_3}{\bf q}_\sT;c_0, \bk_3-{\bf q}_\sT\Big)\Big]\Big\rgroup
\nonumber\\
&
\hspace{-0.6cm}
-\frac{1}{N_c^2}
\Big\lgroup 
{\tilde A}^{\bar\lambda}(\bk_3)
\Big[ 
{\cal \tilde A}^{\bar\eta}({\bf Q}-{\bf q}_\sT;c_0, \bk_3)-{\tilde A}^{\bar\eta}\Big({\bf Q}-\frac{\xi_1}{\bar\xi_3}{\bf q}_\sT\Big) \Big]
\nonumber\\
&
\hspace{2.4cm}
+
{\tilde A}^{\bar\lambda}(\bk_3-{\bf q}_\sT)
\Big[ {\tilde A}^{\bar\eta}\Big({\bf Q}-\frac{\xi_1}{\bar\xi_3}{\bf q}_\sT\Big) - {\cal \tilde A}^{\bar \eta}\Big( {\bf Q}-\frac{\xi_1}{\bar\xi_3}{\bf q}_\sT;c_0, \bk_3-{\bf q}_\sT\Big)
\Big]
\Big\rgroup
\nonumber\\
&
\times
\Big\lgroup 
{\tilde A}^{\bar\lambda'}(\bk_3)
\Big[ 
{\cal \tilde A}^{\bar\eta'}({\bf Q}-{\bf q}_\sT;c_0, \bk_3)-{\tilde A}^{\bar\eta'}\Big({\bf Q}-\frac{\xi_1}{\bar\xi_3}{\bf q}_\sT\Big) \Big]
\nonumber\\
&
\hspace{2.4cm}
+
{\tilde A}^{\bar\lambda'}(\bk_3-{\bf q}_\sT)
\Big[ {\tilde A}^{\bar\eta'}\Big({\bf Q}-\frac{\xi_1}{\bar\xi_3}{\bf q}_\sT\Big) - {\cal \tilde A}^{\bar \eta'}\Big( {\bf Q}-\frac{\xi_1}{\bar\xi_3}{\bf q}_\sT;c_0, \bk_3-{\bf q}_\sT\Big)
\Big]
\Big\rgroup . 
\end{align}
Using the relation
\begin{align}
\mathcal{F}_{g/A}(x_{\scriptscriptstyle{A}},\mathbf{q}_\sT)&=\frac{\mathbf{q}_\sT^{2}}{(2\pi)^2}\int\mathrm{d}x^{+}\mathrm{d}x'^{+}\int_{\mathbf{bb'}}e^{-i\mathbf{q}_\sT\cdot(\mathbf{\mathbf{b}}-\mathbf{b'})}\langle \alpha_{a}^{-}(x^{+},\mathbf{\mathbf{b}})\alpha_{a}^{-}(x'^{+},\mathbf{\mathbf{b'}})\rangle_{x_{\scriptscriptstyle{A}}}\;,
\end{align} 
between the correlator of the semiclassical gauge fields and the unintegrated gluon PDF $\mathcal{F}_{g/A}$, we finally obtain:
\begin{align}
\langle \, I_{qq}\, \rangle_{x_{\scriptscriptstyle{A}}}&=  N_c \,g_s^2  \pi^3
{\cal M}^{\bar\lambda\bar\lambda';\bar\eta\bar\eta'}_{qq}\Big(\bar\xi_3,\frac{\xi_2}{\bar\xi_3}\Big)\,
{\cal W}^{\bar\lambda\bar\lambda';  \bar\eta\bar\eta'}_{qq} \, \frac{1}{\pi{\bf q}^2_{\sT}} \, \mathcal{F}_{g/A}(x_{\scriptscriptstyle{A}},{\bf q}_{\sT})\;.
\end{align}

The $\langle \, I_{\bar q\bar q} \, \rangle_{x_{\scriptscriptstyle{A}}}$ contribution can be computed in a similar manner. Using the dilute expansion of the reduced antiquark amplitude given in Eq.~\eqref{T_qbar_final}, together with the definition of the unintegrated gluon distribution, one obtains:
\begin{align}
\langle \, I_{\bar q\bar q}\, \rangle_{x_{\scriptscriptstyle{A}}}&=N_c \,g_s^2  \pi^3 
{\cal M}^{\bar\lambda\bar\lambda';\bar\eta\bar\eta'}_{\bar q\bar q}\Big(\xi_1,\frac{\xi_2}{\bar\xi_1}\Big)\,
{\cal W}^{\bar\lambda\bar\lambda';  \bar\eta\bar\eta'}_{\bar q\bar q} \, \frac{1}{\pi{\bf q}^2_{\sT}} \, \mathcal{F}_{g/A}(x_{\scriptscriptstyle{A}},{\bf q}_{\sT})\;,
\end{align}
with the Weizs\"acker-Williams field structure ${\cal W}^{\bar\lambda\bar\lambda';  \bar\eta\bar\eta'}_{\bar q\bar q}$:
\begin{align}
\label{def_WW_barqbarq}
{\cal W}^{\bar\lambda\bar\lambda';  \bar\eta\bar\eta'}_{\bar q\bar q} &= 
\Big\lgroup
{\tilde A}^{\bar\lambda}(\bk_1)
\Big[ {\cal \tilde A}^{\bar\eta}({\bf K};\tilde c_0, \bk_1) - {\cal \tilde A}^{\bar\eta}({\bf K}-{\bf q}_\sT;\tilde c_0, \bk_1)\Big]\Big\rgroup
\\
&
\hspace{2.4cm}
\times
\Big\lgroup 
{\tilde A}^{\bar\lambda'}(\bk_1)
\Big[  {\cal \tilde A}^{\bar\eta'}({\bf K};\tilde c_0,\bk_1) - {\cal \tilde A}^{\bar\eta'}({\bf K}-{\bf q}_\sT;\tilde c_0,\bk_1)\Big]\Big\rgroup  
\nonumber\\
&
+
\Big\lgroup
{\tilde A}^{\bar\lambda}(\bk_1)
\Big[  {\tilde A}^{\bar\eta}\Big({\bf K}-\frac{\xi_3}{\bar\xi_1}{\bf q}_\sT\Big) -  {\cal \tilde A}^{\bar\eta}({\bf K};\tilde c_0,\bk_1)\Big]
\nonumber\\
&
\hspace{2.4cm}
+
{\tilde A}^{\bar\lambda}(\bk_1-{\bf q}_\sT)
\Big[ {\cal \tilde A}^{\bar\eta}\Big( {\bf K}-\frac{\xi_3}{\bar\xi_1}{\bf q}_\sT;\tilde c_0, \bk_1-{\bf q}_\sT\Big) - {\tilde A}^{\bar\eta}\Big({\bf K}-\frac{\xi_3}{\bar\xi_1}{\bf q}_\sT\Big) \Big]\Big\rgroup
\nonumber\\
&
\times
\Big\lgroup
{\tilde A}^{\bar\lambda'}(\bk_1)
\Big[ {\tilde A}^{\bar\eta'}\Big({\bf K}-\frac{\xi_3}{\bar\xi_1}{\bf q}_\sT\Big) - {\cal \tilde A}^{\bar\eta'}({\bf K};\tilde c_0,\bk_1)\Big]
\nonumber\\
&
\hspace{2.4cm}
+
{\tilde A}^{\bar\lambda'}(\bk_1-{\bf q}_\sT)
\Big[ {\cal \tilde A}^{\bar\eta'}\Big( {\bf K}-\frac{\xi_3}{\bar\xi_1}{\bf q}_\sT;\tilde c_0, \bk_1-{\bf q}_\sT\Big) -  {\tilde A}^{\bar\eta'}\Big({\bf K}-\frac{\xi_3}{\bar\xi_1}{\bf q}_\sT\Big)\Big]\Big\rgroup
\nonumber\\
&
\hspace{-0.6cm}
-\frac{1}{N_c^2}
\Big\lgroup 
{\tilde A}^{\bar\lambda}(\bk_1)
\Big[ 
{\tilde A}^{\bar\eta}\Big({\bf K}-\frac{\xi_3}{\bar\xi_1}{\bf q}_\sT\Big) - {\cal \tilde A}^{\bar\eta}({\bf K}-{\bf q}_\sT;\tilde c_0, \bk_1)\Big]
\nonumber\\
&
\hspace{2.4cm}
+
{\tilde A}^{\bar\lambda}(\bk_1-{\bf q}_\sT)
\Big[{\cal \tilde A}^{\bar \eta}\Big( {\bf K}-\frac{\xi_3}{\bar\xi_1}{\bf q}_\sT;\tilde c_0, \bk_1-{\bf q}_\sT\Big) - {\tilde A}^{\bar\eta}\Big({\bf K}-\frac{\xi_3}{\bar\xi_1}{\bf q}_\sT\Big)  
\Big]
\Big\rgroup
\nonumber\\
&
\times
\Big\lgroup 
{\tilde A}^{\bar\lambda'}(\bk_1)
\Big[ 
{\tilde A}^{\bar\eta'}\Big({\bf K}-\frac{\xi_3}{\bar\xi_1}{\bf q}_\sT\Big) -{\cal \tilde A}^{\bar\eta'}({\bf K}-{\bf q}_\sT;\tilde c_0, \bk_1) \Big]
\nonumber\\
&
\hspace{2.4cm}
+
{\tilde A}^{\bar\lambda'}(\bk_1-{\bf q}_\sT)
\Big[ {\cal \tilde A}^{\bar \eta'}\Big( {\bf K}-\frac{\xi_3}{\bar\xi_1}{\bf q}_\sT;\tilde c_0, \bk_1-{\bf q}_\sT\Big) - {\tilde A}^{\bar\eta'}\Big({\bf K}-\frac{\xi_3}{\bar\xi_1}{\bf q}_\sT\Big)  
\Big]
\Big\rgroup . 
\nonumber
\end{align}
Not surprisingly, the above expression is the same as the one we obtained for the quark-quark contribution in Eq.~\eqref{def_WW_qq}, up to the change of the subscripts $1\leftrightarrow 3$. Since, as discussed previously, the functions that encode the product of splitting amplitudes follow the same rule, we have that
\beq
\langle \, I_{\bar q\bar q} \, \rangle_{x_{\scriptscriptstyle{A}}}= \langle \, I_{qq} \, \rangle_{x_{\scriptscriptstyle{A}}}(1\leftrightarrow 3)\;,
\eeq
reflecting the invariance of QCD under charge conjugation.

It turns out that the HEF limits of all the remaining contributions, i.e. $\langle \, I_{q\bar q} \, \rangle_{x_{\scriptscriptstyle{A}}}$, $\langle \, I_{CC} \, \rangle_{x_{\scriptscriptstyle{A}}}$, $\langle \, I_{Cq} \, \rangle_{x_{\scriptscriptstyle{A}}}$ and $\langle \, I_{C\bar q} \, \rangle_{x_{\scriptscriptstyle{A}}}$, which can be calculated with the help of the expansions Eqs.~\eqref{T_q_final},~\eqref{T_q_final} and~\eqref{T_inst_final} on the amplitude level, can be cast in a similar form:
\beq
\langle \, I_{...}\, \rangle_{x_{\scriptscriptstyle{A}}}=N_c \,g_s^2  \pi^3\big({\cal M}(\xi_1,\xi_2) \,\otimes
{\cal W} \big) \, \frac{1}{\pi{\bf q}^2_{\sT}} \, \mathcal{F}_{g/A}(x_{\scriptscriptstyle{A}},{\bf q}_{\sT})\;.
\eeq
where the symbol $\otimes$ denotes a contraction over all open vector indices. The corresponding Weizs\"acker-Williams field structures $\mathcal{W}$ are given by, respectively:
\begin{align}
\label{def_WW_qbarq}
{\cal W}^{\bar\lambda\bar\lambda';  \bar\eta\bar\eta'}_{q\bar q} &= -
\Big\lgroup
{\tilde A}^{\bar\lambda}(\bk_3)
\Big[ {\cal \tilde A}^{\bar\eta}({\bf Q}-{\bf q}_\sT;c_0, \bk_3)-{\cal \tilde A}^{\bar\eta}({\bf Q};c_0, \bk_3)\Big]\Big\rgroup
\\
&
\hspace{0.35cm}
\times
\Big\lgroup 
{\tilde A}^{\bar\lambda'}(\bk_1)
\Big[ {\cal \tilde A}^{\bar\eta'}({\bf K}-{\bf q}_\sT;\tilde c_0,\bk_1)-{\tilde A}^{\bar\eta'}\Big({\bf K}-\frac{\xi_3}{\bar\xi_1}{\bf q}_\sT\Big)\Big]  
\nonumber\\
&
\hspace{2.4cm}
+
{\tilde A}^{\bar\lambda'}(\bk_1-{\bf q}_\sT)
\Big[ {\tilde A}^{\bar\eta'}\Big({\bf K}-\frac{\xi_3}{\bar\xi_1}{\bf q}_\sT\Big) - {\cal \tilde A}^{\bar \eta'}\Big( {\bf K}-\frac{\xi_3}{\bar\xi_1}{\bf q}_\sT;\tilde c_0, \bk_1-{\bf q}_\sT\Big)\Big]
\Big\rgroup
\nonumber\\
&
\hspace{0.35cm}
-
\Big\lgroup 
{\tilde A}^{\bar\lambda}({\bf k}_3)
\Big[ {\cal \tilde A}^{\bar\eta}({\bf Q};c_0,\bk_3)-{\tilde A}^{\bar\eta}\Big({\bf Q}-\frac{\xi_1}{\bar\xi_3}{\bf q}_\sT\Big)\Big]
\nonumber\\
&
\hspace{2.4cm}
+
{\tilde A}^{\bar\lambda}(\bk_3-{\bf q}_\sT)
\Big[ {\tilde A}^{\bar\eta}\Big({\bf Q}-\frac{\xi_1}{\bar\xi_3}{\bf q}_\sT\Big) - {\cal \tilde A}^{\bar\eta}\Big( {\bf Q}-\frac{\xi_1}{\bar\xi_3}{\bf q}_\sT;c_0,\bk_3-{\bf q}_\sT\Big)\Big]
\Big\rgroup
\nonumber\\
&
\hspace{0.35cm}
\times 
\Big\lgroup 
{\tilde A}^{\bar\lambda'}(\bk_1)
\Big[ {\cal \tilde A}^{\bar\eta'}({\bf K}-{\bf q}_\sT;\tilde c_0, \bk_1) - {\cal \tilde A}^{\bar\eta'}({\bf K};\tilde c_0,\bk_1)\Big]
\Big\rgroup
\nonumber\\
&
\hspace{-0.3cm}
+
\frac{1}{N_c^2}
\Big\lgroup
{\tilde A}^{\bar\lambda}(\bk_3)
\Big[ {\cal \tilde A}^{\bar\eta}({\bf Q}-{\bf q}_\sT;c_0, bk_3)-{\tilde A}^{\bar\eta}\Big({\bf Q}-\frac{\xi_1}{\bar\xi_3}{\bf q}_\sT\Big)\Big]
\nonumber\\
&
\hspace{2.4cm}
+
{\tilde A}^{\bar\lambda}(\bk_3-{\bf q}_\sT)
\Big[ {\tilde A}^{\bar\eta}\Big({\bf Q}-\frac{\xi_1}{\bar\xi_3}{\bf q}_\sT\Big) - {\cal \tilde A}^{\bar\eta}\Big( {\bf Q}-\frac{\xi_1}{\bar\xi_3}{\bf q}_\sT;c_0,\bk_3-{\bf q}_\sT\Big)\Big]
\Big\rgroup
\nonumber\\
&
\hspace{0.35cm}
\times 
\Big\lgroup 
{\tilde A}^{\bar\lambda'}(\bk_1)
\Big[ {\cal \tilde A}^{\bar\eta'}({\bf K}-{\bf q}_\sT;\tilde c_0, \bk_1) - {\tilde A}^{\bar\eta'}\Big({\bf K}-\frac{\xi_3}{\bar\xi_1}{\bf q}_\sT\Big)\Big]
\nonumber\\
&
\hspace{2.4cm}
+
{\tilde A}^{\bar\lambda'}(\bk_1-{\bf q}_\sT)
\Big[ {\tilde A}^{\bar\eta'}\Big({\bf K}-\frac{\xi_3}{\bar\xi_1}{\bf q}_\sT\Big) - {\cal \tilde A}^{\bar \eta'}\Big( {\bf K}-\frac{\xi_3}{\bar\xi_1}{\bf q}_\sT;\tilde c_0, \bk_1-{\bf q}_\sT\Big)\Big]
\Big\rgroup
\nonumber
\end{align}
\begin{align}
\label{def_WW_CC}
 {\cal W}_{CC}&=
\Big\lgroup 
\tilde{C}\big[  \xi_3,\bar \xi_3( {\bf Q}-{\bf q}_\sT); \xi_1\xi_2,\bk_3\big] - \tilde{C}\big[\xi_3,\bar \xi_3 {\bf Q}; \xi_1\xi_2, \bk_3\big]
\Big\rgroup^2
\nonumber\\
&
+
\Big\lgroup 
\tilde{C}\Big[ \xi_3, \bar \xi_3\Big({\bf Q}-\frac{\xi_1}{\bar\xi_3}{\bf q}_\sT\Big); \xi_1\xi_2, \bk_3-{\bf q}_\sT\Big] 
-
{\tilde C}\big[\xi_3, \bar \xi_3{\bf Q}; \xi_1\xi_2, \bk_3\big]
\Big\rgroup^2
\nonumber\\
&
\hspace{-0.6cm}
-
\frac{1}{N^2_c}
\Big\lgroup 
\tilde{C}\big[ \xi_3, \bar \xi_3({\bf Q}-{\bf q}_\sT); \xi_1\xi_2, \bk_3\big]-{\tilde C}\Big[\xi_3, \bar \xi_3\Big( {\bf Q}-\frac{\xi_1}{\bar\xi_3}{\bf q}_\sT\Big); \xi_1\xi_2, \bk_3-{\bf q}_\sT\Big]
\Big\rgroup^2\;,
\end{align} 

\begin{align}
\label{def_WW_Cq}
{\cal W}^{\bar\lambda \bar\eta}_{Cq} &=
\Big\lgroup
{\tilde A}^{\bar\lambda}(\bk_3)\Big[ {\tilde {\cal A}}^{\bar\eta}({\bf Q}-{\bf q}_\sT; c_0,\bk_3) - {\tilde {\cal A}}^{\bar\eta}({\bf Q};c_0,\bk_3)\Big] \Big\rgroup
\\
&
\times \, 
\Big\lgroup
{\tilde C}\big[\xi_3,\bar \xi_3({\bf Q}-{\bf q}_\sT); \xi_1\xi_2,\bk_3\big] - {\tilde C}\big[\xi_3,\bar \xi_3{\bf Q}; \xi_1\xi_2,\bk_3\big] 
\Big\rgroup
\nonumber\\
&
+
\Big\lgroup 
{\tilde A}^{\bar\lambda}(\bk_3)\Big[ {\cal \tilde A}^{\bar\eta}({\bf Q};c_0, \bk_3) - {\tilde A}^{\bar\eta}\Big({\bf Q}-\frac{\xi_1}{\bar\xi_3}{\bf q}_\sT\Big)\Big] 
\nonumber\\
&
\hspace{4.5cm}
+
{\tilde A}^{\bar\lambda}(\bk_3-{\bf q}_\sT) \Big[ {\tilde A}^{\bar\eta}\Big({\bf Q}-\frac{\xi_1}{\bar\xi_3}{\bf q}_\sT\Big)-{\tilde {\cal A}}^{\bar\eta}\Big({\bf Q}-\frac{\xi_1}{\bar\xi_3}{\bf q}_\sT;c_0,\bk_3-{\bf q}_\sT\Big)\Big]
\Big\rgroup
\nonumber\\
&
\times
\Big\lgroup
{\tilde C}\big[\xi_3,\bar \xi_3{\bf Q}; \xi_1\xi_2,\bk_3\big] - {\tilde C}\Big[\xi_3,\bar \xi_3\Big({\bf Q}-\frac{\xi_1}{\bar\xi_3}{\bf q}_\sT\Big); \xi_1\xi_2,\bk_3-{\bf q}_\sT\Big] 
\Big\rgroup
\nonumber\\
&
-\frac{1}{N^2_c}
\Big\lgroup {\tilde A}^{\bar\lambda}(\bk_3)
\Big[ {\cal \tilde A}^{\bar\eta}({\bf Q}-{\bf q}_\sT;c_0, \bk_3) - {\tilde A}^{\bar\eta}\Big({\bf Q}-\frac{\xi_1}{\bar\xi_3}{\bf q}_\sT\Big)\Big]
\nonumber\\
&
\hspace{4.25cm}
+{\tilde A}^{\bar\lambda}(\bk_3-{\bf q}_\sT)
\Big[ {\tilde A}^{\bar\eta}\Big({\bf Q}-\frac{\xi_1}{\bar\xi_3}{\bf q}_\sT\Big)-{\tilde {\cal A}}^{\bar\eta}\Big({\bf Q}-\frac{\xi_1}{\bar\xi_3}{\bf q}_\sT; c_0; \bk_3-{\bf q}_{\sT}\Big)\Big]\Big\rgroup
\nonumber\\
&
\hspace{0.6cm}
\times
\Big\lgroup
{\tilde C}\big[\xi_3,\bar \xi_3({\bf Q}-{\bf q}_\sT); \xi_1\xi_2,\bk_3\big] - {\tilde C}\Big[\xi_3,\bar \xi_3\Big({\bf Q}-\frac{\xi_1}{\bar\xi_3}{\bf q}_\sT\Big); \xi_1\xi_2,\bk_3-{\bf q}_\sT\Big] 
\Big\rgroup \nonumber\;,
\end{align}
and lastly:

\begin{align}
\label{def_WW_Cbarq}
{\cal W}^{\bar\lambda \bar\eta}_{C\bar q} &=
-\Big\lgroup
{\tilde A}^{\bar\lambda}(\bk_1)\Big[ {\tilde {\cal A}}^{\bar\eta}({\bf K}-{\bf q}_\sT; \tilde c_0,\bk_1) - {\tilde {\cal A}}^{\bar\eta}({\bf K};\tilde c_0,\bk_1)\Big] \Big\rgroup
\\
&
\hspace{0.3cm}
\times \, 
\Big\lgroup
{\tilde C}\big[\xi_3,\bar \xi_3{\bf Q}; \xi_1\xi_2,\bk_3\big] - {\tilde C}\Big[\xi_3,\bar \xi_3\Big({\bf Q}-\frac{\xi_1}{\bar\xi_3}{\bf q}_\sT\Big); \xi_1\xi_2,\bk_3-{\bf q}_\sT\big] 
\Big\rgroup
\nonumber\\
&
\hspace{0.3cm}
-
\Big\lgroup 
{\tilde A}^{\bar\lambda}(\bk_1)\Big[ {\cal \tilde A}^{\bar\eta}({\bf K};\tilde c_0, \bk_1) - {\tilde A}^{\bar\eta}\Big({\bf K}-\frac{\xi_3}{\bar\xi_1}{\bf q}_\sT\Big)\Big] 
\nonumber\\
&
\hspace{3cm}
+
{\tilde A}^{\bar\lambda}(\bk_1-{\bf q}_\sT) \Big[ {\tilde A}^{\bar\eta}\Big({\bf K}-\frac{\xi_3}{\bar\xi_1}{\bf q}_\sT\Big)-{\tilde {\cal A}}^{\bar\eta}\Big({\bf K}-\frac{\xi_3}{\bar\xi_1}{\bf q}_\sT;\tilde c_0,\bk_1-{\bf q}_\sT\Big)\Big]
\Big\rgroup
\nonumber\\
&
\hspace{0.3cm}
\times
\Big\lgroup
{\tilde C}\big[\xi_3,\bar \xi_3({\bf Q}-{\bf q}_\sT); \xi_1\xi_2,\bk_3\big] - {\tilde C}\big[\xi_3,\bar \xi_3{\bf Q}; \xi_1\xi_2,\bk_3\big] 
\Big\rgroup
\nonumber\\
&
\hspace{-0.28cm}
+\frac{1}{N^2_c}
\Big\lgroup {\tilde A}^{\bar\lambda}(\bk_1)
\Big[ {\cal \tilde A}^{\bar\eta}({\bf K}-{\bf q}_\sT;\tilde c_0, \bk_1) - {\tilde A}^{\bar\eta}\Big({\bf K}-\frac{\xi_3}{\bar\xi_1}{\bf q}_\sT\Big)\Big]
\nonumber\\
&
\hspace{3cm}
+{\tilde A}^{\bar\lambda}(\bk_1-{\bf q}_\sT)
\Big[ {\tilde A}^{\bar\eta}\Big({\bf K}-\frac{\xi_3}{\bar\xi_1}{\bf q}_\sT\Big)-{\tilde {\cal A}}^{\bar\eta}\Big({\bf K}-\frac{\xi_3}{\bar\xi_1}{\bf q}_\sT; \tilde c_0; \bk_1-{\bf q}_{\sT}\Big)\Big]\Big\rgroup
\nonumber\\
&
\hspace{0.6cm}
\times
\Big\lgroup
{\tilde C}\big[\xi_3,\bar \xi_3({\bf Q}-{\bf q}_\sT); \xi_1\xi_2,\bk_3\big] - {\tilde C}\Big[\xi_3,\bar \xi_3\Big({\bf Q}-\frac{\xi_1}{\bar\xi_3}{\bf q}_\sT\Big); \xi_1\xi_2,\bk_3-{\bf q}_\sT\Big] 
\Big\rgroup \nonumber\;,
\end{align}

We can now combine all the contributions that have been computed separately, and write down the final factorized expression for the photoproduction of three jets in the HEF limit as:
\beq
\label{total_HEF}
(2\pi)^9 \frac{\mathrm{d}\sigma^{\gamma A\to q\bar q g+X}}{\mathrm{d}^3{\vec k}_1 \, \mathrm{d}^3{\vec k}_2 \, \mathrm{d}^3{\vec k}_3}\bigg|_{\rm HEF \,  limit}
= 
2\pi 
\delta\big( p^+-\sum_{i=1}^3k_i^+\big)
\, {\cal W}^{\rm total} \;  \frac{1}{\pi{\bf q}^2_{\sT}} \, \mathcal{F}_{g/A}(x_{\scriptscriptstyle{A}},{\bf q}_{\sT})\;,
\eeq
where the total  Weizs\"acker-Williams field structure ${\cal W}^{\rm total}$ reads 
\beq
{\cal W}^{\rm total}&=& \pi^3 N_c\,g_e^2 g_s^4\,\frac{1}{k_2^+\, p^+} \, 
 \Big\{ 
 {\cal M}^{\bar\lambda\bar\lambda';\bar\eta\bar\eta'}_{qq}\Big( \bar\xi_3, \frac{\xi_2}{\bar\xi_3}\Big) 
 {\cal W}^{\bar\lambda\bar\lambda';\bar\eta\bar\eta'}_{qq}
 \nonumber\\
 &&+\, 
 {\cal M}^{\bar\eta\bar\eta';\bar\lambda\bar\lambda'}_{\bar q\bar q}\Big( \xi_1, \frac{\xi_2}{\bar\xi_1}\Big) 
 {\cal W}^{\bar\lambda\bar\lambda';\bar\eta\bar\eta'}_{\bar q\bar q}
 +\,
 2 \, 
 {\cal M}^{\bar\lambda\bar\lambda';\bar\eta\bar\eta'}_{q\bar q}( \xi_1,\xi_2) 
 {\cal W}^{\bar\lambda\bar\lambda';\bar\eta\bar\eta'}_{q\bar q}
 \nonumber\\
 &&
 +
 {\cal M}_{CC}(\xi_1,\xi_2) 
 {\cal W}_{CC} 
+
2 \, 
{\cal M}_{Cq}^{\bar\lambda\bar\eta}(\xi_1,\xi_2) 
{\cal W}_{Cq}^{\bar\lambda\bar\eta}
+
2\, 
{\cal M}_{C\bar q}^{\bar\lambda\bar\eta}(\xi_1,\xi_2)
{\cal W}_{C\bar q}^{\bar\lambda\bar\eta} \; 
\Big\}\;.\label{eq:HEFhardparts}
\eeq
The products of splitting functions and the Weizs\"acker-Williams field structures are given in Eqs. \eqref{Mqq_def}-\eqref{def_WW_qq}, \eqref{def_M_barqbarq}-\eqref{def_WW_barqbarq}, \eqref{def_M_qbarq}-\eqref{def_WW_qbarq}, \eqref{def_M_CC}-\eqref{def_WW_CC}, \eqref{def_M_Cq}-\eqref{def_WW_Cq} and \eqref{def_M_Cbarq}-\eqref{def_WW_Cbarq}, respectively. Just like in the CGC result and in the correlation limit, the HEF cross section is fully symmetric under the exchange of the quark with the antiquark, and relations similar to Eq.~(\ref{eq:hardpartssymmetry}) hold between the hard parts in (\ref{eq:HEFhardparts}). Note that the above HEF factorization formula differs fundamentally from the TMD one derived in the correlation limit (\ref{total_Fact}). The latter is valid in the limit of small-$q_{\scriptscriptstyle{T}}$ and features a factorization between collinear hard parts and a gluon correlator that contains both an unpolarized and a linearly polarized gluon TMD. The HEF cross section which we derived here, however, is valid in the kinematic region $q_{\scriptscriptstyle{T}} \ll Q_s $, features a single unintegrated PDF, and involves hard parts $\mathcal{W}$ that are $q_{\scriptscriptstyle{T}}$-dependent.

\subsection{\label{TMDHEF}Relation between the TMD and HEF limits of the cross section}
An important feature of the CGC, that can be used as a powerful consistency check, is the fact that the correlation and HEF limits of the CGC cross section formally share a further limit, in which they become the same. Indeed, using the well-known fact that when $q_{\scriptscriptstyle{T}}\gg Q_s$, the unpolarized and linearly polarized gluon TMDs coincide with $1/\pi$ times the unintegrated gluon PDF (see appendix \ref{app:unintegrated}), the cross section in the correlation limit can be simplified to:
\beq
(2\pi)^{9}\frac{\mathrm{d}\sigma^{\gamma A\to q\bar{q}g+X}}{\mathrm{d}^{3}\vec{k}_{1}\mathrm{d}^{3}\vec{k}_{2}\mathrm{d}^{3}\vec{k}_{3}}\bigg|_{\mathrm{corr.},q_\sT \gg Q_s}
&&= 
2\pi\delta\big(p^{+}-\sum_{i=1}^3k_{i}^{+}\big) 
\, \big[ {\rm H} \big]_{ij}^{\rm total} \label{eq:corr2dilute}\nonumber \\ 
&&
\times
\bigg[\frac{1}{2}\delta^{ij}\mathcal{F}_{WW}(x_{\scriptscriptstyle{A}},{\bf q}_{\sT})+\frac{1}{2}\Big(2 \frac{q_{\sT}^{i}q_{\sT}^{j}}{{\bf q}_{\sT}^{2}}-\delta^{ij}\Bigr)\mathcal{H}_{WW}(x_{\scriptscriptstyle{A}},{\bf q}_{\sT})\bigg]\Bigg|_{q_\sT \gg Q_s}\;, \nonumber\\
&&= 
2\pi\delta\big(p^{+}-\sum_{i=1}^3k_{i}^{+}\big) 
\, \big[ {\rm H} \big]_{ij}^{\rm total}
\times
\frac{q_{\sT}^{i}q_{\sT}^{j}}{\pi {\bf q}_{\sT}^{2}}\mathcal{F}_{g/A}(x_{\scriptscriptstyle{A}},{\bf q}_{\sT})\;.\label{eq:corrdilute}
\eeq
We observe that the above expression is exactly the same as the following limit of the cross section in the HEF limit Eq.~(\ref{total_HEF}):
\beq
(2\pi)^9 \frac{\mathrm{d}\sigma^{\gamma A\to q\bar q g+X}}{\mathrm{d}^3{\vec k}_1 \, \mathrm{d}^3{\vec k}_2 \, \mathrm{d}^3{\vec k}_3}\bigg|_{\mathrm{HEF}, q_\sT \ll |\mathbf{k}_i|}
&&=2\pi \delta\big( p^+-\sum_{i=1}^3k_i^+\big) \mathcal{F}_{g/A}(x_{\scriptscriptstyle{A}},{\bf q}_{\sT}) \\
&&
\times \lim_{q_\sT \to 0}\, \frac{1}{\pi {\bf q}^2_{\sT}}\, {\cal W}^{\rm total} \;   \, \;,\nonumber\\
&&=(\ref{eq:corrdilute})\;.\nonumber
\eeq
In fact, this correspondence holds even on the level of the individual channels:
\beq
\mathcal{M}\otimes\big[{\rm H}_n\big]_{ij}\frac{q_{\sT}^{i}q_{\sT}^{j}}{{\bf q}_{\sT}^{2}}=\lim_{q_\sT \to 0}\mathcal{M}\otimes \mathcal{W}_n\frac{1}{{\bf q}^2_{\sT}}\;,
\label{eq:TMDHEFcorrespondence}
\eeq
where $n\in \{qq,\bar{q}\bar{q},q\bar{q},CC,Cq,C\bar{q}\}$ and where $\otimes$ stands for the contraction of the appropriate transverse indices.
This remarkable connection between hard parts $H_n$ and $\mathcal{W}_n$, that are derived from two very different limits of the CGC, is in fact expected, and can be explained as follows.
Let $q=x_{\scriptscriptstyle{A}}p_{\scriptscriptstyle{A}}+q_\perp$ be the momentum of the incoming gluon with polarization $\epsilon^{\mu}_{\lambda}$, with $q_\perp=(0^+,0^-,\mathbf{q}_{\sT})$ and $p_{\scriptscriptstyle{A}}=\frac{1}{x_{\scriptscriptstyle{A}}}(0^+,q^{-},\mathbf{0})$, and $A_\mu$ the Feynman amplitude for a certain channel with an open Lorentz index such that $H=\sum_\lambda H_{\mu\nu} \epsilon^{\mu}_{\lambda}\epsilon^{\nu*}_{\lambda}=\sum_\lambda A_{\mu}A_{\nu}^\dagger \epsilon^{\mu}_{\lambda}\epsilon^{\nu*}_{\lambda}$. We then have from the Ward identity $H_{\mu\nu}q^\mu= 0$:
\beq
H^{ij}q^i_{\sT}q^j_{\sT}=H_{\mu\nu}q_\perp^\mu q_\perp^\nu=x_{\scriptscriptstyle{A}}^2 H_{\mu\nu}p_{\scriptscriptstyle{A}}^{\mu}p_{\scriptscriptstyle{A}}^{\nu}\;.
\eeq
In the way the unintegrated gluon PDF is coupled to the hard part, in the last line of Eq.~(\ref{eq:corr2dilute}), one can therefore recognize the so-called nonsense polarization \cite{Catani:1990eg,ITMD}: 
\beq
\sum_\lambda\epsilon_{\mu}^{\lambda}\epsilon_{\nu}^{\lambda*}\simeq\epsilon_{\mu}^{0}\epsilon_{\nu}^{0*}=2x_{\scriptscriptstyle{A}}^2\frac{p_{\scriptscriptstyle{A}}^{\mu}p_{\scriptscriptstyle{A}}^{\nu}}{\mathbf{q}_\sT^2}\;.
\eeq
The above prescription for the sum over the polarization vectors of the incoming gluon, is a part of the procedure used in HEF (from a diagrammatic point of view, so without resorting to the CGC) to allow this gluon to be off-shell, with a virtuality $q^2=-\mathbf{q}_\sT^2$. The cross section Eq.~(\ref{total_HEF}) that we obtained from the weak-field expansion of CGC is the same as one would have obtained from QCD in the HEF approach. Hence, Eq.~(\ref{eq:corr2dilute}) only differs from Eq.~(\ref{total_HEF}) in that the hard parts in the former are collinear, since they are obtained from a TMD-like factorization. Correcting for this by placing the incoming gluon on-shell in the HEF hard parts $\mathcal{W}$, by means of the limiting procedure, yields the relation \eqref{eq:TMDHEFcorrespondence}. Note that in general, many different TMDs can play a role in the TMD cross section, not only the Weizs\"acker-Williams one. In that case, each gluon TMD will be coupled to their own hard part \cite{ITMD,Bomhof:2006dp}, and all these contributions need to be summed to match the $q_\sT\to0$ limit of their HEF counterparts channel per channel.

\section{\label{numerics}Numerical study}
In this section, we will perform a numerical study of the cross section in both the correlation and the HEF limit, using analytical and numerical models of the gluon distribution in the target proton.

Before doing so, let us make some estimates of the relevant phase space for our process. Photoproduction can take place as the low-$Q^2$ limit of deep-inelastic lepton-proton/ion scattering, for instance in the future Electron-Ion Collider, or in ultra-peripheral collisions (UPCs) involving protons or heavy ions.
Let us denote our process as $l(\ell)+A(p_{\scriptscriptstyle{A}})\to q(k_1)+g(k_2)+\bar{q}(k_3)+X$, with $A$ the target proton or ion, and where $l$ is the source of the real photon flux; either a charged lepton, either a proton or ion. The momenta of the photon source and the target are:
\begin{align}
\ell & = (\ell^+,0^-,\mathbf{0})\qquad \mathrm{and}\qquad p_{\scriptscriptstyle{A}}=(0^+,p_{\scriptscriptstyle{A}}^-,\mathbf{0})\;,
\end{align}
hence the center-of-mass energy is equal to $s=(\ell+p_{\scriptscriptstyle{A}})^2\simeq 2\ell^+ p_{\scriptscriptstyle{A}}^-$. The four-momenta of the real photon $p$ and the incoming gluon $q$ are then given by
\begin{align}
p & = y\ell=(y \ell^+,0^-,\mathbf{0})\qquad \mathrm{and}\qquad q=(0^+,k_1^-+k_2^-+k_3^-, \mathbf{k}_1+\mathbf{k}_2+\mathbf{k}_3)\;,
\end{align}
and the value of $x$ reached in the experiment is:
\begin{align}
\xA & = \frac{q^-}{p_{\scriptscriptstyle{A}}^-}=\frac{1}{s y}\Big(\frac{\mathbf{k}_1^2}{\xi_1}+\frac{\mathbf{k}_2^2}{\xi_2}+\frac{\mathbf{k}_3^2}{\xi_3}\Big)\;.
\end{align}
It follows that for realistic EIC center-of-mass (c.o.m.) energies $\sqrt{s}\sim\,100\,\mathrm{GeV}$, and for values of $y$ as large as possible, demanding that $x\lesssim10^{-2}$ means that the transverse momenta of the jets should not exceed $3\,\mathrm{GeV}$. On the other hand, in ultra-peripheral proton-proton or lead-lead collisions at the LHC one is able to reach c.o.m. energies of the order of $\sqrt{s}\sim3\,\mathrm{TeV}$ and $\sqrt{s}\sim7\cdot10^2\,\mathrm{GeV}$, respectively (see e.g. Ref.~\cite{Bertulani:2005ru}). In such collisions, values of $x$ down to $x\sim10^{-3} -10^{-4}$ are attainable, with transverse jet momenta of more workable magnitudes $|\mathbf{k}_i|\sim10\,\mathrm{GeV}$.

\subsection{Correlation limit}
We showed in section \ref{correlation} that, in the correlation limit $| \mathbf{q}_{\scriptscriptstyle{T}}|\ll |\mathbf{k}_{i}|$, the CGC cross section can be written in the TMD factorized form of Eq.~(\ref{total_Fact}). As is usual in TMD factorization, the hard part $\big[H\big]^{\mathrm{total}}$ is collinear, in the sense that the incoming gluon is on-shell and does not carry any transverse momentum w.r.t. its parent proton or ion. The only $q_{\scriptscriptstyle{T}}$-dependence is in the gluon correlator, which contains the transverse momentum dependent PDFs $\mathcal{F}_{WW}(x,\mathbf{q}_\sT^2)$ and  $\mathcal{H}_{WW}(x,\mathbf{q}_\sT^2)$.

In order to numerically simulate the cross section in the correlation limit, we resort to a model for $\mathcal{F}_{WW}(x,\mathbf{q}_\sT^2)$ and  $\mathcal{H}_{WW}(x,\mathbf{q}_\sT^2)$ since there are at present no fits for gluon TMDs available. For the kinematics under consideration, it is a reasonable choice to use the nonperturbative McLerran-Venugopalan (MV) model \cite{MV}, which is expected to provide a reliable description of the gluon distribution inside an unpolarized nucleus or proton at small enough $x$. In the MV model, the unpolarized and linearly polarized Weizs\"acker-Williams gluon TMDs read \cite{firstlowxTMDs}:
\begin{align}
\mathcal{F}_{WW}(x,\mathbf{q}_\sT^2)=\frac{S_\perp C_F}{\alpha_s \pi^3} \int \mathrm{d}r \frac{J_0(|q_\sT r|)}{r}\left (1-e^{-\frac{r^2}{4}Q_{sg}^2\ln{(1/r^2 \Lambda^2)} }\right )\,,\nonumber\\
\mathcal{H}_{WW}(x,\mathbf{q}_\sT^2)=\frac{S_\perp C_F}{\alpha_s \pi^3} \int \mathrm{d}r \frac{J_2(|q_\sT r|)}{r \ln \frac{1}{r^2 \Lambda^2}}\left (1-e^{-\frac{r^2}{4}Q_{sg}^2\ln{(1/r^2 \Lambda^2)} }\right )\,.
\label{eq:MVmodel}
\end{align}
In the above formulas, $S_\perp$ is the transverse size of the proton, and $\Lambda$ is an infrared cutoff, which we take to be $\Lambda_\mathrm{QCD}\simeq 0.2\,\mathrm{GeV}$. Furthermore, $Q_{sg}$ is the gluon saturation scale for which we choose the value $Q_{sg}=0.6\,\mathrm{GeV}$ for $x=10^{-2}$ \cite{Marquet:2016cgx,Marquet:2017xwy}. A factor $e$ is added inside the logarithms to guarantee the convergence of the expressions Eq.~(\ref{eq:MVmodel}).

\begin{figure}[t]
\begin{centering}
\includegraphics[clip,scale=0.3]{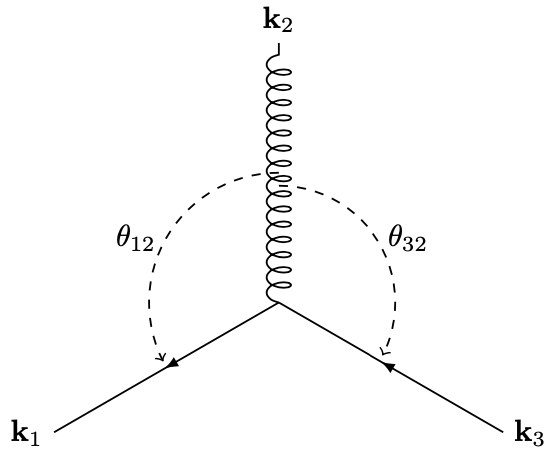}
\par\end{centering}
\caption{\label{fig:angles}Definition of the angles in function of which we plot the cross section.}
\end{figure}

\begin{figure}[t]
\begin{centering}
\includegraphics[clip,scale=0.38]{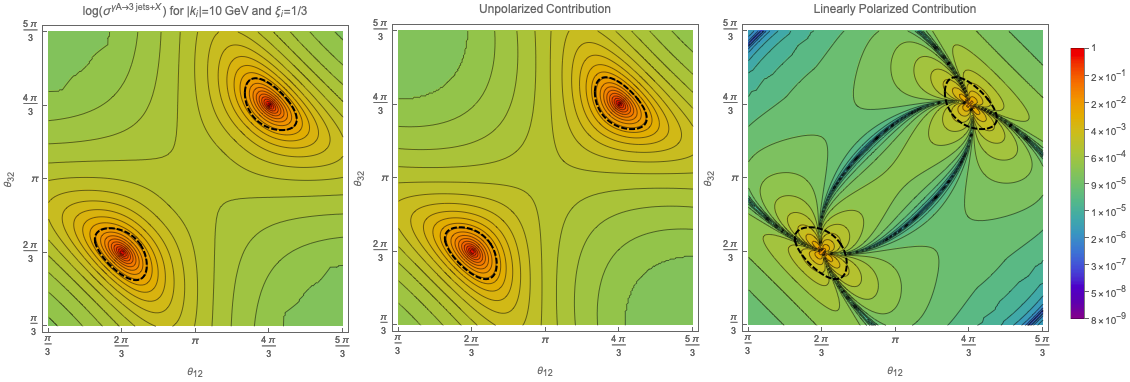}
\par\end{centering}
\caption{\label{fig:MVfull}Logarithm of the cross section in the correlation limit. The limit of validity is demarcated by the black dashed line.}
\end{figure}

With these expressions at hand, we plot the logarithm of the differential real photon-proton cross section in the correlation limit, Eq.~\eqref{total_Fact}, but expressed in slightly different variables and with the overall azimuthal dependence integrated out:
\beq
\ln\,\frac{\mathrm{d}\sigma^{\gamma A \to q \bar{q} g+X}}{\mathrm{d}\theta_{12} \mathrm{d}\theta_{32} \Pi_i  \mathrm{d}\xi_i  \mathrm{d}|\mathbf{k}_i|}\bigg|_{\rm corr. \,  limit}\;.
\eeq
In the above, the angles $\theta_{12}$ and $\theta_{32}$ are defined as in Fig.~\ref{fig:angles}. We present the result in Fig.~\ref{fig:MVfull} for the choice $|\mathbf{k}_i|=10\,\mathrm{GeV}$ and $\xi_i=1/3$ with $i=1,2,3$, and using the WW gluon TMDs introduced in Eq.~\eqref{eq:MVmodel}. Since our calculation is valid in the correlation limit $|\mathbf{q}_\sT|\ll |\mathbf{k}_i |\sim 10\,\mathrm{GeV}$, the maximum value of $q_\sT$ where our calculation can be trusted is estimated to be $q_{\sT,\mathrm{max}}=\sqrt{10}\,\mathrm{GeV}$, and is illustrated by a black dashed line. Note that we choose to normalize the cross section by its maximal value over the kinematic range, so that we are not sensitive to prefactors such as $S_\perp$. For illustrative purposes, we separately plot the contributions to the cross section from the unpolarized and from the linearly polarized gluon TMDs. The contribution to the total cross section due to $\mathcal{H}_{WW}$ is very small, since its corresponding hard part is one to two orders of magnitude smaller than the hard part corresponding to $\mathcal{F}_{WW}$. Nevertheless, as is clearly visible from the plots, the contribution from the linearly polarized gluon TMD exhibits azimuthal modulations, which can be exploited to extract this TMD experimentally (see e.g. Ref.~\cite{Boer:2016fqd}).

\begin{figure}[t]
\begin{centering}
\includegraphics[clip,scale=0.38]{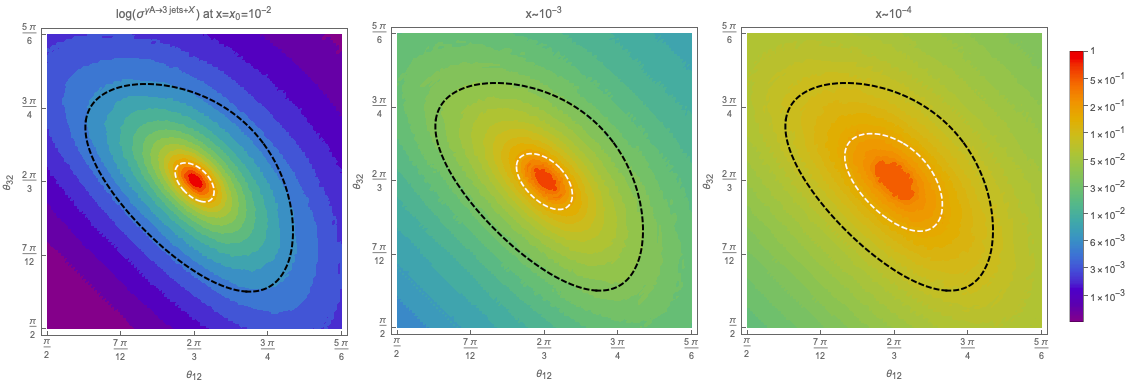}
\par\end{centering}
\caption{\label{fig:total_evolution}JIMWLK evolution of the logarithm of the total cross section in the correlation limit, for the values $\xi_i=1/3$ and $|\mathbf{k}_i|=10\,\mathrm{GeV}$. The limit of validity is demarcated by the black dashed line, and the approximate values of the saturation scale (resp. $Q_{sg}=0.60\,\mathrm{GeV}$, $Q_{sg}=0.86\,\mathrm{GeV}$, and $Q_{sg}=1.5\,\mathrm{GeV}$) by the white dashed line.}
\end{figure}

\begin{figure}[t]
\begin{centering}
\includegraphics[clip,scale=0.38]{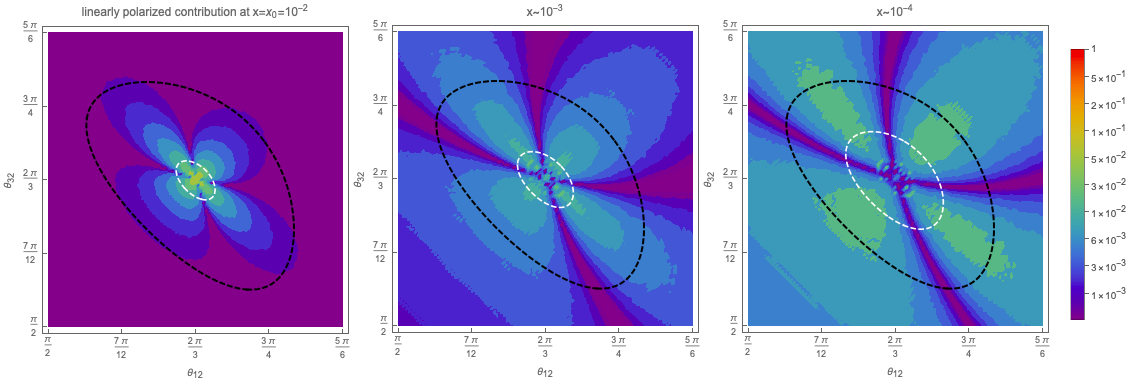}
\par\end{centering}
\caption{\label{fig:lin_evolution}JIMWLK evolution of the logarithm of the linearly polarized gluon contribution to the cross section in the correlation limit, for the values $\xi_i=1/3$ and $|\mathbf{k}_i|=10\,\mathrm{GeV}$. The limit of validity is demarcated by the black dashed line, and the approximate values of the saturation scale (resp. $Q_{sg}=0.60\,\mathrm{GeV}$, $Q_{sg}=0.86\,\mathrm{GeV}$, and $Q_{sg}=1.5\,\mathrm{GeV}$) by the white dashed line.}
\end{figure}

Furthermore, in Figs.~\ref{fig:total_evolution} and \ref{fig:lin_evolution}, we show the nonlinear low-$x$ evolution of the cross section. For this, we make use of the results of Ref.~\cite{Marquet:2017xwy}, in which the numerical JIMWLK evolution of different gluon TMDs was performed, including the unpolarized and linearly polarized Weizs\"acker-Williams distribution. Its initial condition is a numerical implementation of the MV model, which slightly differs from the analytical one in Eqs.~\eqref{eq:MVmodel}. Once again, we identify the starting point of the evolution to be $x=x_0=10^{-2}$ with an associated saturation scale $Q_{sg}=0.6\,\mathrm{GeV}$. The evolution is performed to values of $x\simeq 10^{-3}$ and $x\simeq 10^{-4}$, with corresponding saturation scales of $Q_{sg}=0.86\,\mathrm{GeV}$ and $Q_{sg}=1.5\,\mathrm{GeV}$, respectively \cite{Marquet:2016cgx,Marquet:2017xwy}. Clearly, the peak of the cross section around $q_\sT=0$ broadens quickly with the evolution, which is expected from the behavior of the saturation scale $Q_{sg}$ around which the gluons inside the target are distributed. Likewise, the angular modulations, which appear in the contribution to the cross section from linearly polarized gluons, are suppressed and pushed further away from the center at $q_\sT=0$. 

\subsection{HEF limit}
We perform a similar numerical calculation of the cross section Eq.~(\ref{total_HEF}), fixing the momenta of the outgoing particles as $|\mathbf{k}_i|=10\,\mathrm{GeV}$ and $\xi_i=1/3$ with $i=1,2,3$ and choosing again to vary the angles $\theta_{12}$ and $\theta_{32}$ defined in Fig.~\ref{fig:angles}. Opting again for a relative overall normalization, it is sufficient to model the $q_{\scriptscriptstyle{T}}$-dependence of radiative tail of the unintegrated gluon PDF $\mathcal{F}_{g/A}(x_{\scriptscriptstyle{A}},\mathbf{q}_{\scriptscriptstyle{T}})$. From general considerations, at large $q_{\scriptscriptstyle{T}}$ we can take
\beq
\mathcal{F}_{g/A}(x_{\scriptscriptstyle{A}},\mathbf{q}_{\scriptscriptstyle{T}})\propto \frac{1}{\mathbf{q}_{\scriptscriptstyle{T}}^2}\;.
\eeq

\begin{figure}[t]
\begin{centering}
\includegraphics[clip,scale=0.48]{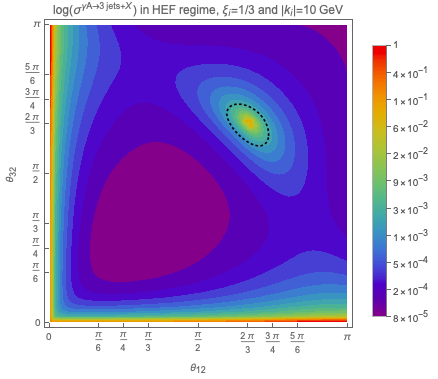}
\par\end{centering}
\caption{\label{fig:HEF}The $\gamma p\to q\bar{q}g$ cross section HEF factorization, for the values $\xi_i=1/3$ and $|\mathbf{k}_i|=10\,\mathrm{GeV}$. The black dotted line corresponds to the values $q_\sT= \sqrt{10}\,Q_{sg}$, and demarcates the domain where the HEF region of validity is expected to break down.}
\end{figure}

The result of the calculation is shown in Fig.~\ref{fig:HEF}. As we would expect from the fundamental features of QCD, the cross section peaks in the corners, where the gluon is emitted collinear to the quark or the antiquark. Another peak is situated in the back-to-back limit where $q_\sT \simeq 0$. However, this peak is outside of the validity domain of our HEF calculation for which we require $q_\sT \gg Q_s$.

\section{\label{outlook}Summary and outlook}
We calculated the partonic cross section for the forward inclusive production of a quark-antiquark pair plus a gluon in the scattering of a real photon with a proton or nucleus. The computation was performed at leading order in the CGC effective theory, valid at low values of $x$, taking the multiple rescatterings of the partons off the semiclassical gluon fields in the target proton or nucleus into account. From our partonic cross section, the cross section for three-jet production in the low-$Q^2$ limit of deep-inelastic scattering, or in ultra-peripheral collisions, can be easily obtained by convolving the result with the relevant real photon flux. Note that the cross section for three-jet production in deep-inelastic scattering at low-$x$ has been calculated in a helicity framework in \cite{Ayala:2016lhd}.

Our main result is the correlation limit of this cross section, corresponding to the kinematical configuration where the total transverse momentum $q_\sT$ of the three outgoing particles is much smaller than their individual transverse momenta. We demonstrated how in this limit the cross section simplifies and factorizes into a partonic hard part on the one hand, independent of $q_\sT$, and a gluon correlator on the other hand, which parametrizes the proton or nucleus target in terms of the unpolarized and linearly polarized gluon TMDs $\mathcal{F}_{WW}(\xA,q_\sT)$ and $\mathcal{H}_{WW}(\xA,q_\sT)$. Using earlier results of the JIMWLK evolution of these two gluon TMDs, we numerically studied the nonlinear small-$x$ evolution of this cross section.

In addition, we calculated the dilute limit of the CGC cross section by performing the weak-field expansion. Once again, a simple factorization formula was obtained, this time in terms of the unintegrated gluon PDF, and with a hard part corresponding to the $\gamma g^* \to q\bar q g$ off-shell matrix element in the language of high-energy factorization (or $k_{\sT}$-factorization). We commented on how and why, when placing the incoming gluon on-shell, this hard part corresponds to the large-$q_\sT$ limit of the hard part of the TMD factorization formula found in the correlation limit. We verified that this correspondence even holds channel per channel, which provides us with a powerful consistency check of our results.

Our calculation provides an important contribution towards establishing the phenomenology of forward particle production at low-$x$, in the line of \cite{Ayala:2016lhd,Iancu:2018aa,Bury:2018kvg}. In particular, recently there has been much interest in processes with photons in the final state as a clean probe for saturation effects \cite{Benic:2016uku,Benic:2017znu,monster,Benic:2018hvb,Roy:2018jxq}, and it can be noted that the cross section for the photoproduction of two jets and a photon can very easily be adapted from the results of the present paper, with a simplified Wilson line structure.

More importantly, we demonstrated for the first time that the CGC-TMD correspondence still holds, at least at leading order, when considering three outgoing colored particles. Finally, within the framework set up in this work and in \cite{monster}, we are currently pursuing the computation of photon-jet production in the CGC at NLO, to investigate whether this correspondence holds beyond tree level as well.

\section*{Acknowledgments}
We thank Guillaume Beuf for stimulating discussions. We thank the French-Polish Collaboration Agreement POLONIUM for support. RB and PT are grateful to the National Centre for Nuclear Research (NCBJ) Warsaw for hospitality and support. TA expresses his gratitude to CPHT at Ecole Polytechnique for their hospitality and support during a visit when part of this work was done. The work of TA is supported by Grant No. 2018/31/D/ST2/00666 (SONATA 14 - National Science Centre, Poland) and MSCA RISE 823947 `Heavy ion collisions: collectivity and precision in saturation physics' (HIEIC). The work of RB is supported by the U.S. Department of Energy, Office of Science, Office of Nuclear Physics, under contract No. DE- SC0012704, and in part by Laboratory Directed Research and Development (LDRD) funds from Brookhaven Science Associates. The work of CM is supported by the Agence Nationale de la Recherche under the project ANR-16-CE31-0019-02. This work has received funding from the European Union's Horizon 2020 research and innovation programme under grant agreement No. 82409.

\appendix

\section{\label{appendices}Appendices}
\subsection{Fock states and wave functions \label{app:Fock}}

The perturbative time evolution of an asymptotic eigenstate $|\omega_0\rangle$ of the Hamiltonian, into a physical Fock state $|\psi(0)\rangle$ at the time of interaction $x^+=0$, can be computed as follows:
\begin{equation}
\begin{aligned}|\psi(0)\rangle & =T\exp\Bigl(-i\int_{-\infty}^{0}\mathrm{d}x^+\mathcal{H}_\mathrm{int}(x^+)\Bigr)|\omega_{0}\rangle\\
 & =|\omega_{0}\rangle+\underset{\omega}{\sum}|\omega\rangle\frac{\langle\omega|H_{\mathrm{int}}|\omega_{0}\rangle}{q^-_{0}-q^-}+\underset{\omega,\omega'}{\sum}|\omega\rangle\frac{\langle\omega|H_{\mathrm{int}}|\omega'\rangle\langle\omega'|H_{\mathrm{int}}|\omega_{0}\rangle}{(q^-_{0}-q'^-)(q^-_{0}-q^-)}+...
\end{aligned}
\end{equation}
In the above, $\mathcal{H}_\mathrm{int}(x^+)$ is the interaction part of the Hamiltonian, which evolves in light-cone time as $\mathcal{H}_\mathrm{int}(x^+)=e^{i\hat{E}x^+}H_\mathrm{int}e^{-i\hat{E}x^+}$ where $\hat{E}|\omega\rangle=q^- |\omega\rangle$.
Adapting the above formula to the perturbative dressing of a real photon state, we obtain the following expression to
order $g_{s}g_{e}$ in the coupling (using the notation $\vec{k}=(k^+,\mathbf{k})$):
\begin{align}
|(\boldsymbol{\gamma})[\vec{p}]_{\lambda}\rangle_{D} & =|(\boldsymbol{\gamma})[\vec{p}]_{\lambda}\rangle_{0}\nonumber\\
 & +\sum_{s,s',i,j}\int\frac{\mathrm{d}^{3}\vec{q}_{1}}{\left(2\pi\right)^{3}2q_{1}^{+}}\frac{\mathrm{d}^{3}\vec{q}_2}{\left(2\pi\right)^{3}2q_2^{+}}|(\mathbf{q})[\vec{q}_{1}]_{s}^{i};(\bar{\mathbf{q}})[\vec{q}_2]_{s'}^{j}\rangle_{0}\frac{\langle(\mathbf{q})[\vec{q}_{1}]_{s}^{i};(\bar{\mathbf{q}})[\vec{q}_2]_{s'}^{j}|H_{\mathrm{int}}|(\boldsymbol{\gamma})[\vec{p}]_{\lambda}\rangle}{p^{-}-q_{1}^{-}-q^{-}_2}\nonumber\\
 & +\sum_{s,s',i,j,\eta,c}\int\frac{\mathrm{d}^{3}\vec{k}_{1}}{\left(2\pi\right)^{3}2k_{1}^{+}}\frac{\mathrm{d}^{3}\vec{k}_{2}}{\left(2\pi\right)^{3}2k_{2}^{+}}\frac{\mathrm{d}^{3}\vec{k}_{3}}{\left(2\pi\right)^{3}2k_{3}^{+}}|(\mathbf{q})[\vec{k}_{1}]_{s}^{i};(\mathbf{g})[\vec{k}_{2}]_{c}^{\eta};(\bar{\mathbf{q}})[\vec{k}_{3}]_{s'}^{j}\rangle_{0}\nonumber\\
 & \times\biggl[\int\frac{\mathrm{d}^{3}\vec{l}}{\left(2\pi\right)^{3}2l^{+}}\frac{\langle(\mathbf{q})[\vec{k}_{1}]_{s}^{i};(\mathbf{g})[\vec{k}_{2}]_{c}^{\eta}|H_{\mathrm{int}}|(\mathbf{q})[\vec{l}]_{\bar{s}}^{\bar{i}}\rangle}{p^{-}-k_{1}^{-}-k_{2}^{-}-k_{3}^{-}}\frac{\langle(\mathbf{q})[\vec{l}]_{\bar{s}}^{\bar{i}};(\bar{\mathbf{q}})[\vec{k}_{3}]_{s'}^{j}|H_{\mathrm{int}}|(\boldsymbol{\gamma})[\vec{p}]_{\lambda}\rangle}{p^{-}-l^{-}-k_{3}^{-}}\nonumber\\
 & +\int\frac{\mathrm{d}^{3}\vec{m}}{\left(2\pi\right)^{3}2m^{+}}\frac{\langle(\mathbf{g})[\vec{k}_{2}]_{c}^{\eta};(\bar{\mathbf{q}})[\vec{k}_{3}]_{s'}^{j}|H_{\mathrm{int}}|(\bar{\mathbf{q}})[\vec{m}]_{\bar{s}}^{\bar{i}}\rangle}{p^{-}-k_{1}^{-}-k_{2}^{-}-k_{3}^{-}}\frac{\langle(\bar{\mathbf{q}})[\vec{m}]_{\bar{s}}^{\bar{i}};(\mathbf{q})[\vec{k}_{1}]_{s}^{i}|H_{\mathrm{int}}|(\boldsymbol{\gamma})[\vec{p}]_{\lambda}\rangle}{p^{-}-m^{-}-k_{1}{}^{-}}\nonumber\\
 & +\frac{\langle(\mathbf{q})[\vec{k}_{1}]_{s}^{i};(\mathbf{g})[\vec{k}_{2}]_{c}^{\eta};(\bar{\mathbf{q}})[\vec{k}_{3}]_{s'}^{j}|H_{\mathrm{int}}|(\boldsymbol{\gamma})[\vec{p}]_{\lambda}\rangle}{p^{-}-k_{1}^{-}-k_{2}^{-}-k_{3}^{-}}\biggr]+\mathcal{O}(g_s^2 g_e,\,g_s g_e^2)\;.\label{eq:photondressed}
\end{align}
In the above expression, the standard properties of light-cone perturbation theory (LCPT) apply. In particular, at each vertex three-momentum is conserved, hence it is understood
that $\vec{q}_{1}=\vec{p}-\vec{q}_{2}$, $\vec{k}_{3}=\vec{p}-\vec{k}_{1}-\vec{k}_{2}$, and so on. Moreover, the momenta satisfy on-shell conditions, implying that in our massless case $k_i^-=\mathbf{k}_i^2/2k_i^+$.
Introducing what we call the ‘wave functions' $F$, the dressed photon state can be cast in a much more compact form:
\begin{align}
|(\boldsymbol{\gamma})[\vec{p}]_{\lambda}\rangle_{D} & =|(\boldsymbol{\gamma})[\vec{p}]_{\lambda}\rangle_{0}+g_{e}\delta^{ij}\int\frac{\mathrm{d}^{3}\vec{q}_{1}}{\left(2\pi\right)^{3}}F_{\gamma}^{(1)}\big[(\mathbf{q})[\vec{q}_{1}];(\bar{\mathbf{q}})[\vec{q}_{2}]\big]_{s's}^{\lambda}\times|(\mathbf{q})[\vec{q}_{1}]_{s}^{i};(\bar{\mathbf{q}})[\vec{q}_{2}]_{s'}^{j}\rangle_{0}\nonumber\\
 & +g_{e}g_{s}t_{ij}^{c}\int\frac{\mathrm{d}^{3}\vec{k}_{1}}{\left(2\pi\right)^{3}}\frac{\mathrm{d}^{3}\vec{k}_{2}}{\left(2\pi\right)^{3}}\biggl\{\Bigl(F_{q}^{(2)}+F_{\bar{q}}^{(2)}+F_{C}^{(2)}\Bigr)\Bigl[(\mathbf{q})[\vec{k}_{1}];(\mathbf{g})[\vec{k}_{2}];(\bar{\mathbf{q}})[\vec{k}_{3}]\Bigr]^{\lambda \eta}_{s s'}\nonumber\\
 & \times|(\mathbf{q})[\vec{k}_{1}]_{s}^{i};(\mathbf{g})[\vec{k}_{2}]_{c}^{\eta};(\bar{\mathbf{q}})[\vec{k}_{3}]_{s'}^{j}\rangle_{0}\;.\label{eq:photondressedfinal}
\end{align}
The wave functions contain the dynamics of the splittings, and their expressions in terms of the matrix elements are obtained by comparing Eqs.~\eqref{eq:photondressed} and  \eqref{eq:photondressedfinal}.  They can be calculated with the help of the Feynman rules in appendix \ref{sec:LCPT}. Since their calculation has been performed explicitly for similar processes in Refs.~\cite{Altinoluk:2018uax,monster}, but with slightly different conventions, we restrict ourselves here to the computation of $F_{\gamma}^{(1)}$: the leading-order splitting of a photon into a quark-antiquark pair. For the other wave functions, we will merely present the final results.

\begin{figure}[t]
\begin{centering}
\includegraphics[clip,scale=0.18]{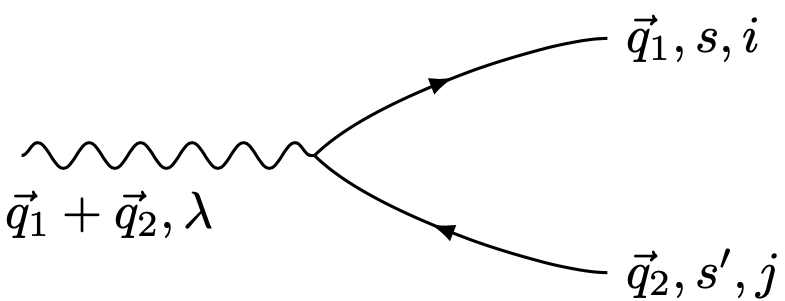}\includegraphics[clip,scale=0.18]{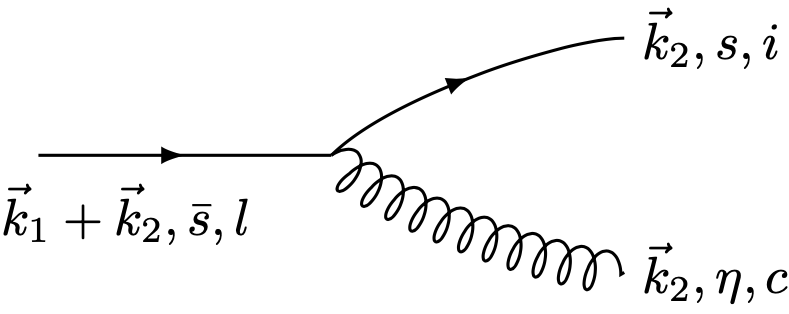}\includegraphics[scale=0.18]{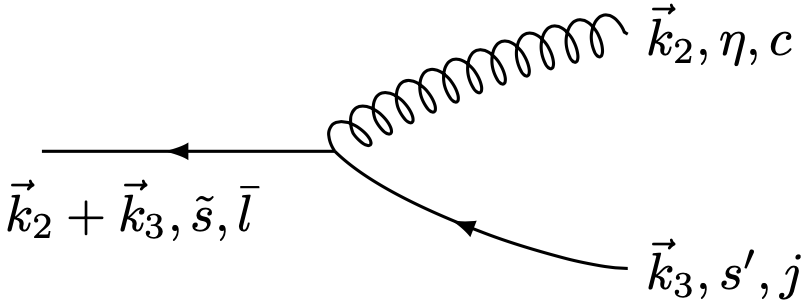}
\par\end{centering}
\caption{\label{fig:LOsplitting}The relevant leading-order splittings.}
\end{figure}

The (dimension $-1$) $\gamma\to q\bar{q}$ wave function (see Fig.~\ref{fig:LOsplitting}) is defined as:
\begin{equation}
\begin{aligned} & F_{\gamma}^{(1)}\big[(\mathbf{q})[\vec{q}_{1}];(\bar{\mathbf{q}})[\vec{q}_{2}]\big]_{ss'}^{\lambda}\\
 & =\frac{1}{2q_{1}^{+}}\sum_{s,s',i,j}\int\frac{\mathrm{d}^{3}\vec{q}_{2}}{(2\pi)^{3}2q_{2}^{+}}\int\mathrm{d}^{3}\vec{x}\frac{\langle0|d_{j}^{s'}(\vec{q}_{2})b_{i}^{s}(\vec{q}_{1})\,H_{\mathrm{int}}\,a^{\dagger\lambda}(\vec{p})|0\rangle}{p^{-}-q_{1}^{-}-q_{2}^{-}}\;,
\end{aligned}
\label{eq:Fqqbar}
\end{equation}
where one should be careful to note that an arrow over a momentum vector indicates $\vec{k}=(k^+,\mathbf{k})$, while an arrow over a coordinate means $\vec{x}=(x^-,\mathbf{x})$.
Making use of the LCPT Feynman rules as well as the conventions
for the \mbox{(anti-)commutation} relations listed in appendix \ref{sec:LCPT}:
\begin{equation}
\begin{aligned} & F_{\gamma}^{(1)}\big[(\mathbf{q})[\vec{q}_{1}];(\bar{\mathbf{q}})[\vec{q}_{2}]\big]_{ss'}^{\lambda}\\
 & =\frac{1}{2q_{1}^{+}}\sum_{s,s',i,j}\int\frac{\mathrm{d}^{3}\vec{q}_{2}}{(2\pi)^{3}2q_{2}^{+}}\int\mathrm{d}^{3}\vec{x}\frac{\langle0|d_{j}^{s'}(\vec{q}_{2})b_{i}^{s}(\vec{q}_{1}):\bar{q}(\vec{x})\cancel{A}(\vec{x})q(\vec{x}):a^{\dagger\lambda}(\vec{p})|0\rangle}{p^{-}-q_{1}^{-}-q_{2}^{-}}\;,\\
 & =\frac{1}{2q_{1}^{+}}\sum_{s,s',i,j}\int\frac{\mathrm{d}^{3}\vec{q}_{2}}{2q_{2}^{+}}\frac{\delta^{(3)}(\vec{p}-\vec{q}_{1}-\vec{q}_{2})}{p^{-}-q_{1}^{-}-q_{2}^{-}}\bar{u}^{s}(\vec{q}_{1})\cancel{\epsilon}_{\lambda}(\vec{p})v^{s'}(\vec{q}_{2})\;.
\end{aligned}
\end{equation}
Note that, with a minor abuse of notation, we extracted the factor
$g_{e}\delta^{ij}$ from the interaction Hamiltonian and placed it
in front of the wave function in Eq. \eqref{eq:photondressedfinal}. After some algebra, we find that:
\begin{equation}
\begin{aligned}p^{-}-q_{1}^{-}-q_{2}^{-} & =\frac{-p^{+}}{2q_{1}^{+}(p^{+}-q_{1}^{+})}\big(\frac{q_1^+}{p^+}\mathbf{p}-\mathbf{q}_{1}\big)^{2}\;,\end{aligned}
\label{eq:den}
\end{equation}
such that we obtain:
\begin{equation}
\begin{aligned}F_{\gamma}^{(1)}\big[(\mathbf{q})[\vec{q}_{1}];(\bar{\mathbf{q}})[\vec{q}_{2}]\big]_{ss'}^{\lambda} & =\frac{-1}{2p^{+}}\sum_{s,s',i,j}\frac{\bar{u}^{s}(\vec{q}_{1})\cancel{\epsilon}_{\lambda}(\vec{p})v^{s'}(\vec{q}_{2})}{\big(\frac{q_1^+}{q^+}\mathbf{p}-\mathbf{q}_{1}\big)^{2}}\;.\end{aligned}
\end{equation}
As we will show in appendix \ref{sec:genericsplitting}, the numerator in the above formula can be calculated in terms of the so-called \emph{good spinors}, on which we will elaborate later. In the massless case under consideration, the result is generic and does not depend on whether one considers quark or antiquark spinors. We can therefore read off the result from the general formula \eqref{eq:BeufGeneral}, and adapt it to the kinematics of the $\gamma\to q\bar{q}$ splitting under consideration:
\begin{equation}
\begin{aligned}\bar{u}(\vec{q}_{1})\cancel{\epsilon}_{\lambda}(\vec{p})v(\vec{q}_{2}) & =\frac{p^{+}}{2q_{1}^{+}q_{2}^{+}}(\frac{q_1^+}{p^+} p^{j}-q_{1}^{j})\epsilon_{\lambda}^{i}\bar{u}_{G}(\vec{q}_{1})\gamma^{+}\Bigl[(1-2\frac{q_1^+}{p^+})\delta^{ij}-i\sigma^{ij}\Bigr]v_{G}(\vec{q}_{2})\;.\end{aligned}
\end{equation}
To make this result more explicit, we choose a set of polarization vectors and Dirac spinors, for which we refer to appendix \ref{sec:explicit}. With this choice, our final result for the $\gamma\to q\bar{q}$ wave function is:
\begin{equation}
\begin{aligned}F_{\gamma}^{(1)}\big[(\mathbf{q})[\vec{q}_{1}];(\bar{\mathbf{q}})[\vec{q}_{2}]\big]_{ss'}^{\lambda} & =-\frac{q_1^+ + q_2^+}{2\sqrt{q^+_1q^+_2}}\frac{q_{1}^{+}q_{2}^{\bar{\lambda}}-q_{2}^{+}q_{1}^{\bar{\lambda}}}{\big(q_{1}^{+}\mathbf{q}_{2}-q_{2}^{+}\mathbf{q}_{1}\big)^{2}}\Psi_{ss'}^{\lambda\bar{\lambda}}\big(\frac{q^+_1}{q_1^+ + q_2^+}\big)\;,\end{aligned}
\label{eq:F1}
\end{equation}
where $\Psi_{ss'}^{\lambda\bar{\lambda}}(\xi)$ is the dimensionless splitting function, defined as:
\begin{equation}
\begin{aligned}\Psi_{ss'}^{\lambda\bar{\lambda}}(\xi) & =(1-2\xi)\delta^{\lambda\bar{\lambda}}\delta_{s,-s'}-i\epsilon^{\lambda\bar{\lambda}}\sigma_{s,-s'}^{3}\;.\end{aligned}
\label{eq:PsiTolga}
\end{equation}
For the other wave functions, the computation is analogous, and we refer to Refs.~\cite{Altinoluk:2018uax,monster} for the explicit calculation of similar second-order functions. Apart from the $\gamma\to q \bar{q}$ case, the $q\to qg$ and $\bar{q}\to \bar{q}g$ leading-order wave functions (see Fig.~\ref{fig:LOsplitting}) will be important in this work:
\begin{align}
F_{q}^{(1)}\big[(\mathbf{q})[\vec{k}_{1}];(\mathbf{g})[\vec{k}_{2}]\big]_{s\bar{s}}^{\eta} & =\frac{(k_{1}^{+}+k_{2}^{+})^{3/2}}{2k_{2}^{+}\sqrt{k_{1}^{+}}}\phi_{s\bar{s}}^{\eta\bar{\eta}}\big(\frac{k_{2}^{+}}{k_{1}^{+}+k_{2}^{+}}\big)\frac{k_{2}^{+}k_{1}^{\bar{\eta}}-k_{1}^{+}k_{2}^{\bar{\eta}}}{\big(k_{2}^{+}\mathbf{k}_{1}-k_{1}^{+}\mathbf{k}_{2}\big)^{2}}\;,\nonumber\\
F_{\bar{q}}^{(1)}\big[(\mathbf{g})[\vec{k}_{2}];(\bar{\mathbf{q}})[\vec{k}_{3}]\big]_{s'\bar{s}}^{\eta} & =\frac{-(k_{2}^{+}+k_{3}^{+})^{3/2}}{2k_{2}^{+}\sqrt{k_{3}^{+}}}\phi_{s'\bar{s}}^{\eta\bar{\eta}}\big(\frac{k_{2}^{+}}{k_{2}^{+}+k_{3}^{+}}\big)\frac{k_{2}^{+}k_{3}^{\bar{\eta}}-k_{3}^{+}k_{2}^{\bar{\eta}}}{\big(k_{2}^{+}\mathbf{k}_{3}-k_{3}^{+}\mathbf{k}_{2}\big)^{2}}\;,
\end{align}
with:
\begin{align}
\phi_{\alpha\beta}^{\lambda\bar{\lambda}}(z) & =(2-z)\delta^{\lambda\bar{\lambda}}\delta_{\alpha\beta}-iz\epsilon^{\lambda\bar{\lambda}}\sigma_{\alpha\beta}^{3}\;.
\end{align}
Once again, with some abuse of notation the factors $g_s t^c_{ij}$ were extracted.
The results for the second-order splitting functions (see Fig.~\ref{fig:NLOsplitting}) are:
\begin{align}
F_{q}^{(2)}\Bigl[(\mathbf{q})[\vec{k}_{1}];(\mathbf{g})[\vec{k}_{2}];(\bar{\mathbf{q}})[\vec{k}_{3}]\Bigr]_{ss'}^{\lambda\eta} & =\frac{-1}{4k_{2}^{+}\sqrt{k_{1}^{+}k_{3}^{+}}}\Psi_{\bar{s}s'}^{\lambda\bar{\lambda}}(\bar{\xi}_{3})\phi_{s\bar{s}}^{\eta\bar{\eta}}\bigl(\frac{\xi_{2}}{\bar{\xi}_{3}}\bigr)\frac{\xi_3 p^{\bar{\lambda}}-k_3^{\bar{\lambda}}}{\bigl(\xi_{3}\mathbf{p}-\mathbf{k}_{3}\bigr){}^{2}}\\
 & \times\frac{\xi_3\big(\bar{\xi}_{3}k_{2}^{\bar{\eta}}-\xi_{2}(p^{\bar{\eta}}-k_{3}^{\bar{\eta}})\big)}{\xi_{2}(\xi_{1}\mathbf{p}-\mathbf{k}_{1})^{2}+\xi_{1}(\xi_{2}\mathbf{p}-\mathbf{k}_{2})^{2}-(\xi_{2}\mathbf{k}_{1}-\xi_{1}\mathbf{k}_{2})^{2}}\;,\nonumber\\
F_{\bar{q}}^{(2)}\Bigl[(\mathbf{q})[\vec{k}_{1}];(\mathbf{g})[\vec{k}_{2}];(\bar{\mathbf{q}})[\vec{k}_{3}]\Bigr]_{ss'}^{\lambda\eta} & =\frac{1}{4k_{2}^{+}\sqrt{k_{1}^{+}k_{3}^{+}}}\Psi_{\bar{s}s}^{\lambda\bar{\lambda}}(\bar{\xi}_{1})\phi_{s'\bar{s}}^{\eta\bar{\eta}}\bigl(\frac{\xi_{2}}{\bar{\xi}_{1}}\bigr)\frac{\xi_{1}p^{\bar{\lambda}}-k_{1}^{\bar{\lambda}}}{(\xi_{1}\mathbf{p}-\mathbf{k}_{1})^{2}}\nonumber\\
 & \times\frac{\xi_1\big(\bar{\xi}_{1}k_{2}^{\bar{\eta}}-\xi_{2}(p^{\bar{\eta}}-k_{1}^{\bar{\eta}})\big)}{\xi_{2}(\xi_{1}\mathbf{p}-\mathbf{k}_{1})^{2}+\xi_{1}(\xi_{2}\mathbf{p}-\mathbf{k}_{2})^{2}-(\xi_{2}\mathbf{k}_{1}-\xi_{1}\mathbf{k}_{2})^{2}}\;,\nonumber\\
F_{C}^{(2)}\Big[(\mathbf{q})[\vec{k}_{1}];(\mathbf{g})[\vec{k}_{2}];(\bar{\mathbf{q}})[\vec{k}_{3}]\Big]_{ss'}^{\lambda\eta} & =\frac{1}{4k_{2}^{+}\sqrt{k_{1}^{+}k_{3}^{+}}}\varphi_{ss'}^{\lambda\eta}(\xi_{1},\xi_{2})\nonumber\\
&\times \frac{1}{\xi_{2}(\xi_{1}\mathbf{p}-\mathbf{k}_{1})^{2}+\xi_{1}(\xi_{2}\mathbf{p}-\mathbf{k}_{2})^{2}-(\xi_{2}\mathbf{k}_{1}-\xi_{1}\mathbf{k}_{2})^{2}}  \;,\nonumber
\end{align}
where the notations $\xi_1=k_1^+/p^+$, $\xi_2=k_2^+/p^+$, $\xi_3=(p^+ -k_1^+ -k_2^+)/p^+$ were introduced. The splitting function of the instantaneous splitting of a photon into a gluon, a quark and an antiquark, is given by:
\begin{align}
\varphi_{\alpha\beta}^{\lambda\eta}(\xi_1,\xi_2) & =\frac{\xi_{1}\xi_{2}\xi_{3}}{\bar{\xi}_{1}\bar{\xi}_{3}}\big[(\bar{\xi}_{3}-\bar{\xi}_{1})\delta^{\lambda\eta}\delta_{\alpha,-\beta}+i(\bar{\xi}_{3}+\bar{\xi}_{1})\epsilon^{\lambda\eta}\sigma_{\alpha,-\beta}^{3}\big]\;.
\end{align}
In mixed Fourier space, the wave functions are found to be:
\begin{align}
F_{\gamma}^{(1)}\left[(\mathbf{q})[q_{1}^{+},\mathbf{x}_{1}];(\bar{\mathbf{q}})[q_{2}^{+},\mathbf{x}_{3}]\right]_{ss'}^{\lambda}&=\frac{-iq_{2}^{+}(q_{1}^{+}+q_{2}^{+})}{2\sqrt{q_{1}^{+}q_{2}^{+}}}\Psi_{ss'}^{\lambda\bar{\lambda}}\big(\frac{q_{1}^{+}}{q_{1}^{+}+q_{2}^{+}}\big)\nonumber\\
&\qquad\qquad\times\delta^{(2)}(q_{1}^{+}\mathbf{x}_{1}+q_{2}^{+}\mathbf{x}_{3})A^{\bar{\lambda}}(\mathbf{x}_{1})\;,\nonumber\\
F_{q}^{(1)}\left[(\mathbf{q})[k_{1}^{+},\mathbf{x}_1];(\mathbf{g})[k_{2}^{+},\mathbf{x}_2]\right]_{ss'}^{\eta} &=\frac{i\sqrt{k_{1}^{+}}(k_{1}^{+}+k_{2}^{+})^{\frac{3}{2}}}{2k_{2}^{+}}\phi_{ss'}^{\eta\bar{\eta}}\big(\frac{k_{2}^{+}}{k_{1}^{+}+k_{2}^{+}}\big)\nonumber\\
&\qquad\qquad\times\delta^{(2)}(k_{2}^{+}\mathbf{x}_{2}+k_{1}^{+}\mathbf{\mathbf{x}}_{1})A^{\bar{\eta}}(\mathbf{x}_{2})\;,\nonumber\\
F_{\bar{q}}^{(1)}\left[(\mathbf{g})[k_{2}^{+},\mathbf{x}_2];(\bar{\mathbf{q}})[k_{3}^{+},\mathbf{x}_3]\right]_{s'\tilde{s}}^{\eta} &=\frac{-i\sqrt{k_{3}^{+}}(k_{2}^{+}+k_{3}^{+})^{\frac{3}{2}}}{2k_{2}^{+}}\phi_{s'\tilde{s}}^{\eta\bar{\eta}}\big(\frac{k_{2}^{+}}{k_{2}^{+}+k_{3}^{+}}\big)\nonumber\\
&\qquad\qquad\times\delta^{(2)}(k_{2}^{+}\mathbf{x}_{2}+k_{3}^{+}\mathbf{x}_{3})A^{\bar{\eta}}(\mathbf{x}_{2})\;,\label{eq:f1Fourier}
\end{align}
and
\begin{align} 
& F_{q}^{(2)}\Bigl[(\mathbf{q})[k_{1}^{+},\mathbf{w}-\mathbf{x}_{1}];(\mathbf{g})[k_{2}^{+},\mathbf{w}-\mathbf{x}_{2}];(\bar{\mathbf{q}})[k_{3}^{+},\mathbf{w}-\mathbf{x}_{3}]\Bigr]_{ss'}^{\lambda\eta}\nonumber\\
 & =\frac{1}{4k_{2}^{+}\sqrt{k_{1}^{+}k_{3}^{+}}}\Psi_{\bar{s}s'}^{\lambda\bar{\lambda}}(\bar{\xi}_{3})\phi_{s\bar{s}}^{\eta\bar{\eta}}\bigl(\frac{\xi_{2}}{\bar{\xi}_{3}}\bigr)\int_{\mathbf{v}}\delta^{(2)}\big(\mathbf{w}-\xi_i \mathbf{x}_{i}\big)\delta^{(2)}\Big(\mathbf{v}-\frac{\xi_{1}}{\bar{\xi}_{3}}\mathbf{x}_{1}-\frac{\xi_{2}}{\bar{\xi}_{3}}\mathbf{x}_{2}\Big)\nonumber\\
 & \qquad\qquad\qquad\qquad\qquad\qquad \times A^{\bar{\eta}}(\mathbf{x}_{1}-\mathbf{x}_{2})\mathcal{A}^{\bar{\lambda}}\bigl(\xi_{3},\mathbf{x}_{3}-\mathbf{v};\frac{\xi_{1}}{\bar{\xi}_{3}},\mathbf{x}_{1}-\mathbf{x}_{2}\bigr)\;,\nonumber\\
 & F_{\bar{q}}^{(2)}\Bigl[(\mathbf{q})[k_{1}^{+},\mathbf{w}-\mathbf{x}_{1}];(\mathbf{g})[k_{2}^{+},\mathbf{w}-\mathbf{x}_{2}];(\bar{\mathbf{q}})[k_{3}^{+},\mathbf{w}-\mathbf{x}_{3}]\Bigr]_{ss'}^{\lambda\eta}\nonumber\\
 & =\frac{-1}{4k_{2}^{+}\sqrt{k_{1}^+ k_{3}^{+}}}\Psi_{\bar{s}s}^{\lambda\bar{\lambda}}(\bar{\xi}_{1})\phi_{\bar{s}s'}^{\eta\bar{\eta}}\bigl(\frac{\xi_{2}}{\bar{\xi}_{1}}\bigr)\int_{\mathbf{u}}\delta^{(2)}\big(\mathbf{w}-\xi_i \mathbf{x}_{i}\big)\delta^{(2)}\Big(\mathbf{u}-\frac{\xi_{2}}{\bar{\xi}_{1}}\mathbf{x}_{2}-\frac{\xi_{3}}{\bar{\xi}_{1}}\mathbf{x}_{3}\Big)\nonumber\\
 & \qquad\qquad\qquad\qquad\qquad\qquad \times A^{\bar{\eta}}(\mathbf{x}_{3}-\mathbf{x}_{2})\mathcal{A}^{\bar{\lambda}}\bigl(\xi_{1},\mathbf{x}_{1}-\mathbf{u};\frac{\xi_{3}}{\bar{\xi}_{1}},\mathbf{x}_{3}-\mathbf{x}_{2}\bigr)\;,\nonumber\\
& F_{C}^{(2)}\Bigl[(\mathbf{q})[k_{1}^{+},\mathbf{w}-\mathbf{x}_{1}];(\mathbf{g})[k_{2}^{+},\mathbf{w}-\mathbf{x}_{2}];(\bar{\mathbf{q}})[k_{3}^{+},\mathbf{w}-\mathbf{x}_{3}]\Bigr]_{ss'}^{\lambda\eta}\nonumber\\
 & =\varphi_{ss'}^{\lambda\eta}(\xi_{1},\xi_{2})\delta^{(2)}\big(\mathbf{w}-\xi_i \mathbf{x}_{i}\big)\times\mathcal{C}\bigl(\xi_{3},\mathbf{w}-\mathbf{x}_{3};\xi_{1}\xi_{2},\mathbf{x}_{2}-\mathbf{x}_{1}\bigr)\;.\label{eq:f2Fourier}
\end{align}
The expressions for the non-Abelian Weizs\"acker-Williams field $A$, the modified Weizs\"acker-Williams field $\mathcal{A}$ and the Coulomb field $\mathcal{C}$ can be found in Eqs.~\eqref{eq:WWfield}, \eqref{eq:modWWfield} and \eqref{eq:Coulombfield}.

\subsection{\label{app:outgoingFock}Outgoing Fock state}

The first step towards the computation of the outgoing Fock state,
is performing the Fourier transform of the transverse coordinates.
For the zeroth-order term of the dressed photon state (\ref{eq:photondressedfinal}), this is trivial and yields:
\begin{align}
|(\boldsymbol{\gamma})[\vec{p}]_{\lambda}\rangle_{D}\Bigr|_{0} & =\int_{\mathbf{w}}e^{-i\mathbf{p}\cdot\mathbf{w}}|(\boldsymbol{\gamma})[p^{+},\mathbf{w}]_{\lambda}\rangle_{0}\;.
\end{align}
For the order $g_{e}$ term, we have that:
\begin{align}
|(\boldsymbol{\gamma})[\vec{p}]_{\lambda}\rangle_{D}\Bigr|_{g_{e}} 
 & =g_e \delta^{ij}\int\frac{\mathrm{d}^{3}\vec{q}_{1}}{(2\pi)^{3}}\int_{\mathbf{yzy'z'}}e^{-i\mathbf{q}_{1}\cdot(\mathbf{z}+\mathbf{z}')}e^{-i\mathbf{q}_{2}\cdot(\mathbf{y}+\mathbf{y}')} F_{\gamma}^{(1)}\Big[(\mathbf{q})[q_{1}^{+},\mathbf{z}'];(\bar{\mathbf{q}})[q_{2}^{+},\mathbf{y}']\Big]_{s's}^{\lambda}\nonumber\\
 & \times|(\mathbf{q})[q_{1}^{+},\mathbf{z}]_{s}^{i};(\bar{\mathbf{q}})[q_{2}^{+},\mathbf{y}]_{s'}^{j}\rangle_{0}\Bigr|_{\vec{q}_{2}=\vec{p}-\vec{q}_{1}}\;.
\end{align}
Renaming $\mathbf{y}'=\mathbf{w}-\mathbf{y}$, using that $\vec{q}_{2}=\vec{p}-\vec{q}_{1}$,
and integrating over the momentum $\mathbf{q}_{1}$, one obtains:
\begin{align}
|(\boldsymbol{\gamma})[\vec{p}]_{\lambda}\rangle_{D}\Bigr|_{g_{e}} 
 & =g_e \delta^{ij}\int\frac{\mathrm{d}q_{1}^{+}}{2\pi}\int_{\mathbf{wyz}}e^{-i\mathbf{p}\cdot\mathbf{w}}F_{\gamma}^{(1)}\Big[(\mathbf{q})[q_{1}^{+},\mathbf{w}-\mathbf{z}];(\bar{\mathbf{q}})[p^{+}-q_{1}^{+},\mathbf{w}-\mathbf{y}]\Big]_{s's}^{\lambda}\nonumber\\
 & \times|(\mathbf{q})[q_{1}^{+},\mathbf{z}]_{s}^{i};(\bar{\mathbf{q}})[p^{+}-q_{1}^{+},\mathbf{y}]_{s'}^{j}\rangle_{0}\;.
\label{eq:F1Fourier}
\end{align}
Likewise, for any of the $g_{e}g_{s}$ contributions the Fourier transform looks as follows:
\begin{align}
|(\boldsymbol{\gamma})[\vec{p}]_{\lambda}\rangle_{D}\Bigr|_{g_{e}g_{s}} & =g_{e}g_{s}t_{ij}^{c}\int\frac{\mathrm{d}^{3}\vec{k}_{1}}{(2\pi)^{3}}\frac{\mathrm{d}^{3}\vec{k}_{2}}{(2\pi)^{3}}\prod_{i=1}^{3}\int_{\mathbf{x}_{i}\mathbf{x}'_{i}}e^{-i\mathbf{k}_{i}\cdot(\mathbf{x}_{i}+\mathbf{x}'_{i})}\nonumber\\
&\times F^{(2)}\Big[(\mathbf{q})[k_{1}^{+},\mathbf{x}'_{1}];(\mathbf{g})[k_{2}^{+},\mathbf{x}'_{2}];(\bar{\mathbf{q}})[k_{3}^{+},\mathbf{x}'_{3}]\Big]_{s's}^{\lambda\eta}\nonumber\\
 & \times|(\mathbf{q})[k_{1}^{+},\mathbf{x}_{1}]_{s}^{i};(\mathbf{g})[k_{2}^{+},\mathbf{x}_{2}]_{c}^{\eta};(\bar{\mathbf{q}})[k_{3}^{+},\mathbf{x}_{3}]_{s'}^{j}\rangle_{0}\;.
\label{eq:F2Fourier}
\end{align}
After the substitution $\mathbf{x}'_{3}\to\mathbf{w}-\mathbf{x}_{3}$,
and integrating over the momenta $\mathbf{k}_{1}$ and $\mathbf{k}_{2}$:
\begin{align}|(\boldsymbol{\gamma})[\vec{p}]_{\lambda}\rangle_{D}\Bigr|_{g_{e}g_{s}} 
 & =\int\frac{\mathrm{d}k_{1}^{+}}{2\pi}\frac{\mathrm{d}k_{2}^{+}}{2\pi}\int_{\mathbf{w}\mathbf{x}_{1}\mathbf{x}_{2}\mathbf{x}_{3}}e^{-i\mathbf{p}\cdot\mathbf{w}}\nonumber\\
&\times F^{(2)}\Big[(\mathbf{q})[k_{1}^{+},\mathbf{w}-\mathbf{x}_{1}];(\mathbf{g})[k_{2}^{+},\mathbf{w}-\mathbf{x}_{2}];(\bar{\mathbf{q}})[k_{3}^{+},\mathbf{w}-\mathbf{x}_{3}]\Big]_{s's}^{\lambda\eta}\nonumber\\
 & \times|(\mathbf{q})[k_{1}^{+},\mathbf{x}_{1}]_{s}^{i};(\mathbf{g})[k_{2}^{+},\mathbf{x}_{2}]_{c}^{\eta};(\bar{\mathbf{q}})[k_{3}^{+},\mathbf{x}_{3}]_{s'}^{j}\rangle_{0}\;.
\end{align}
Combining the above elements, the complete dressed photon state in
mixed Fourier space reads (where we set the transverse momentum $\mathbf{p}$
of the incoming photon equal to zero):
\begin{align}
|(\boldsymbol{\gamma})[p^+,\mathbf{p}=0]_{\lambda}\rangle_{D} & =\int_{\mathbf{w}}|(\boldsymbol{\gamma})[p^{+},\mathbf{w}]_{\lambda}\rangle_{0}\nonumber\\
 &\hspace{-5em} +g_{e}\delta^{ij}\int\frac{\mathrm{d}q_{1}^{+}}{2\pi}\int_{\mathbf{wyz}}F_{\gamma}^{(1)}\Big[(\mathbf{q})[q_{1}^{+},\mathbf{w}-\mathbf{z}];(\bar{\mathbf{q}})[p^{+}-q_{1}^{+},\mathbf{w}-\mathbf{y}]\Big]_{s's}^{\lambda}\nonumber\\
 &\hspace{-5em} \qquad\times|(\mathbf{q})[q_{1}^{+},\mathbf{z}]_{s}^{i};(\bar{\mathbf{q}})[p^{+}-q_{1}^{+},\mathbf{y}]_{s'}^{j}\rangle_{0}\nonumber\\
 &\hspace{-5em} +g_{e}g_{s}t_{ij}^{c}\int\frac{\mathrm{d}k_{1}^{+}}{2\pi}\frac{\mathrm{d}k_{2}^{+}}{2\pi}\int_{\mathbf{w}\mathbf{x}_{1}\mathbf{x}_{2}\mathbf{x}_{3}}\nonumber\\
 &\hspace{-5em}\times \bigl(F_{q}^{(2)}+F_{\bar{q}}^{(2)}+F_{C}^{(2)}\bigr)\Big[(\mathbf{q})[k_{1},\mathbf{w}-\mathbf{x}_{1}];(\mathbf{g})[k_{2},\mathbf{w}-\mathbf{x}_{2}];(\bar{\mathbf{q}})[k_{3},\mathbf{w}-\mathbf{x}_{3}]\Big]_{s's}^{\lambda\eta}\nonumber\\
 &\hspace{-5em} \qquad\times|(\mathbf{q})[k_{1}^{+},\mathbf{x}_{1}]_{s}^{i};(\mathbf{g})[k_{2}^{+},\mathbf{x}_{2}]_{c}^{\eta};(\bar{\mathbf{q}})[k_{3}^{+},\mathbf{x}_{3}]_{s'}^{j}\rangle_{0}\;.
\end{align}
Now, the outgoing state is obtained by acting on the bare components
with the appropriate Wilson lines (defined in Eq.~\eqref{eq:defWL}, the subscripts $F$ and $A$ denote whether the generators of $SU(N_c)$ are in the fundamental or adjoint representation, respectively):
\begin{align}
|(\boldsymbol{\gamma})[p^+,\mathbf{p}=0]_{\lambda}\rangle_{D} & =\int_{\mathbf{w}}|(\boldsymbol{\gamma})[p^{+},\mathbf{w}]_{\lambda}\rangle_{0}\nonumber\\
 & \hspace{-5em}+g_{e}\int\frac{\mathrm{d}q_{1}^{+}}{2\pi}\int_{\mathbf{wyz}}\bigl[S_{F}(\mathbf{z})S_{F}^{\dagger}(\mathbf{y})\bigr]_{ij}F_{\gamma}^{(1)}\Big[(\mathbf{q})[q_{1}^{+},\mathbf{w}-\mathbf{z}];(\bar{\mathbf{q}})[p^{+}-q_{1}^{+},\mathbf{w}-\mathbf{y}]\Big]_{s's}^{\lambda}\nonumber\\
 &\hspace{-5em} \qquad\times|(\mathbf{q})[q_{1}^{+},\mathbf{z}]_{s}^{i};(\bar{\mathbf{q}})[p^{+}-q_{1}^{+},\mathbf{y}]_{s'}^{j}\rangle_{0}\nonumber\\
 & \hspace{-5em}+g_{e}g_{s}\int\frac{\mathrm{d}k_{1}^{+}}{2\pi}\frac{\mathrm{d}k_{2}^{+}}{2\pi}\int_{\mathbf{w}\mathbf{x}_{1}\mathbf{x}_{2}\mathbf{x}_{3}}\bigl[S_{F}(\mathbf{x}_{1})t^{d}S_{F}^{\dagger}(\mathbf{x}_{3})\bigr]_{ij}S_{A}(\mathbf{x}_{2})^{dc}\nonumber\\
 &\hspace{-5em} \qquad\times\bigl(F_{q}^{(2)}+F_{\bar{q}}^{(2)}+F_{C}^{(2)}\bigr)\Bigl[(\mathbf{q})[k_{1},\mathbf{w}-\mathbf{x}_{1}];(\mathbf{g})[k_{2},\mathbf{w}-\mathbf{x}_{2}];(\bar{\mathbf{q}})[k_{3},\mathbf{w}-\mathbf{x}_{3}]\Bigr]_{s's}^{\eta}\nonumber\\
 & \hspace{-5em}\qquad\times|(\mathbf{q})[k_{1}^{+},\mathbf{x}_{1}]_{s}^{i};(\mathbf{g})[k_{2}^{+},\mathbf{x}_{2}]_{c}^{\lambda\eta};(\bar{\mathbf{q}})[k_{3}^{+},\mathbf{x}_{3}]_{s'}^{j}\rangle_{0}\;.
\end{align}

Finally, the outgoing state needs to be written as a function of the
\emph{dressed} states, not the bare ones. Schematically, the procedure
goes as follows: first, observe that all the dressed states are,
up to $g_{e}g_{s}$ accuracy, related to the bare ones as:
\begin{align}
|\gamma\rangle_{D} & =|\gamma\rangle_{0}+F_{\gamma}^{(1)}|q\bar{q}\rangle_{0}+\bigl(F_{q}^{(2)}+F_{\bar{q}}^{(2)}+F_{C}^{(2)}\bigr)|q\bar{q}g\rangle_{0}\;,\nonumber\\
|q\bar{q}\rangle_{D} & =|q\bar{q}\rangle_{0}+\bigl(F_{q}^{(1)}+F_{\bar{q}}^{(1)}\bigr)|q\bar{q}g\rangle_{0}\;,\nonumber\\
|q\bar{q}g\rangle_{D} & =|q\bar{q}g\rangle_{0}\;.
\end{align}
The outgoing state is then given by:
\begin{align}
|\gamma\rangle_{\mathrm{out}} & =|\gamma\rangle_{0}+\bigl[S_{F}(\mathbf{z})S_{F}^{\dagger}(\mathbf{y})\bigr]F_{\gamma}^{(1)}|q\bar{q}\rangle_{0}\nonumber\\
&\qquad+\bigl[S_{F}(\mathbf{x}_{1})t^{c}S_{F}^{\dagger}(\mathbf{x}_{3})\bigr]S_{A}(\mathbf{x}_{2})\bigl(F_{q}^{(2)}+F_{\bar{q}}^{(2)}+F_{C}^{(2)}\bigr)|q\bar{q}g\rangle_{0}\;,\nonumber\\
 & =|\gamma\rangle_{D}+\big(\bigl[S_{F}(\mathbf{z})S_{F}^{\dagger}(\mathbf{y})\bigr]-\mathbb{1}\big)F_{\gamma}^{(1)}|q\bar{q}\rangle_{D}\nonumber\\
 & \qquad+\Bigl\{\big(\bigl[S_{F}(\mathbf{x}_{1})t^{c}S_{F}^{\dagger}(\mathbf{x}_{3})\bigr]S_{A}(\mathbf{x}_{2})-\mathbb{1}\big)\bigl(F_{q}^{(2)}+F_{\bar{q}}^{(2)}+F_{C}^{(2)}\bigr)\nonumber\\
& \qquad-\big(\bigl[S_{F}(\mathbf{z})S_{F}^{\dagger}(\mathbf{y})\bigr]-\mathbb{1}\big)\bigl(F_{q}^{(1)}+F_{\bar{q}}^{(1)}\bigr)\Bigr\}|q\bar{q}g\rangle_{D}\;.
\end{align}
The above procedure can be performed explicitly from a closer look at the $|q\bar{q}\rangle\to|q\bar{q}g\rangle$ splittings. It is easy to see that, in mixed Fourier space:
\begin{align}
&|(\mathbf{q})[n^{+},\mathbf{b}]_{\bar{s}}^{l};(\bar{\mathbf{q}})[m^{+},\mathbf{c}]_{\tilde{s}}^{\bar{l}}\rangle_{0}  =|(\mathbf{q})[n^{+},\mathbf{b}]_{\bar{s}}^{l};(\bar{\mathbf{q}})[m^{+},\mathbf{c}]_{\tilde{s}}^{\bar{l}}\rangle_{D}\nonumber\\
 & -g_{s}t_{il}^{c}\int\frac{\mathrm{d}k_{2}^{+}}{2\pi}\int_{\mathbf{x}_{1}\mathbf{x}_{2}}F_{q}^{(1)}\Bigl[(\mathbf{q})[k_{1}^{+}=n^{+}-k_{2}^{+},\mathbf{b}-\mathbf{x}_{1}];(\mathbf{g})[k_{2}^{+},\mathbf{b}-\mathbf{x}_{2}]\Bigr]_{s\bar{s}}^{\eta}\nonumber\\
 & \quad\times|(\mathbf{q})[k_{1}^{+}=n^{+}-k_{2}^{+},\mathbf{x}_{1}]_{s}^{i};(\mathbf{g})[k_{2}^{+},\mathbf{x}_{2}]_{c}^{\eta};(\bar{\mathbf{q}})[k_{3}^{+}=m^{+},\mathbf{x}_{3}=\mathbf{a}]_{s'=\tilde{s}}^{j=\bar{l}}\rangle_{D}\nonumber\\
 & -g_{s}t_{\bar{l}j}^{c}\int\frac{\mathrm{d}k_{2}^{+}}{2\pi}\int_{\mathbf{x}_{2}\mathbf{x}_{3}}F_{\bar{q}}^{(1)}\Bigl[(\mathbf{g})[k_{2}^{+},\mathbf{c}-\mathbf{x}_{2}];(\bar{\mathbf{q}})[k_{3}^{+}=m^{+}-k_{2}^{+},\mathbf{c}-\mathbf{x}_{3}]\Bigr]_{s'\tilde{s}}^{\eta}\nonumber\\
 & \quad\times|(\mathbf{q})[k_{1}^{+}=n^{+},\mathbf{x}_{1}=\mathbf{b}]_{\bar{s}}^{l};(\mathbf{g})[k_{2}^{+},\mathbf{x}_{2}]_{c}^{\lambda\eta};(\bar{\mathbf{q}})[k_{3}^{+}=m^{+}-k_{2}^{+},\mathbf{x}_{3}]_{s'}^{j}\rangle_{D}\;.
\end{align}
We thus obtain for the outgoing state (retaining only the terms that
yield the three outgoing particles we are after):
\begin{align}
|(\boldsymbol{\gamma})[p^+,\mathbf{p}=0]_{\lambda}\rangle_{\mathrm{out}} & =g_{e}g_{s}\int\frac{\mathrm{d}k_{1}^{+}}{2\pi}\frac{\mathrm{d}k_{2}^{+}}{2\pi}\int_{\mathbf{w}\mathbf{x}_{1}\mathbf{x}_{2}\mathbf{x}_{3}}\biggl\{\Bigl(\bigl[S_{F}(\mathbf{x}_{1})t^{d}S_{F}^{\dagger}(\mathbf{x}_{3})\bigr]_{ij}S_{A}(\mathbf{x}_{2})^{dc}-t_{ij}^{c}\Bigr)\nonumber\\
 &\hspace{-5em} \qquad\times\bigl(F_{q}^{(2)}+F_{\bar{q}}^{(2)}+F_{C}^{(2)}\bigr)\Bigl[(\mathbf{q})[k_{1}^{+},\mathbf{w}-\mathbf{x}_{1}];(\mathbf{g})[k_{2}^{+},\mathbf{w}-\mathbf{x}_{2}];(\bar{\mathbf{q}})[k_{3}^{+},\mathbf{w}-\mathbf{x}_{3}]\Bigr]_{s's}^{\eta\lambda}\nonumber\\
 &\hspace{-5em} \qquad-\int_{\mathbf{v}}\Big(\bigl[t^{c}S_{F}(\mathbf{v})S_{F}^{\dagger}(\mathbf{x}_{3})\bigr]_{ij}-t_{ij}^c\Big)F_{\gamma}^{(1)}\Big[(\mathbf{q})[k_{1}^{+}+k_{2}^{+},\mathbf{w}-\mathbf{v}];(\bar{\mathbf{q}})[k_{3}^{+},\mathbf{w}-\mathbf{x}_{3}]\Big]_{\bar{s}s'}^{\lambda}\nonumber\\
 &\hspace{-5em} \qquad\times F_{q}^{(1)}\Bigl[(\mathbf{q})[k_{1}^{+},\mathbf{v}-\mathbf{x}_{1}];(\mathbf{g})[k_{2}^{+},\mathbf{v}-\mathbf{x}_{2}]\Bigr]_{s\bar{s}}^{\eta}\nonumber\\
 &\hspace{-5em} \qquad-\int_{\mathbf{v}}\Bigl(\bigl[S_{F}(\mathbf{x}_{1})S_{F}^{\dagger}(\mathbf{v})t^{c}\bigr]_{ij}-t_{ij}^{c}\Bigr)F_{\gamma}^{(1)}\Big[(\mathbf{q})[k_{1}^{+},\mathbf{w}-\mathbf{x}_{1}];(\bar{\mathbf{q}})[k_{2}^{+}+k_{3}^{+},\mathbf{w}-\mathbf{v}]\Big]_{s\tilde{s}}^{\lambda}\nonumber\\
 &\hspace{-5em} \qquad\times F_{\bar{q}}^{(1)}\Bigl[(\mathbf{g})[k_{2}^{+},\mathbf{v}-\mathbf{x}_{2}];(\bar{\mathbf{q}})[k_{3}^{+},\mathbf{v}-\mathbf{x}_{3}]\Bigr]_{s'\tilde{s}}^{\eta}\biggr\}\nonumber\\
 &\hspace{-5em} \qquad\times|(\mathbf{q})[k_{1}^{+},\mathbf{x}_{1}]_{s}^{i};(\mathbf{g})[k_{2}^{+},\mathbf{x}_{2}]_{c}^{\eta};(\bar{\mathbf{q}})[k_{3}^{+},\mathbf{x}_{3}]_{s'}^{j}\rangle_{D}\;.
\end{align}
Using the results for the Fourier transforms of the splitting functions
Eqs. (\ref{eq:f1Fourier}) and (\ref{eq:f2Fourier}), as well as the definitions \eqref{G_q}, \eqref{G_qbar} and \eqref{G_inst} and the identity:
\begin{align}
\bigl[S_{F}(\mathbf{x}_{1})t^{d}S_{F}^{\dagger}(\mathbf{x}_{3})\bigr]_{ij}S_{A}(\mathbf{x}_{2})^{dc}=\bigl[S_{F}(\mathbf{x}_{1})S_{F}^{\dagger}(\mathbf{x}_{2})t^{c}S_{F}(\mathbf{x}_{2})S_{F}^{\dagger}(\mathbf{x}_{3})\bigr]_{ij}\;,
\end{align}
one arrives at the final expression Eq.~\eqref{eq:fulloutgoingstate}.

\subsection{\label{sec:LCPT}LCPT conventions and Feynman rules}

We follow the conventions of Ref.~\cite{Beuf}, in which the quark
and gluon fields are defined as follows in terms of the creation- and annihilation operators:
\begin{equation}
\begin{aligned}\label{quantumfields}q_i(\vec{x}) & =\sum_{s}\int\frac{\mathrm{d}^{3}\vec{k}}{\left(2\pi\right)^{3}2k{}^{+}}\Bigl[e^{-i\vec{k}\vec{x}}b_{i}^{s}(\vec{k})u^{s}(\vec{k})+e^{i\vec{k}\vec{x}}d_{i}^{\dagger s}(\vec{k})v^{s}(\vec{k})\Bigr]\;,\\
A_c^{\mu}(\vec{x}) & =\sum_{\lambda}\int\frac{\mathrm{d}^{3}\vec{k}}{\left(2\pi\right)^{3}2k^{+}}\Bigl[e^{-i\vec{k}\vec{x}}a_c^{\lambda}(\vec{k})\epsilon_{\lambda}^{\mu}(\vec{k})+e^{i\vec{k}\vec{x}}a_c^{\dagger\lambda}(\vec{k})\epsilon_{\lambda}^{\mu*}(\vec{k})\Bigr]\;,
\end{aligned}
\end{equation}
with the following (anti-)commutation relations:
\begin{equation}
\begin{aligned}\bigl\{ b_{i}^{s}(\vec{k}_{1}),b_{j}^{\dagger s'}(\vec{k}_{2})\bigr\} & =2k_{1}^{+}(2\pi)^{3}\delta^{(3)}(\vec{k}_{1}-\vec{k}_{2})\delta^{ss'}\delta_{ij}\;,\\
\bigl\{ d_{i}^{s}(\vec{k}_{1}),d_{j}^{\dagger s'}(\vec{k}_{2})\bigr\} & =2k_{1}^{+}(2\pi)^{3}\delta^{(3)}(\vec{k}_{1}-\vec{k}_{2})\delta^{ss'}\delta_{ij}\;,\\
\bigl[a^{\lambda}(\vec{k}_{1}),a^{\dagger\lambda'}(\vec{k}_{2})\bigr] & =2k_{1}^{+}(2\pi)^{3}\delta^{(3)}(\vec{k}_{1}-\vec{k}_{2})\delta^{\lambda\lambda'}\;.
\end{aligned}
\end{equation}
As is clear from the above definitions, in our conventions all the creation and annihilation
operators have mass dimension $-1$. Hence, it follows that the for the moment unspecified Dirac spinors
are required to have dimension $1/2$, in order for the quark field to have dimension $3/2$. Following the same reasoning,
the polarization vectors are dimensionless, such that the boson
field has dimension $1$. With these definitions, each $n$-particle
Fock state in momentum space has dimension $-n$.

The above conventions are in contrast with the ones used in our earlier work Ref.~\cite{Altinoluk:2018uax,monster},
where all creation and annihilation operators have dimension $-3/2$, while the quark and gluon fields keep their usual dimensions $3/2$ and $1$.

\begin{figure}[t]
\begin{centering}
\includegraphics[clip,scale=0.33]{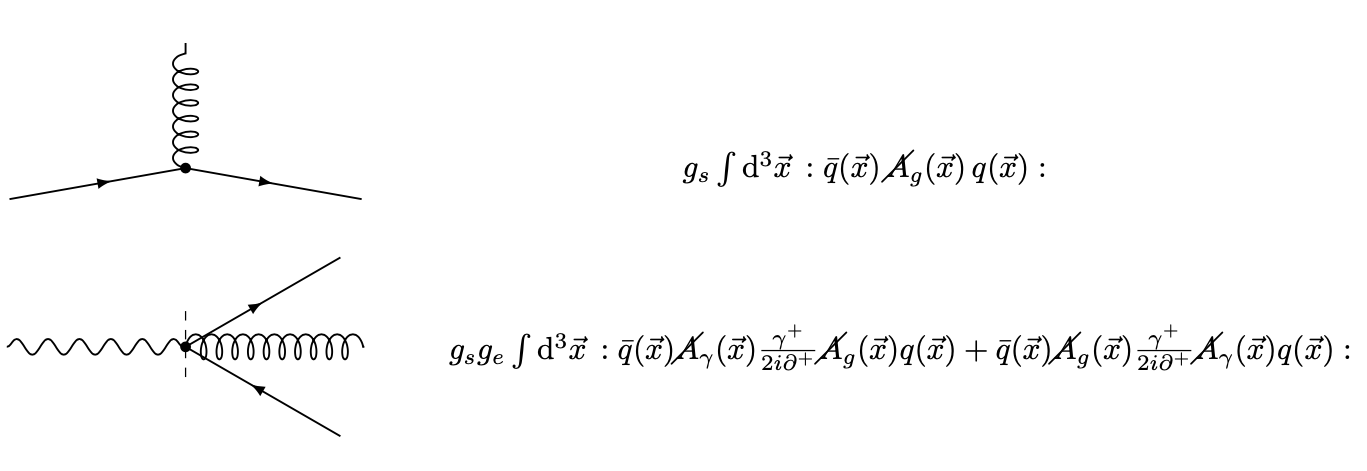}
\par\end{centering}
\caption{\label{fig:Feynman}Relevant LCPT Feynman rules. As in the covariant case, the quark-antiquark-photon vertex is simply obtained from the quark-antiquark-gluon one by replacing $g_s\to g_e$ and $A_g\to A_\gamma$. Note that $:\cal{O}:$ denotes normal ordering, and that we used the notation $\vec{x}=(x^-,\mathbf{x})$.}
\end{figure}

The Feynman rules for LCPT can be found in Refs.~\cite{Beuf,Brodsky:1997de}. The relevant ones for our process are listed in Fig.~\ref{fig:Feynman}, where it is understood that the gluon field is $A_g\equiv A_{g,c}t^c$. The quark-antiquark-photon vertex is then obtained from the quark-antiquark-gluon one by the replacement $g_s\to g_e$ and $A_g\to A_\gamma$.

\subsection{\label{sec:genericsplitting}Generic quark-boson-antiquark splitting
function}

In this subsection, we show how to derive a generic expression for
the quark-boson-antiquark splitting, irregardless of which particle
is the radiator. Without losing generality, let us focus on the structure:
\begin{equation}
\bar{u}(\vec{q}_{2})\cancel{\epsilon}_{\lambda}(\vec{p})v(\vec{q}_{1})\;.\label{eq:gennum0}
\end{equation}
We will decompose the spinors in terms of \emph{good} and \emph{bad}
components, which can be performed by means of the projectors:
\begin{equation}
\mathcal{P}_{B}\equiv\frac{\gamma^{+}\gamma^{-}}{2}=\frac{\gamma^{0}\gamma^{-}}{\sqrt{2}}\quad\mathrm{and}\quad\mathcal{P}_{G}\equiv\frac{\gamma^{-}\gamma^{+}}{2}=\frac{\gamma^{0}\gamma^{+}}{\sqrt{2}}\;.
\end{equation}
In LCPT even intermediate particles follow the classical equations of motion, hence it follows in a straightforward
way from the Dirac equation that the bad components of a spinor depend
on the good ones through the relations (see Ref.~\cite{Beuf}):
\begin{equation}
u_{B}^{s}(\vec{k})=\frac{\gamma^{+}}{2k^{+}}\mathbf{k}\cdot\boldsymbol{\gamma}\,u_{G}^{s}(\vec{k})\quad\mathrm{and}\quad v_{B}^{s}(\vec{k})=\frac{\gamma^{+}}{2k^{+}}\mathbf{k}\cdot\boldsymbol{\gamma}\,v_{G}^{s}(\vec{k})\;,\label{eq:bad1}
\end{equation}
and similarly:
\begin{equation}
\bar{u}_{B}^{s}(\vec{k})=\bar{u}_{G}^{s}(\vec{k})\mathbf{k}\cdot\boldsymbol{\gamma}\frac{\gamma^{+}}{2k^{+}}\quad\mathrm{and}\quad\bar{v}_{B}^{s}(\vec{k})=\bar{v}_{G}^{s}(\vec{k})\mathbf{k}\cdot\boldsymbol{\gamma}\frac{\gamma^{+}}{2k^{+}}\;.\label{eq:bad2}
\end{equation}
With the above relations, as well as the following parameterization
of the polarization vector for the gluon:
\begin{equation}
\epsilon_{\lambda}^{\mu}(\vec{k})=\left(0^{+},\frac{\mathbf{k}\cdot\boldsymbol{\epsilon}^{\lambda}}{k^{+}},\boldsymbol{\epsilon}^{\lambda}\right)\;,
\end{equation}
and using the identities in appendix \ref{sec:UsefulIdentities},
we obtain the following expression for Eq.~\eqref{eq:gennum0}:
\begin{equation}
\begin{aligned}\bar{u}(\vec{q}_{1})\cancel{\epsilon}_{\lambda}(\vec{p})v(\vec{q}_{2}) & =\bigl(\bar{u}_{G}(\vec{q}_{1})+\bar{u}_{B}(\vec{q}_{1})\bigr)\cancel{\epsilon}_{\lambda}(\vec{p})\bigl(v_{G}(\vec{q}_{2})+v_{B}(\vec{q}_{2})\bigr)\;,\\
 & =\epsilon_{\lambda}^{i}\bar{u}_{G}(\vec{q}_{1})\Bigl(1+\mathbf{q}_{1}\cdot\boldsymbol{\gamma}\frac{\gamma^{+}}{2q_{1}^{+}}\Bigr)\Bigl(\frac{p^{i}}{p^{+}}\gamma^{+}-\gamma^{i}\Bigr)\Bigl(1+\frac{\gamma^{+}}{2q_{2}^{+}}\mathbf{q}_{2}\cdot\boldsymbol{\gamma}\Bigr)v_{G}(\vec{q}_{2})\;,\\
 & =\epsilon_{\lambda}^{i}\bar{u}_{G}(\vec{q}_{1})\Bigl(\frac{p^{i}}{p^{+}}\gamma^{+}-\gamma^{i}-\mathbf{q}_{1}\cdot\boldsymbol{\gamma}\frac{\gamma^{+}}{2q_{1}^{+}}\gamma^{i}\Bigr)\Bigl(1+\frac{\gamma^{+}}{2q_{2}^{+}}\mathbf{q}_{2}\cdot\boldsymbol{\gamma}\Bigr)v_{G}(\vec{q}_{2})\;,\\
 & =\epsilon_{\lambda}^{i}\bar{u}_{G}(\vec{q}_{1})\gamma^{+}\Bigl(\frac{p^{i}}{p^{+}}+\mathbf{q}_{1}\cdot\boldsymbol{\gamma}\frac{\gamma^{i}}{2q_{1}^{+}}+\frac{\gamma^{i}}{2q_{2}^{+}}\mathbf{q}_{2}\cdot\boldsymbol{\gamma}\Bigr)v_{G}(\vec{q}_{2})\;.
\end{aligned}
\end{equation}
Furthermore, using the decomposition Eq.~\eqref{eq:gammadecomp}, we obtain the result:
\begin{equation}
\begin{aligned}\bar{u}(\vec{q}_{1})\cancel{\epsilon}_{\lambda}(\vec{p})v(\vec{q}_{2}) & =\epsilon_{\lambda}^{i}\bar{u}_{G}(\vec{q}_{1})\gamma^{+}\Bigl(\frac{p^{i}}{p^{+}}+\frac{q_{1}^{j}}{2q_{1}^{+}}(-\delta^{ji}-i\sigma^{ji})+\frac{q_{2}^{j}}{2q_{2}^{+}}(-\delta^{ij}-i\sigma^{ij})\Bigr)v_{G}(\vec{q}_{2})\;,\\
 & =\epsilon_{\lambda}^{i}\bar{u}_{G}(\vec{q}_{1})\gamma^{+}\Bigl[\Bigl(\frac{p^{i}}{p^{+}}-\frac{q_{1}^{i}}{2q_{1}^{+}}-\frac{q_{2}^{i}}{2q_{2}^{+}}\Bigr)-i\sigma^{ij}\Bigl(\frac{q_{2}^{j}}{2q_{2}^{+}}-\frac{q_{1}^{j}}{2q_{1}^{+}}\Bigr)\Bigr]v_{G}(\vec{q}_{2})\;,
\end{aligned}
\label{eq:BeufGeneral}
\end{equation}
which was obtained earlier in Ref.~\cite{Beuf}. Since in the massless case the good and bad projections (\ref{eq:bad1}) and (\ref{eq:bad2}) are the same for the quark and the antiquark, the above result is general, and one can freely interchange $\bar{u}\to\bar{v}$ or $v\to u$.

For instance, imposing $\vec{p}=\vec{q}_{1}+\vec{q}_{2}$ in Eq.~\eqref{eq:BeufGeneral}, and defining $\xi=q_{1}^{+}/p^{+}$, one obtains the numerator of the $\gamma\to q\bar{q}$ splitting function:
\begin{equation}
\begin{aligned}\bar{u}(\vec{q}_{1})\cancel{\epsilon}_{\lambda}(\vec{p})v(\vec{q}_{2}) & =\frac{p^{+}}{2q_{1}^{+}q_{2}^{+}}\epsilon_{\lambda}^{i}(\xi p^{j}-q_{1}^{j})\bar{u}_{G}(\vec{q}_{1})\gamma^{+}\Bigl[(1-2\xi)\delta^{ij}-i\sigma^{ij}\Bigr]v_{G}(\vec{q}_{2})\;.\end{aligned}
\end{equation}

\subsection{\label{sec:UsefulIdentities}Useful identities}

\paragraph{Gamma matrices}

From the relation:
\begin{equation}
\begin{aligned}\left\{ \gamma^{\mu},\gamma^{\nu}\right\}  & =2g^{\mu\nu}\mathbb{1}_4\;,\end{aligned}
\end{equation}
and the definitions $\gamma^{\pm}=(\gamma^{0}\pm\gamma^{3})/\sqrt{2}$,
one easily obtains the following identities:
\begin{equation}
\begin{aligned}(\gamma^{+})^{2} & =0\;,\\
\left\{ \gamma^{+},\gamma^{i}\right\}  & =0\;,\\
\left\{ \gamma^{+},\gamma^{0}\right\}  & =\sqrt{2}\;.
\end{aligned}
\end{equation}
Moreover, any transverse gamma matrix $\gamma^{i}$ sandwished between
`good' Dirac spinors $u_{G}=\mathcal{P}_{G}u$ or $v_{G}=\mathcal{P}_{G}v$
disappears:
\begin{equation}
\bar{u}_{G}^{s}(\vec{k})\gamma^{i}u_{G}^{s'}(\vec{p})=\bar{v}_{G}^{s}(\vec{k})\gamma^{i}v_{G}^{s'}(\vec{p})=\bar{v}_{G}^{s}(\vec{k})\gamma^{i}u_{G}^{s'}(\vec{p})=\bar{u}_{G}^{s}(\vec{k})\gamma^{i}v_{G}^{s'}(\vec{p})=0\;.
\end{equation}
The following decomposition will often be useful:
\begin{equation}
\begin{aligned}\gamma^{i}\gamma^{j} & =\frac{1}{2}\{\gamma^{i},\gamma^{j}\}+\frac{1}{2}[\gamma^{i},\gamma^{j}]\;,\\
 & =-\delta^{ij}-i\sigma^{ij}\;,\label{eq:gammadecomp}
\end{aligned}
\end{equation}
where we defined $\sigma^{ij}=(i/2)[\gamma^{i},\gamma^{j}]$.

\paragraph{Dirac sigma}

The Dirac sigma is Hermitian $\left(\sigma^{ij}\right)^{\dagger}=\sigma^{ij}$,
and commutes with the non-transverse gamma matrices: $\left[\sigma^{ij},\gamma^{0}\right]=\left[\sigma^{ij},\gamma^{3}\right]=0$,
and by extension $\left[\sigma^{ij},\gamma^{+}\right]=\left[\sigma^{ij},\mathcal{P}_{G}\right]=0$.
Moreover, we have that:
\begin{equation}
\begin{aligned}\sigma^{il}\sigma^{jl} & =\delta^{ij}\;,\end{aligned}
\end{equation}
and 
\begin{equation}
\begin{aligned}\mathrm{Tr}(\mathcal{P}_{G}\sigma^{ij}) & =0\;,\\
\mathrm{Tr}(\mathcal{P}_{G}\sigma^{ij}\sigma^{kl}) & =2\epsilon^{ij}\epsilon^{kl}\;.
\end{aligned}
\end{equation}

\subsection{\label{sec:explicit}Explicit spinor representation}
One can simplify the wave functions considerably by choosing an explicit parameterization of the Dirac spinors. We follow the so-called `Kogut-Soper'
conventions \cite{Kogut:1969xa,Brodsky:1997de}, where the gauge $A^+ =0$ is chosen, such that the transverse polarization vectors of the gluon can be written as:
\begin{equation}
\epsilon_{\lambda}^{\mu}(\vec{k})=\Bigl(0,\frac{\mathbf{k}\cdot\boldsymbol{\epsilon}^{\lambda}}{k^{+}},\boldsymbol{\epsilon}^{\lambda}\Bigr)\;,
\end{equation}
and the longitudinal polarization vector (with $Q^{2}=k^{2}$):
\begin{equation}
\epsilon_{L}^{\mu}(\vec{k})=\Bigl(0,\frac{Q}{k^{+}},0\Bigr)\;.
\end{equation}
However, in contrast to Kogut and Soper, we choose linear polarization
vectors: 
\begin{equation}
\epsilon_{\lambda}^{i}=\delta^{i\lambda}\;,
\end{equation}
with $\epsilon_{\lambda}^{i\dagger}\epsilon_{\lambda}^{j}=\delta^{ij}$.
We work in the chiral representation of the gamma matrices:
\begin{equation}
\gamma^{0}=\begin{pmatrix}0 & \mathbb{1}_2\\
\mathbb{1}_2 & 0
\end{pmatrix}\;,\quad\gamma^{i}=\begin{pmatrix}0 & -\sigma^{i}\\
\sigma^{i} & 0
\end{pmatrix}\;,
\end{equation}
with the usual Pauli matrices:
\begin{align}
\sigma^{1} & =\begin{pmatrix}0 & 1\\
1 & 0
\end{pmatrix}\;,\qquad\sigma^{2}=\begin{pmatrix}0 & -i\\
i & 0
\end{pmatrix}\;,\qquad\sigma^{3}=\begin{pmatrix}1 & 0\\
0 & -1
\end{pmatrix}\;,
\end{align}
such that:
\begin{align}
\gamma^{+} & =\begin{pmatrix}0 & 0 & 0 & 0\\
0 & 0 & 0 & \sqrt{2}\\
\sqrt{2} & 0 & 0 & 0\\
0 & 0 & 0 & 0
\end{pmatrix}\;,\quad\gamma^{1}=\begin{pmatrix}0 & 0 & 0 & -1\\
0 & 0 & -1 & 0\\
0 & 1 & 0 & 0\\
1 & 0 & 0 & 0
\end{pmatrix}\;,\quad\gamma^{2}=\begin{pmatrix}0 & 0 & 0 & i\\
0 & 0 & -i & 0\\
0 & -i & 0 & 0\\
i & 0 & 0 & 0
\end{pmatrix}\;.
\end{align}
The Dirac sigma is explicitly:
\begin{equation}
\begin{aligned}\sigma^{ij} & =\frac{i}{2}\left[\gamma^{i},\gamma^{j}\right]\;,\\
\sigma^{12} & =\begin{pmatrix}1 & 0 & 0 & 0\\
0 & -1 & 0 & 0\\
0 & 0 & 1 & 0\\
0 & 0 & 0 & -1
\end{pmatrix}\;,\quad\sigma^{21}=-\sigma^{12}\;,\quad\sigma^{11}=\sigma^{22}=0\;.
\end{aligned}
\end{equation}
With these conventions, the Dirac spinors are given by ($1$ corresponds
to spin up, $-1$ to spin down):
\begin{equation}
u^{1}(\vec{k})=\frac{1}{2^{1/4}\sqrt{k^{+}}}\left(\begin{array}{c}
\sqrt{2}k^{+}\\
k_{1}+ik_{2}\\
m\\
0
\end{array}\right)\;,\qquad u^{-1}(\vec{k})=\frac{1}{2^{1/4}\sqrt{k^{+}}}\left(\begin{array}{c}
0\\
m\\
-k_{1}+ik_{2}\\
\sqrt{2}k^{+}
\end{array}\right)\;,
\end{equation}
and
\begin{equation}
v^{1}(\vec{k})=\frac{1}{2^{1/4}\sqrt{k^{+}}}\left(\begin{array}{c}
0\\
-m\\
-k_{1}+ik_{2}\\
\sqrt{2}k^{+}
\end{array}\right)\;,\qquad v^{-1}(\vec{k})=\frac{1}{2^{1/4}\sqrt{k^{+}}}\left(\begin{array}{c}
\sqrt{2}k^{+}\\
k_{1}+ik_{2}\\
-m\\
0
\end{array}\right)\;.
\end{equation}
From the above expressions and in the massless case, it is very easy to obtain the good spinors
$u_{G}^{s}(\vec{k})=\mathcal{P}_{G}u^{s}(\vec{k})$ and $v_{G}^{s}(\vec{k})=\mathcal{P}_{G}v^{s}(\vec{k})$:
\begin{equation}
u_{G}^{1}(\vec{k})=2^{1/4}\sqrt{k^{+}}\left(\begin{array}{c}
1\\
0\\
0\\
0
\end{array}\right)\;,\qquad u_{G}^{-1}(\vec{k})=2^{1/4}\sqrt{k^{+}}\left(\begin{array}{c}
0\\
0\\
0\\
1
\end{array}\right)\;,
\end{equation}
and
\begin{equation}
v_{G}^{1}(\vec{k})=u_{G}^{-1}(\vec{k})=2^{1/4}\sqrt{k^{+}}\left(\begin{array}{c}
0\\
0\\
0\\
1
\end{array}\right)\;,\qquad v_{G}^{-1}(\vec{k})=u_{G}^{1}(\vec{k})=2^{1/4}\sqrt{k^{+}}\left(\begin{array}{c}
1\\
0\\
0\\
0
\end{array}\right)\;.
\end{equation}
The complex conjugates are:
\begin{equation}
\bar{u}_{G}^{1}(\vec{k})=2^{1/4}\sqrt{k^{+}}\begin{pmatrix}0, & 0, & 1, & 0\end{pmatrix}\;,\qquad\bar{u}_{G}^{-1}(\vec{k})=2^{1/4}\sqrt{k^{+}}\begin{pmatrix}0, & 1, & 0, & 0\end{pmatrix}\;,
\end{equation}
and
\begin{equation}
\bar{v}_{G}^{1}(\vec{k})=\bar{u}_{G}^{-1}(\vec{k})=2^{1/4}\sqrt{k^{+}}\begin{pmatrix}0, & 1, & 0, & 0\end{pmatrix}\;,\qquad\bar{v}_{G}^{-1}(\vec{k})=\bar{u}_{G}^{1}(\vec{k})=2^{1/4}\sqrt{k^{+}}\begin{pmatrix}0, & 0, & 1, & 0\end{pmatrix}\;.
\end{equation}
Useful identities in the computation of the squared splitting function
are:
\begin{equation}
\sigma_{\alpha\beta}^{3}\sigma_{\gamma\beta}^{3}=\epsilon^{\alpha\beta}\epsilon^{\gamma\beta}=\delta_{\alpha\gamma}\;,\label{eq:sigmaepsilon}
\end{equation}
\begin{equation}
\sigma_{-\alpha,-\beta}^{3}=-\sigma_{\alpha\beta}^{3}\quad\mathrm{and}\quad\sigma_{\alpha\alpha}^{3}=0\;.
\end{equation}

\subsection{\label{app:unintegrated}Weizsäcker-Williams gluon TMD and unintegrated gluon distribution}
We define the gluon density inside a left-moving proton or nucleus
as the gluon number operator evaluated in the target's nonperturbative
Fock state \cite{IancuReport}:
\begin{equation}
\frac{\mathrm{d}N_{g}}{\mathrm{d}\,\mathrm{ln}(1/x_{\scriptscriptstyle{A}})\mathrm{d}^{2}\mathbf{k}}\equiv\frac{1}{2(2\pi)^3}\langle a_{c}^{\dagger\lambda}(\vec{k})a_{c}^{\lambda}(\vec{k})\rangle_{x_{\scriptscriptstyle{A}}}\;,
\end{equation}
where $x_{\scriptscriptstyle{A}}=k^{-}/p_{A}^{-}$ is the fraction of the target's longitudinal
momentum carried by the gluon. For the present discussion we briefly switch to the target light-cone gauge $\tilde{A}^-=0$, which is more convenient for our purposes. Fields in this gauge will be denoted by adding a tilde. From the Fourier expansion of the gluon quantum field, keeping in mind that the integration over the longitudinal momentum
component is restricted to $k^{-}>0$, we easily obtain that:
\begin{equation}
\tilde{A}_{c}^{i}(\vec{k})=\frac{1}{2k^{-}}a_{c}^{i}(\vec{k}),\qquad\mathrm{and}\qquad \tilde{A}_{c}^{i}(-\vec{k})=\frac{1}{2k^{-}}a_{c}^{i\dagger}(\vec{k})\;,
\end{equation}
where we chose linear polarization vectors $\epsilon_{\lambda}^{i}=\delta_{\lambda}^{i}$. The color field strengths in the present gauge are equal to $\tilde{F}_c^{i-}(\vec{k})=-ik^{-}\tilde{A}_c^{i}(\vec{k})$,
such that:
\begin{align}
\frac{\mathrm{d}N_{g}}{\mathrm{d}\,\mathrm{ln}(1/x_{\scriptscriptstyle{A}})\mathrm{d}^{2}\mathbf{k}}\equiv \frac{1}{4\pi^3}\big\langle \tilde{F}_c^{i-}(-\vec{k})\tilde{F}_c^{i-}(\vec{k})\big\rangle_{x_{\scriptscriptstyle{A}}}\;.\label{eq:gluondensity}
\end{align}

\paragraph{Weizsäcker-Williams gluon TMD}

The gauge-invariant operator definition of the Weizsäcker-Williams gluon TMD is
\begin{align}
\mathcal{F}_{WW}(x_{\scriptscriptstyle{A}},\mathbf{k}) & =\frac{2}{p_{A}^{-}}\int\frac{\mathrm{d}\xi^{+}\mathrm{d}^{2}\boldsymbol{\xi}}{(2\pi)^3}e^{-ik^{-}\xi^{+}}e^{i\mathbf{k}\cdot\boldsymbol{\xi}}\nonumber\\
&\mathrm{Tr}\,\langle P|F^{i+}(\xi^{+},\boldsymbol{\xi})U^{[+]}(\xi^{+},\boldsymbol{\xi})F^{i-}(0^{+},\mathbf{0})U^{[+]\dagger}(\xi^{+},\boldsymbol{\xi})|P\rangle\;,
\end{align}
where the color field strengths are evaluated in the proton states
$|P\rangle$, and where a gauge-invariant definition is guaranteed
by the inclusion of the following gauge links:
\begin{equation}
U^{[+]}(\xi^{+},\boldsymbol{\xi})=U(0^{+},\mathbf{0};+\infty,\mathbf{0})U(+\infty,\mathbf{0},+\infty,\boldsymbol{\xi})U(+\infty,\boldsymbol{\xi};\xi^{+},\boldsymbol{\xi})\;.
\end{equation}
These links can be eliminated by choosing the target light-cone gauge\footnote{The WW distribution is the only gluon TMD for which \emph{all} gauge
links can be eliminated simultaneously, and for which therefore the interpretation
of counting gluon states holds.} $\tilde{A}^{-}=0$, with the additional condition that
$\tilde A^{i}(\xi^{+}=+\infty)=0$, such that,
\begin{equation}
\mathcal{F}_{WW}(x_{\scriptscriptstyle{A}},\mathbf{k})=\frac{1}{p_{A}^{-}}\int\frac{\mathrm{d}\xi^{+}\mathrm{d}^{2}\boldsymbol{\xi}}{(2\pi)^3}e^{-ik^{-}\xi^{+}}e^{i\mathbf{k}\cdot\boldsymbol{\xi}}\langle P|\tilde F_{c}^{i-}(\xi^{+},\boldsymbol{\xi})\tilde F_{c}^{i-}(0^{+},\mathbf{0})|P\rangle\;,\label{eq:WWLC}
\end{equation}
where we also evaluated the trace, using $\mathrm{Tr}(t^a t^b) = \delta^{ab}/2$. Furthermore, making use of translational invariance and Fourier transforming
the field strengths (we use the notation $\vec{v}=(v^+,\mathbf{v})$\,):
\begin{align}
\mathcal{F}_{WW}(x_{\scriptscriptstyle{A}},\mathbf{k}) & =\frac{1}{p_{A}^{-}}\frac{1}{(2\pi)^{3}\delta^{(3)}(0)}\int\frac{\mathrm{d}^3\vec{v}\mathrm{d}^3\vec{w}}{(2\pi)^3}e^{-i\vec{k}\cdot(\vec{v}-\vec{w})}\langle P|\tilde F_{c}^{i-}(\vec{v})\tilde F_{c}^{i-}(\vec{w})|P\rangle\;,\nonumber\\
 & =\frac{1}{p_{A}^{-}}\frac{1}{(2\pi)^{6}\delta^{(3)}(0)}\langle P|\tilde F_{c}^{i-}(-\vec{k})\tilde F_{c}^{i-}(\vec{k})|P\rangle\;.
\end{align}
Finally, the Fock states of the nuclear target differ from the usual
hadronic states by a normalization:
\begin{equation}
\langle\mathcal{O}\rangle_{x_{\scriptscriptstyle{A}}}=\frac{\langle P|\mathcal{O}|P\rangle}{\langle P|P\rangle},\qquad\mathrm{with}\qquad\langle P|P\rangle=(2\pi)^{3}2p_{A}^{-}\delta^{(3)}(0)\;,
\end{equation}
such that:
\begin{align}
\mathcal{F}_{WW}(x_{\scriptscriptstyle{A}},\mathbf{k}) & =\frac{1}{4\pi^3}\langle \tilde F_{c}^{i-}(-\vec{k})\tilde F_{c}^{i-}(\vec{k})\rangle_{x_{\scriptscriptstyle{A}}}=\frac{\mathrm{d}N_{g}}{\mathrm{d}\,\mathrm{ln}(1/x_{\scriptscriptstyle{A}})\mathrm{d}^{2}\mathbf{k}}\;.
\end{align}
Therefore, in a judicious choice of gauge, the Weizs\"acker-Williams gluon TMD can be interpreted directly as the number operator of gluons in Fock space.

\paragraph{Unintegrated gluon distribution}
The unintegrated gluon distribution $\mathcal{F}_{g/A}(x_{\scriptscriptstyle{A}},\mathbf{k})$, on the other
hand, is commonly defined via the dipole cross section in the dilute
limit of the CGC, to which it is linearly related \cite{Kutak:2004ym}:
\begin{equation}
\sigma_{\mathrm{dip}}(x_{\scriptscriptstyle{A}},\mathbf{r})=\frac{4\pi\alpha_{s}}{N_{c}}\int\frac{\mathrm{d}^{2}\mathbf{k}}{\mathbf{k}^{2}}\mathcal{F}_{g/A}(x_{\scriptscriptstyle{A}},\mathbf{k})\big(1-e^{i\mathbf{k}\cdot\mathbf{r}}\big)\;.\label{eq:dipolecrosssection}
\end{equation}
To compute the dipole cross section in the dilute limit, we need to expand the Wilson lines to second order in the semiclassical gauge fields $\alpha^-$ (we now go back to the projectile light-cone gauge $A^+=0$ used throughout the paper):
\begin{equation}
S_{F}(\mathbf{x})\simeq1+ig_{s}\int\mathrm{d}z^{+}\alpha_{a}^{-}(z^{+},\mathbf{x})t^{a}+\frac{(ig_{s})^{2}}{2!}\mathcal{P}\int\mathrm{d}z_{1}^{+}\mathrm{d}z_{2}^{+}\alpha_{a}^{-}(z_{1}^{+},\mathbf{x})t^{a}\alpha_{b}^{-}(z_{2}^{+},\mathbf{x})t^{b}\;.
\end{equation}
Note that, to the present accuracy the path ordering does not matter
(since $\mathrm{Tr}(t^{a}t^{b})=\mathrm{Tr}(t^{b}t^{a})=\delta^{ab}/2$),
and there is thus a strict separation between transverse and longitudinal
dynamics. With this expansion, the dipole cross section reads: 
\begin{align}\sigma_{\mathrm{dip}}(x_{\scriptscriptstyle{A}},\mathbf{r}) & =2S_{\perp}\big\langle1-\frac{1}{N_{c}}\mathrm{Tr}\,S_{F}(\mathbf{r})S_{F}^{\dagger}(\mathbf{0})\,\big\rangle_{x_{\scriptscriptstyle{A}}}\;,\\
 & =\frac{2g_{s}^{2}S_{\perp}}{N_{c}}\int\mathrm{d}v^{+}\mathrm{d}w^{+} \mathrm{Tr}\,\big\langle \alpha_{}^{-}(v^{+},\mathbf{r})\alpha_{}^{-}(w^{+},\mathbf{r})-\alpha_{}^{-}(v^{+},\mathbf{r})\alpha_{}^{-}(w^{+},\mathbf{0})\big\rangle_{x_{\scriptscriptstyle{A}}}\;,\nonumber\\
 & =\frac{2g_{s}^{2}}{N_{c}}\int\mathrm{d}^3\vec{v}\mathrm{d}^3\vec{w}\big(\delta^{(2)}(\mathbf{\mathbf{v}}-\mathbf{w})-\delta^{(2)}(\mathbf{\mathbf{v}}-\mathbf{w}-\mathbf{r})\big) \mathrm{Tr}\,\big\langle \alpha_{}^{-}(\vec{v})\alpha_{}^{-}(\vec{w})\big\rangle_{x_{\scriptscriptstyle{A}}}\;,\nonumber\\
 & =\frac{2g_{s}^{2}}{N_{c}}\int\mathrm{d}^3\vec{v}\mathrm{d}^3\vec{w} \int\frac{\mathrm{d}^{2}\mathbf{k}}{(2\pi)^{2}}\big(e^{-i\mathbf{k}\cdot(\mathbf{\mathbf{v}}-\mathbf{w})}-e^{-i\mathbf{k}\cdot(\mathbf{\mathbf{v}}-\mathbf{w}-\mathbf{r})}\big)\mathrm{Tr}\,\big\langle \alpha_{}^{-}(\vec{v})\alpha_{}^{-}(\vec{w})\big\rangle_{x_{\scriptscriptstyle{A}}}\;,\nonumber
\end{align}
which is equal to Eq.~\eqref{eq:dipolecrosssection} when we define:
\begin{align}
\mathcal{F}_{g/A}(x_{\scriptscriptstyle{A}},\mathbf{k})&=\frac{\mathbf{k}^{2}}{2\pi^2}\int\mathrm{d}^3\vec{v}\mathrm{d}^3\vec{w}\,e^{-i\mathbf{k}\cdot(\mathbf{\mathbf{v}}-\mathbf{w})}\mathrm{Tr}\,\big\langle \alpha_{}^{-}(\vec v)\alpha_{}^{-}(\vec w)\big\rangle_{x_{\scriptscriptstyle{A}}}\;,\label{eq:uPDF} \\
&\overset{\mathrm{P.I.}}{=}\frac{1}{2\pi^2}\int\mathrm{d}^3\vec{v}\mathrm{d}^3\vec{w}\,e^{-i\mathbf{k}\cdot(\mathbf{\mathbf{v}}-\mathbf{w})}\mathrm{Tr}\,\big\langle \partial^i \alpha_{}^{-}(\vec v)\partial^i \alpha_{}^{-}(\vec w)\big\rangle_{x_{\scriptscriptstyle{A}}}\;,\nonumber\\
&=\frac{1}{2\pi^2}\int\mathrm{d}^3\vec{v}\mathrm{d}^3\vec{w}\,e^{-i\mathbf{k}\cdot(\mathbf{\mathbf{v}}-\mathbf{w})}\mathrm{Tr}\,\big\langle F^{i-}(\vec v) F^{i-}(\vec w)\big\rangle_{x_{\scriptscriptstyle{A}}}\;.\nonumber
\end{align}
Restoring the factor $e^{-ik^{-}(v^{+}-w^{+})}=e^{-i x_{\scriptscriptstyle{A}}p_{\scriptscriptstyle{A}}^{-}(v^{+}-w^{+})}\simeq1$, performing a gauge transformation to the target light-cone gauge, we finally obtain:
\begin{align}
\mathcal{F}_{g/A}(x_{\scriptscriptstyle{A}},\mathbf{k}) & =\frac{1}{(2\pi)^{2}}\big\langle \tilde F_{a}^{i-}(-\vec{k})\tilde F_{a}^{i-}(\vec{k})\big\rangle_{x_{\scriptscriptstyle{A}}}\;,\nonumber\\
 & =\pi\frac{\mathrm{d}N_{g}}{\mathrm{d}\,\mathrm{ln}(1/x_{\scriptscriptstyle{A}})\mathrm{d}^{2}\mathbf{k}}\;.
\end{align}
Hence, in the dilute limit at low-$x$ where Eq.~\eqref{eq:dipolecrosssection} holds, the unintegrated gluon PDF and the Weizsäcker-Williams gluon TMD are equal up to a factor $\pi$ \cite{Collinsrant}. In the appropriate gauge, both can be shown to be equal (up to the same factor $\pi$) to the gluon number counting operator evaluated in the CGC average.


\begin{thebibliography}{99}
\bibitem{bfkl} 
  E.~A.~Kuraev, L.~N.~Lipatov and V.~S.~Fadin,
  Sov.\ Phys.\ JETP {\bf 45} (1977) 199 
  [Zh.\ Eksp.\ Teor.\ Fiz.\  {\bf 72} (1977) 377];
  I.~I.~Balitsky and L.~N.~Lipatov,
  Sov.\ J.\ Nucl.\ Phys.\  {\bf 28} (1978) 822 
  [Yad.\ Fiz.\  {\bf 28} (1978) 1597].
 
 \bibitem{balitsky} I. Balitsky,
{ Nucl. Phys.}  B {\bf 463} (1996) 99;
{ Phys. Rev. Lett.} {\bf 81} (1998) 2024;
{Phys. Rev.} D {\bf 60} (1999) 014020.

 \bibitem{kovchegov}
  Y.~V.~Kovchegov,
  Phys.\ Rev.\ D {\bf 60} (1999) 034008;
  Phys.\ Rev.\ D {\bf 61} (2000) 074018.
 
\bibitem{JIMWLK} J. Jalilian Marian, A. Kovner, A. Leonidov and H.
Weigert,
{ Nucl. Phys.} B {\bf  504} (1997) 415;
%
{ Phys. Rev.}  D {\bf 59} (1999) 014014;
J. Jalilian Marian, A. Kovner and H. Weigert,
{Phys. Rev.} D {\bf 59} (1999) 014015;
 A. Kovner and J.G. Milhano,
{ Phys. Rev.} D {\bf 61} (2000) 014012;
A. Kovner, J.G. Milhano and H. Weigert,
{Phys. Rev.} D {\bf 62} (2000) 114005 ;
 H. Weigert,
  { Nucl. Phys.} A {\bf  703} (2002) 823.

\bibitem{cgc}  E. Iancu, A. Leonidov and L. McLerran,
{Nucl. Phys.} A {\bf  692} (2001) 583;
{Phys. Lett.} B {\bf  510} (2001) 133;
 E. Ferreiro, E. Iancu, A. Leonidov, L. McLerran,
{Nucl. Phys.} A {\bf 703} (2002) 489. 

\bibitem{firstlowxTMDs}
  F.~Dominguez, B.~W.~Xiao and F.~Yuan,
  Phys.\ Rev.\ Lett.\  {\bf 106} (2011) 022301;
  F.~Dominguez, C.~Marquet, B.~W.~Xiao and F.~Yuan,
  Phys.\ Rev.\ D {\bf 83} (2011) 105005;

\bibitem{gluonTMDs}
P.~J.~Mulders and J.~Rodrigues, 
Phys.~Rev.~D~{\bf 63} (2001) 094021;
S.~Meissner, A.~Metz, and K.~Goeke, Phys.~Rev.~D~{\bf 76} (2007) 034002.

\bibitem{ITMD} 
  P.~Kotko, K.~Kutak, C.~Marquet, E.~Petreska, S.~Sapeta and A.~van Hameren,
  JHEP {\bf 1509} (2015) 106;
  A.~van Hameren, P.~Kotko, K.~Kutak, C.~Marquet, E.~Petreska and S.~Sapeta,
  JHEP {\bf 1612} (2016) 034.

\bibitem{proofITMD}
T.~Altinoluk, R.~Boussarie and P.~Kotko, JHEP {\bf1905} (2019) 156;
T.~Altinoluk and R.~Boussarie, JHEP {\bf1910} (2019) 208. 

  \bibitem{Adrian2015}
  A.~Dumitru, T.~Lappi and V.~Skokov,
  Phys.\ Rev.\ Lett.\  {\bf 115} (2015) 252301.  
  
   \bibitem{Marquet:2016cgx}
  C.~Marquet, E.~Petreska and C.~Roiesnel,
  JHEP {\bf 1610} (2016) 065.

\bibitem{Marquet:2017xwy}
  C.~Marquet, C.~Roiesnel and P.~Taels,
  Phys.\ Rev.\ D {\bf 97} (2018) 014004.
\bibitem{MV}
L.~D.~McLerran and R.~Venugopalan, 
Phys.\ Rev.\ D.\ {\bf49} (1994) 2233; Phys.\ Rev.\ D.\ {\bf49} (1994) 3352; Phys.\ Rev.\ D.\ {\bf50} (1994) 2225.  

\bibitem{TMDevolution}
J.~C.~Collins, D.~E.~Soper and G.~F.~Sterman, 
Nucl.~Phys.~B~{\bf250} (1985) 199;
M.~G.~Echevarria, A.~Idilbi and I.~Scimemi, JHEP \textbf{07} (2012) 002;
J.~C. Collins, Cambridge University Press, 2013.

\bibitem{lowxSudakov}
A.~H.~Mueller, B.~W.~Xiao and F.~Yuan,
Phys.~Rev.~Lett.~{\bf110} (2013) 082301; Phys.~Rev.~D~{\bf88} (2013) 114010.

\bibitem{ShuYi}
A.~Stasto, S.~Y.~Wei, B.~W.~Xiao and F.~Yuan, Phys.~Lett.~B~{\bf784} (2018) 301;
C.~Marquet, S.~Y.~Wei and B.~W.~Xiao, arXiv:1909.08572 [hep-ph].

\bibitem{massivedijet}
 A.~Metz and J.~Zhou,
  Phys.\ Rev.\ D {\bf 84} (2011) 051503;
  F.~Dominguez, J.~W.~Qiu, B.~W.~Xiao and F.~Yuan,
  Phys.\ Rev.\ D {\bf 85} (2012) 045003;
  E.~Akcakaya, A.~Schäfer and J.~Zhou, Phys.~Rev.~D~{\bf87} (2013) 054010.
  
\bibitem{Petreska:2018cbf}
E. Petreska, Int. J. Mod. Phys. E \textbf{27} (2018) 830003.
 
\bibitem{monster}
T.~Altinoluk, R.~Boussarie, C.~Marquet and P.~Taels, JHEP \textbf{1907} (2019) 079. 

\bibitem{Bury:2018kvg}
M.~Bury, P.~Kotko and K.~Kutak,
Eur.~Phys.~J.~C~{\bf79} (2019) 152.
  
\bibitem{Altinoluk:2018uax}
T.~Altinoluk, N.~Armesto, A.~Kovner, M.~Lublinsky and E.~Petreska,
JHEP {\bf1804} (2018) 063.
  
\bibitem{WW}
S.~Frixione, M.~L.~Mangano, P.~Nason and G.~Ridolfi,
Phys.\ Lett.\ B {\bf 319} (1993) 339.

\bibitem{Boer:2016fqd}
D.~Boer, P.~J.~Mulders, C.~Pisano and J.~Zhou,
JHEP {\bf0816}, 001 (2016).

\bibitem{Ayala:2016lhd}
A.~Ayala, M.~Hentschinski, J.~Jalilian-Marian and M.~E.~Tejeda-Yeomans,
Phys.~Lett.~B~{\bf761} (2016) 229.

\bibitem{Beuf}
G.~Beuf, Phys.~Rev.~D~{\bf94} (2016) 054016; Phys.~Rev.~D~{\bf 96} (2017) 074033.

\bibitem{Catani:1990eg}
S.~Catani, M.~Ciafaloni and F.~Hautmann,
Nucl. Phys. B \textbf{366} (1991) 135.

\bibitem{Bomhof:2006dp}
C.~Bomhof, P.~Mulders and F.~Pijlman,
Eur. Phys. J. C \textbf{47} (2006) 147.


\bibitem{Bertulani:2005ru}
C.~A.~Bertulani, S.~R.~Klein and J.~Nystrand,
Ann. Rev. Nucl. Part. Sci. {\bf 55} (2005) 271.

\bibitem{Kogut:1969xa}
J.~B.~Kogut and D.~E.~Soper, Phys.~Rev.~D~{\bf 1} (1970) 2901;
J.~D.~Bjorken, J.~B.~Kogut and D.~E.~Soper,
Phys.~Rev.~D~{\bf3} (1971) 1382.

\bibitem{Brodsky:1997de}
S.~J.~Brodsky, H.~Pauli, and S.~S.~Pinsky,
Phys.~Rept.~{\bf 301} (1998) 299.

\bibitem{IancuReport}
E.~Iancu, Report No. INSPIRE-1494642, 2005

\bibitem{Kutak:2004ym}
K.~Kutak and A.~M.~Stasto, 
Eur.~Phys.~J.~C~{\bf41} (2005) 343.

\bibitem{Collinsrant}
E.~Avsar and J.~C.~Collins, arXiv:1209.1675 [hep-ph].

\bibitem{Iancu:2018aa}
E.~Iancu and Y.~Mulian, 
Nucl.~Phys.~A~{\bf985} (2019) 66.
 

\bibitem{Benic:2016uku} 
  S.~Beni\'c, K.~Fukushima, O.~Garcia-Montero and R.~Venugopalan,
  JHEP {\bf 1701} (2017) 115.
  
 \bibitem{Benic:2017znu}
  S.~Beni\'c and A.~Dumitru,
  Phys.\ Rev.\ D {\bf 97} (2018) 014012.

\bibitem{Benic:2018hvb} 
  S.~Beni\'c, K.~Fukushima, O.~Garcia-Montero and R.~Venugopalan,
 Phys.~Lett.~B~{\bf791} (2019) 11.

\bibitem{Roy:2018jxq} 
  K.~Roy and R.~Venugopalan,
  JHEP {\bf 1805} (2018) 013; 
Phys. Rev. D \textbf{101} (2020) 034028.





\end{thebibliography}
\end{document}